\documentclass[11pt, a4paper, oneside]{Thesis} 

\graphicspath{{Pictures/}} 

\usepackage[square,sort,compress,numbers]{natbib}
\usepackage{cite}
\usepackage{amsmath,latexsym} 
\usepackage{float}

\title{\ttitle} 
\DeclareUnicodeCharacter{2212}{-}
\DeclareUnicodeCharacter{2212}{+}
\begin{document}

\setstretch{1.3} 

\fancyhead{} 
\rhead{\thepage} 
\lhead{} 

%

\thesistitle{Accelerating Expansion of the Universe in Modified Gravity with Quadratic Term}
\documenttype{\bf{THESIS}}
\supervisor{\bf{Prof. Pradyumn Kumar Sahoo}}
\supervisorposition{Professor}
\supervisorinstitute{BITS-Pilani, Hyderabad Campus}
\examiner{}
\degree{Ph.D. Research Scholar}
\coursecode{\textbf{DOCTOR OF PHILOSOPHY}}
\coursename{THESIS}
\authors{\textbf{AAQID MOHI UD DIN BHAT}}
\IDNumber{2021PHXF0456H}
\addresses{}
\subject{}
\keywords{}
\university{\texorpdfstring{\href{http://www.bits-pilani.ac.in/} 
                {Birla Institute of Technology and Science, Pilani}} 
                {Birla Institute of Technology and Science, Pilani}}
\UNIVERSITY{\texorpdfstring{\href{http://www.bits-pilani.ac.in/} 
                {\textbf{BIRLA INSTITUTE OF TECHNOLOGY AND SCIENCE, PILANI}}} 
                {BIRLA INSTITUTE OF TECHNOLOGY AND SCIENCE, PILANI}}



\department{\texorpdfstring{\href{http://www.bits-pilani.ac.in/pilani/Mathematics/Mathematics} 
                {Mathematics}} 
                {Mathematics}}
\DEPARTMENT{\texorpdfstring{\href{http://www.bits-pilani.ac.in/pilani/Mathematics/Mathematics} 
                {Mathematics}} 
                {Mathematics}}
\group{\texorpdfstring{\href{Research Group Web Site URL Here (include http://)}
                {Research Group Name}} 
                {Research Group Name}}
\GROUP{\texorpdfstring{\href{Research Group Web Site URL Here (include http://)}
                {RESEARCH GROUP NAME (IN BLOCK CAPITALS)}}
                {RESEARCH GROUP NAME (IN BLOCK CAPITALS)}}
\faculty{\texorpdfstring{\href{Faculty Web Site URL Here (include http://)}
                {Faculty Name}}
                {Faculty Name}}
\FACULTY{\texorpdfstring{\href{Faculty Web Site URL Here (include http://)}
                {FACULTY NAME (IN BLOCK CAPITALS)}}
                {FACULTY NAME (IN BLOCK CAPITALS)}}

\maketitle

\addtocontents{toc}{\vspace{2em}} 
\frontmatter 
\Certificate

\Declaration

\begin{acknowledgements}

I would like to express my deepest gratitude to my supervisor, \textbf{Prof. Pradyumn Kumar Sahoo}, Professor, Department of Mathematics, BITS-Pilani, Hyderabad Campus, Hyderabad, Telangana. His dedication towards research, expertise in executing research projects, attention towards new research and developments, and last but not least, his contribution to the development of cosmology and to attracting young mathematicians in the field of cosmology are the root motivation of my work, as well as many others.

I sincerely thank the members of the Doctoral Advisory Committee (DAC), \textbf{Prof. Bivudutta Mishra} and \textbf{Prof. Nirman Ganguly}, for their valuable suggestions and constant encouragement to improve my research work.

It is my privilege to thank the HoD, the DRC convener, faculty members, my colleagues, and the Department of Mathematics staff for supporting this amazing journey of my Ph.D. career.

I acknowledge \textbf{BITS-Pilani, Hyderabad Campus} for providing me with the necessary facilities, and the University
Grants Commission (UGC) Maulana Azad National Fellowship (MANF), New Delhi, India, for awarding a Junior Research Fellowship (UGC-Ref.No.: 211610222082) to carry out my research work.

I would like to express my gratitude to my research team: Dr. Raja Solanki, Dr. Sanjay Mandal, Dr. Simran Arora, Sai Swagat Mishra, Dr. Zinnat Hassan, and Dheeraj Singh.

I would like to express my heartiest thanks to my parents \textbf{Mrs. Jana Nohi ud din } and \textbf{Mr. Gh Mohi ud din Bhat}, my wife \textbf{Shazia Tariq} and my daughter \textbf{Abeeha Aaqid Bhat}.

\vspace{1.2 cm}
Aaqid Mohi ud din Bhat,\\
ID: 2021PHXF0456H.

\end{acknowledgements}

\begin{abstract}
In the last century, theoretical and observational developments have established the General Relativity (GR) theory as the most successful theory for describing the gravitational phenomenon. On the other hand, in the last two decades, multiple observational probes have strongly favored the discovery of the acceleration of cosmic expansion. The observational enhancement and development in precision cosmology indicate a requirement to go beyond GR and to search for an alternate description that can resolve the persistent issues.

In Chapter \ref{Chapter1}, we begin our introductory phase with the description of the accelerating Universe. Then, we briefly discussed the fundamentals of GR and its alternative formalism such as the teleparallel equivalent of GR (TEGR) and symmetric teleparallel equivalent of GR (STEGR). Moreover, we highlighted the pros and cons of the standard model of cosmology. Lastly, we have introduced the fundamentals of some important non-Riemannian spacetime geometry, such as modified teleparallel and symmetric teleparallel gravity and its extensions. 
 
In Chapter \ref{Chapter2}, we analyze the cosmological implications of $f(T,\mathcal{T})$ theory by considering the squared-torsion model $f(T,\mathcal{T})= \alpha \mathcal{T}+ \beta T^2$, where $T$ represents torsion and $\mathcal{T}$ is the trace of the energy-momentum tensor. We derive the solutions to the modified Friedmann equations and utilize recent observational data sets to constrain the free parameters of the model. Our analysis reveal the evolution of the deceleration parameter, which explicitly transitions from a decelerating phase to an accelerating one, effectively accounting for the late-time expansion of the Universe.

In Chapter \ref{Chapter3}, we explore the inflationary scenario within the framework of torsion-trace coupling gravity, utilizing a Lagrangian density derived from a function denoted as $f(T,\mathcal{T})$. By applying the slow-roll conditions to a specific $f(T,\mathcal{T})$ model, we compute key inflationary parameters, such as the tensor-to-scalar perturbation ratio $r$, 
the spectral index $n_s$, the running of the spectral index $\alpha_s$, the tensor spectral index $n_t$ and the number of e-folds. The numerical results align well with current observational data.

In Chapter \ref{Chapter4}, we aim to investigate the dark sector of the Universe, which encompasses the enigmatic components of the Dark Matter (DM) and Dark Energy (DE). We consider an extended form of the Equation of State (EoS)  for DM, widely known as the Extended Bose-Einstein Condensation (EBEC) EoS for DM. To describe undetected DM, we consider the modified theories of gravitation within a flat spacetime geometry, dependent solely on non-metricity, particularly, we consider the power law $f(Q)$ Lagrangian $f(Q)= \gamma \left(\frac{Q}{Q_0}\right)^n$. 

\newpage

We derive analytical solutions for the corresponding equations and determine the best-fit values of the parameters. Finally, we examine the thermodynamic stability of the assumed model by analyzing the sound speed parameter to assess its implications and behavior within the framework of the model. We find that the considered $f(Q)$ model can efficiently address the late-time expansion phase of the Universe with the observed transition epoch in a stable way. 

In Chapter \ref{Chapter5}, we propose an extended formulation of symmetric teleparallel gravity by generalizing the gravitational Lagrangian through the inclusion of an arbitrary function of  $f(Q,\mathcal{T_{\mu\nu}} \mathcal{T^{\mu\nu}})$. We derived the FRW equations for a flat, homogeneous, and isotropic spacetime. We find the analytical solution for the barotropic fluid case $p=\omega \rho$ for the model $f(Q, \mathcal{T_{\mu\nu}} \mathcal{T^{\mu\nu}})  = Q + \eta(\mathcal{T_{\mu\nu}} \mathcal{T^{\mu\nu}})$. Additionally, we constrain the parameters of the derived solution $H(z)$ by utilizing observational data from Cosmic Chronometers (CC), Baryon Acoustic Oscillations (BAO), and the latest Supernova (SN) samples. This is achieved through the application of the Markov Chain Monte Carlo (MCMC) sampling technique combined with Bayesian statistical analysis. In addition, we employ the Om diagnostic test to assess the behavior of supporting DM. We find that the behavior of Om diagnostic parameter favors the quintessence type DM model. 

In Chapter \ref{Chapter6}, we delved into the dark sector of the Universe, specifically focusing on DM and DE. We examine an extended version of the EoS for DM, commonly referred to EBEC EoS for DM, given as $p=\alpha \rho + \beta \rho^2$, alongside the modified $f(Q)$ Lagrangian given by $f(Q)= \gamma Q^2$. We introduce a set of dimensionless phase space variables. These phase space variables allow us to reframe the dynamics of the given cosmological system into an autonomous system. We conclude that the considered cosmological scenario successfully describes the evolutionary phase of the Universe from a decelerated matter era to the accelerated expansion epoch in a stable way. Finally, Chapter \ref{Chapter7},  provides a concise summary of the findings of this thesis and outlines potential directions for future research.

\end{abstract} 

\clearpage
\setstretch{1.3} 

\pagestyle{empty} 
\pagenumbering{gobble}
\Dedicatory{ \begin{LARGE}
\textbf{Dedicated to}
\end{LARGE} 
\\
\vspace{3cm}
\textbf {My Parents} And
\\
My Wife \textbf{Shazia Tariq Rather}\\}



\lhead{\emph{Contents}} 
\tableofcontents 
\addtocontents{toc}{\vspace{1em}}
\lhead{\emph{List of tables}}
\listoftables 
\addtocontents{toc}{\vspace{1em}}
\lhead{\emph{List of figures}}
\listoffigures 
\addtocontents{toc}{\vspace{1em}}




\lhead{\emph{List of symbols}}
\listofsymbols{ll}{

$g_{\mu\nu}$: \,\,\,\,\,\, Metric Tensor\\
$g$: \,\,\,\,\,\,\,\,\,\,\, Determinant of $g_{\mu\nu}$\\
$\Gamma^{\lambda}_{\mu\nu}$: \,\,\,\,\, General Affine Connection\\
$\left\lbrace{^{\lambda}}_{\mu\nu}\right\rbrace$: \,\,\,\,\, Levi-Civita Connection\\
$\nabla_{\mu}$: \,\,\,\,\,\, Covariant derivative w.r.t. Levi-Civita Connection\\
$(\mu\nu)$: \,\,\,\,\, Symmetrization over the indices $\mu$ and $\nu$\\
$[\mu\nu]$: \,\,\,\,\,\, Anti-Symmetrization over the indices $\mu$ and $\nu$\\
GR: \,\,\,\,\,\,\,\,\, General Relativity\\
$\Lambda$CDM:\,  $\Lambda$ Cold Dark Matter\\
ECs: \,\,\,\,\,\, Energy Conditions\\
EoS:\,\,\,\,\,\,\,\,\,\,  Equation of State \\
SN Ia:\,\,\,\,\,\,\,  Type Ia Supernovae\\
CMB: \,\,\,\,\,\,\,Cosmic Microwave Background\\
BAO: \,\,\,\,\,\,\, Baryon Acoustic Oscillations\\
DE:\,\,\,\,\,\,\,\,\,\, \,\,\,\,Dark Energy\\
EBEC:\,\,\,\,\,\,\, Extended Bose-Einstien Condensate\\
MCMC:\,\,\,\,\,\,  Markov Chain Monte Carlo}

\addtocontents{toc}{\vspace{2em}}

%
%


\clearpage 





\mainmatter 

\pagestyle{fancy} 


\chapter{Introduction} 
\label{Chapter1}

\lhead{Chapter 1. \emph{General Introduction}} 

\clearpage
\pagebreak

\section{Introduction}

We live in an expanding Universe filled with billions of stars, galaxies, and several mysterious objects; all the galaxies are now rushing away from each other. Cosmology is the large-scale scientific study of the Universe, its elements, and its past and future.
In the past two decades, cosmology has advanced significantly as a science, with unexpectedly fast-paced information regarding the creation, structure, and evolution of the Universe. 

\subsection{The expanding Universe}

In the period 1910-1930, the cosmic distance ladder extended beyond 100 kpc. When observing a galaxy at visible wavelengths, its spectrum usually displays absorption lines. Note that the wavelength of light obtained from a receding object is elongated, leading to what is known as redshift that measures this stretching factor or wavelength shift as follows
\begin{equation}\label{e1}
 z \equiv \frac{\lambda_{obs}-\lambda_{emit}}{\lambda_{emit}}\text{.}
\end{equation}
For non-relativistic motion, i.e. low redshift case, the standard Doppler's effect gives $z \approx \frac{v}{c}$, where $v$ is the velocity of the object receding from us and $c$ is the speed of light. Over a decade of observations of around 40 galaxies, Vesto Slipher at Arizona's Lowell Observatory found that nearly all galaxies except for a few in the Local Group exhibited redshift. Edwin Hubble at Mt. Wilson Observatory in California then attempted to find the distances to these galaxies by correlating these observed redshifts. He discovered that this observed redshift was directly proportional to the distance $d$ from the galaxy that emits the light. In the year 1929, Hubble presented his following groundbreaking results \cite{R0a}
\begin{equation}\label{e2}
z=\frac{H_0}{c}d\text{.}
\end{equation}
Utilizing the Doppler's relation $z \approx \frac{v}{c}$, we have 
\begin{equation}\label{e3}
v=H_0 d\text{.}
\end{equation}
This proportionality constant $H_0$ is well known as the Hubble constant, and it represents the recession speed per unit separation. One of the most prominent tensions of the observational as well as theoretical cosmology is the $H_0$ tension, which is the disagreement in $H_0$ values obtained from the local measurements via the SH0ES collaboration, for instance, R19 \cite{R6a} report $H_0 = 74.03 \pm 1.42 \: km/s/Mpc $ at a $68$\% confidence level (CL), or R20 \cite{R7a} when incorporating Gaia EDR3 parallax measurements, which give $H_0 = 73.2 \pm 1.3 \: km/s/Mpc $ at the same CL, and that of the Planck \cite{R2a} estimate $H_0 = 67.27 \pm 0.60 \: km/s/Mpc$ at $68$\% CL, incorporating the $\Lambda$CDM as the base cosmological model.

\subsection{The accelerating Universe}

In 1998, observational studies of Type Ia supernovae (SNeIa) revealed that the expansion of the Universe is accelerating faster than previously believed. Two leading groups in this field—Riess et al. \cite{R9a}  of the High-Z Supernova and Perlmutter et al. \cite{R10a} of the Supernova Cosmology Project (SCP) found that galaxies and their clusters are moving apart at an accelerating pace. Additionally, various observations, including data from the Planck satellite, led to the remarkable discovery that only 4\% to 5\% of the observable Universe is made up of conventional matter, such as baryons and electrons. The remaining balance of the Universe comprises two fundamentally unidentified components: Dark Matter (DM) (25\%) and Dark Energy (DE) (70\%). The $\Lambda$ Cold Dark Matter ($\Lambda$CDM)  framework, based on the idea that cold DM dominates the Universe and that its late-time behavior is governed by Einstein's cosmological constant, emerged as a result of groundbreaking discoveries in modern cosmology. Consequently, the accelerated expansion of the Universe has remained one of the most puzzling and significant phenomena in cosmology over the last twenty five years.

After the groundbreaking discovery of the accelerating Universe, surveys of SN Ia have received much attention by the research community in the last two decades. Several observational projects were initiated in this area, such as the Nearby Supernova Factory \cite{R11a}, the Sloan Digital Sky Survey \cite{R12a}, the Lick Observatory Supernova Search \cite{R13a}, the Supernova Legacy Survey (SNLS) \cite{R14a}, and the SCP. In 2008, SCP published the Union dataset with $307$ SN Ia samples \cite{R15a}. The Union data set was updated in 2010 to Union 2 \cite{R16a}, which included $557$ sample points, and then was further revised to the Union2.1 dataset consisting of $580$ SN Ia samples \cite{R17a}. Recently, in 2022, Brout et al. published the Pantheon + SH0ES dataset, an update of the Pantheon 2018 version of $1048$ data points \cite{R18a}, which features $1,701$ light curves from $1,550$ SN Ia in the redshift range $ 0.001 \leq z \leq 2.26$ \cite{R19a}.

\subsection{The Big Bang theory}

In the early 1920s, Russian Mathematician Alexander Friedmann found that Einstein's gravitational field equations included non-static solutions, suggesting the possibility of a Universe that expands and changes in size over time. Friedmann's work revealed that the Universe originated approximately 13 billion years ago from a singular event \cite{friedman/1922}. Therefore, according to Friedmann’s solutions, the Universe, all matter, space, and time itself appeared at once in a single instant. British scientist Fred Hoyle called this theory the 'Big Bang'. Under this term, it came to represent the accepted cosmological paradigm, according to which the Universe was created at a single point at an extremely high density and temperature. After learning about this study, Einstein promptly abandoned the cosmological constant as the most significant error of his life.

The Cosmic Microwave Background (CMB) is the legacy of the cosmic recombination epoch. It provides abundant information about the early Universe. In 1964, Penzias and Wilson \cite{R21a} first detected CMB radiation and won the Nobel Prize in Physics in 1978 for its achievement. Their work strongly supported the Big Bang cosmology of the Universe \cite{R38vv}. In 1989, the Cosmic Background Explore (COBE) research group launched the first generation of the CMB radiation satellite and discovered the CMB anisotropy of the Universe \cite{R39u}. Their discovery helps to explore the dynamics of the Universe more precisely. Two lead researchers of the COBE research group, Mather and Smoot, received the Nobel Prize in Physics in 2006. The BOOMERang and maxima experiments \cite{R40uu,R40uv} were the first to measure acoustic oscillations in the angular spectrum of CMB radiation anisotropy \cite{R41u,R42u}. The Wilkison Microwave Anisotropy Probe (WMAP) is the second generation of the CMB radiation satellite, launched in 2001, and it measured the CMB radiation spectrum and probed the cosmological parameters with higher accuracy \cite{R43u,R43v,R43w,R43x}. Its successor, the Planck satellite, was launched in 2009, and the early result was released \cite{Planck/2011}.

\section{The geometric trinity}

The first step in constructing a theory of gravitation is to establish a mathematical framework that will allow us to verify the laws of physics. As is well known, scalars, vectors, tensors, and mathematical quantities help us to test the properties of physical objects and dynamical systems. These mathematical quantities will be used to construct the framework. Moreover, when we begin to study the gravitational fields, in order to verify the physical laws, we need an arbitrary frame of reference. We need to develop a four-dimensional geometry in an arbitrary system. Hence, this section contains some basic mathematical quantities developed using vectors, tensors, and their differential operators.

\subsection{The metric tensor}
We define a differential manifold as a Riemannian space with the basic property that each point of that manifold can be represented as tensor \cite{dub}
\begin{equation}\label{1}
g_{\mu\nu}(x)=g_{\mu\nu}(x^1,x^2,x^3,...,x^n).
\end{equation}
This is a symmetric, twice covariant, and nondegenerate tensor, and we call it a metric tensor. So 
\begin{equation}\label{2}
g_{\mu\nu}=g_{\nu\mu},\,\,g=\text{det}|g_{\mu\nu}|.
\end{equation}
The fundamental properties of the function $g_{\mu\nu}$ should be continuous and have continuous derivatives with respect to all coordinates $(x^1,x^2,x^3,…,x^n)$. This metric helps us to construct a invariant second-order differential form in a Riemannian space, defined as
\begin{equation}\label{3}
ds^2=g_{\mu\nu}dx^{\mu}dx^{\nu}.
\end{equation}
This quantity is called the interval \cite{lan}. It is well known in linear algebra that a square matrix can be reduced to a diagonal form by determining the eigenvalues of the matrix. Similarly, one can reduce the matrix $g_{\mu\nu}$ to the diagonal form by considering a proper coordinate transformation. In that case, the diagonal elements of $g_{\mu\nu}$ may have different signs, and the difference between the positive and negative numbers is called the Riemannian metric signature. Researchers generally use two types of signature convention to explore the fate of the Universe, such as (+,-,-,-) or (-,+,+,+). Throughout the thesis, we will adopt the former signature for the metric. The differential form in equation \eqref{3} can have any sign in Riemannian space. Based on its invariant property, this interval can be characterized in the following types: spacelike, timelike, or null-like.

\subsection{General affine connection}
The fundamental nature of gravitational fields can be explored by the basic properties of the dynamics of physical objects. This can be done by analyzing the geodesic nature of the motion of the test particle; for this, the affine connection plays an important role. In differential geometry, the general affine connection can be written as

\begin{equation}\label{4}
{\Gamma^{\lambda}}_{\mu\nu}=\left\lbrace{^{\lambda}}_{\mu\nu}\right\rbrace+{K^{\lambda}}_{\mu\nu}+{L^{\lambda}}_{\mu\nu},
\end{equation}
where the Levi-Civita connection/Christoffel symbol of the metric is
\begin{equation}\label{5}
\left\lbrace{^{\lambda}}_{\mu\nu}\right\rbrace\equiv\frac{1}{2}g^{\lambda \beta}\left(\partial_{\mu}g_{\beta \nu}+\partial_{\nu}g_{\beta \mu}-\partial_{\beta}g_{\mu \nu}\right),
\end{equation}
and the contortion is
\begin{equation}\label{6}
{K^{\lambda}}_{\mu\nu}\equiv\frac{1}{2}{T^{\lambda}}_{\mu\nu}+T_{(\mu}{}^{\lambda}{}_{\nu)},
\end{equation}
with the torsion tensor ${T^{\lambda}}_{\mu\nu}\equiv 2{\Gamma^{\lambda}}_{[\mu\nu]}$. The disformation $L^{\lambda}_{\mu\nu}$ in terms of the non-metricity tensor can be read as
\begin{equation}\label{7}
{L^{\lambda}}_{\mu\nu}\equiv \frac{1}{2} g^{\lambda\beta}\left(-Q_{\mu\beta\nu}-Q_{\nu \beta\mu}+Q_{\beta\mu\nu}\right).
\end{equation}
Here, the non-metricity tensor $Q_{\gamma \mu\nu}$ is defined as the covariant derivative of the metric tensor with respect to the Weyl-Cartan connection ${\Gamma^{\lambda}}_{\mu\nu}$, $Q_{\gamma \mu\nu}\equiv \nabla_{\gamma}g_{\mu\nu}$, and it can be written as \cite{hehl/1976}
\begin{equation}\label{8}
Q_{\gamma \mu\nu}=\frac{\partial g_{\mu\nu}}{\partial x^{\gamma}}- g_{\nu \lambda}{\Gamma^{\lambda}}_{\mu\gamma} - g_{\lambda\mu}{\Gamma^{\lambda}}_{\nu\gamma}.
\end{equation}
It is important to note that the non-torsional component of the connection $\Gamma^{\alpha}_{\mu\nu}$ is the Levi-Civita corresponds to the Levi-Civita connection, while the contorsion and disformation tensors exhibit torsional behavior under coordinate transformations. Having assembled these essential geometrical elements, we can now define the structure of spacetime geometry as follows:
\begin{itemize}
\item \textit{Metric}: The connection is metric compatible, that is, $Q_{\alpha \mu\nu}(\Gamma, g)=0$. In metric spaces, the length of the vectors remains preserved during parallel transport. Consequently, non-metricity measures the extent to which the length of vectors changes when they are parallel transported.
\item \textit{Torsionless}: When $T^{\alpha}_{\mu\nu}(\Gamma)=0$  the connection is symmetric. Torsion quantifies the failure of a parallelogram to close when two infinitesimal vectors are parallel transported to each other. Thus, in the presence of torsion, it is generally understood that parallelograms do not close.
\item \textit{Flat}: When $R^{\alpha}_{\beta\mu\nu}=0$ the connection is flat, indicating the absence of curvature. Curvature describes the rotation vector experiences when parallel transported along a closed loop, complicating the comparison of vectors at different points in spacetime. However, in flat spaces, vectors do not rotate during parallel transport, allowing for a more straightforward notion of parallelism across distances. This is why theories formulated in such contexts are referred to as teleparallel.
\end{itemize}
Einstein's General Relativity (GR) is constructed on a spacetime that is metric-compatible and torsion-free, attributing gravity solely to curvature. However, it is natural to question, as Einstein himself later did, whether gravity could also be linked to other properties of spacetime, such as torsion and non-metricity. To date, these three approaches to gravity, curvature, torsion, and non-metricity, have been shown to equivalently describe GR, forming a geometric trinity of gravity. The standard formulation of GR relies on the Levi-Civita connection, which assumes that both torsion and non-metricity are zero. In contrast, its teleparallel equivalent employs the Weitzenböck connection, which assumes zero curvature and non-metricity \cite{R41a}. A gravitational model incorporating Weyl-Cartan spacetime was investigated under the Weitzenböck condition, which requires the combined sum of the curvature and torsion scalars to vanish \cite{Haghani}. Another formulation of GR, called the symmetric teleparallel equivalent of GR, remains a relatively unexplored area. In this framework, gravitational interaction is characterized by the non-metricity tensor $Q$, with both curvature and torsion assumed to be zero. Symmetric teleparallel equivalent to gravity was initially introduced in a concise research paper \cite{R50a}, in which the authors highlighted that this formulation offers a fresh perspective on GR. They emphasized that gravitational interactions, mediated by non-metricity, exhibit characteristics akin to the Newtonian force, arising from a potential, specifically, the metric itself. The formulation, on the other hand, is geometric and covariant. The equivalent descriptions to GR by curvature, torsion, and non-metricity provide the starting point for modified theories of gravity once the respective scalar is replaced by the arbitrary functions. These three fundamental theories are called \textit{`Geometrical Trinity of Gravity'}. We will discuss these three geometries in the following subsections.

\subsection{General Relativity}
One of the remarkable aspects of GR is that spacetime is not only curved but also dynamic. This means that while the motion of matter is influenced by the curvature of spacetime, matter itself also generates curvature in spacetime. This relationship between matter and spacetime curvature is mathematically described by the Einstein field equations
\begin{equation}\label{9}
G_{\mu\nu}\equiv R_{\mu\nu}-\frac{1}{2}g_{\mu\nu}R=8\pi G \mathcal{T_{\mu\nu}}.
\end{equation}
Here, $G_{\mu\nu}$, $R_{\mu\nu}$, and $R$ represent the Einstein tensor, the Riemannian tensor, and the Ricci scalar, respectively. These tensors can be written in terms of metric tensor, Christoffel symbols and their derivatives as
\begin{equation}\label{10}
R^{\rho}_{\sigma\mu\nu}=\delta_{\mu}\left\lbrace^{\rho}_{\nu\sigma}\right\rbrace-\delta_{\nu}\left\lbrace^{\rho}_{\mu\sigma}\right\rbrace+\left\lbrace^{\rho}_{\mu\lambda}\right\rbrace \left\lbrace^{\lambda}_{\nu\sigma}\right\rbrace-\left\lbrace^{\rho}_{\nu\lambda}\right\rbrace \left\lbrace^{\lambda}_{\mu\sigma}\right\rbrace,
\end{equation}
\begin{equation}\label{11}
R_{\mu\nu}=R^{\lambda}_{\mu\lambda\nu},\,\,\,\,\, R=g^{\mu\nu}R_{\mu\nu}.
\end{equation}
Also, $\mathcal{T_{\mu\nu}}$ describes the energy-momentum component of the Universe. We shall discuss $T_{\mu\nu}$ in the upcoming section.

It is worthy of mentioning here that the Einstein equation can be derived using variation principle from varying the Einstein-Hilbert action
\begin{equation}\label{12}
S=\frac{1}{2 \kappa} \int  R \sqrt{-g}  d^4x +\int  L_m\sqrt{-g}d^4x
\end{equation}
where $\kappa = \frac{8\pi G}{c^4}$ . In the above action, the first term is the gravitational part and the second term is the matter part. 

Einstein's GR formulation is based on some fundamental concepts such as

\begin{itemize}
\item General covariance: It reflects the principle of relativity, stating that the laws of physics maintain the same form across all coordinate systems.
\item Equivalence principle: It establishes the equality between gravitational mass and inertial mass.
\item Spacetime curvature: It specifies the mass through which gravitational forces govern the dynamics of a system.
\item Levi-Civita connection: It formulates without presence of the torsion and non-metricity.
\end{itemize}
\subsection{Teleparallel equivalent to GR}
The vierbein fields, $e_{\mu}(x^i)$, act as a dynamical variable for the teleparallel gravity. As usual, $x^i$ is used to run over the spacetime coordinates, and $\mu$ denotes the tangent spacetime coordinates. At every point on the manifold, the vierbein fields create an orthonormal basis for the tangent space, represented by the line element of four-dimensional Minkowski spacetime, i.e., $e_{\mu}e_{\nu}=\eta_{\mu\nu}=$diag$(-1,+1,+1,+1)$. In terms of vector components, the vierbein fields can be written as $e^i_{\mu}\partial_i$, and the metric tensor can be written as
\begin{equation}\label{13}
g_{\mu\nu}=\eta_{ij}e^i_{\mu}(x)e^j_{\nu}(x).
\end{equation}
Moreover, the vierbein basis follow the general relation $e^i_{\mu}e^{\mu}_j=\delta^i_j$ and $e^i_{\mu}e^{\nu}_i=\delta^{\nu}_{\mu}$.
In teleparallel gravity, the curvatureless Weitzenb$\ddot{o}$ck connection \cite{R40a} defined as
\begin{equation}\label{14}
\hat{\Gamma}^{\gamma}_{\mu\nu}\equiv e^{\gamma}_i\partial_{\nu}e^i_{\mu}\equiv -e^i_{\mu}\partial_{\nu}e^{\gamma}_i.
\end{equation}
Using Weitzenb$\ddot{o}$ck connection one can write the nonzero torsion tensor as
\begin{equation}\label{15}
T^{\gamma}_{\mu \nu}\equiv \hat{\Gamma}^{\gamma}_{\mu\nu}-\hat{\Gamma}^{\gamma}_{\nu\mu} \equiv e^{\gamma}_i(\partial_{\mu} e^i_{\nu}-\partial_{\nu} e^i_{\mu}).
\end{equation}
The contracted form of the above torsion tensor can be written as follows \cite{Maluf/1994,Hayashi/1979,Arcos/2004}
\begin{equation}\label{16}
\mathcal{T}\equiv S^{\mu \nu}_{\gamma}T^{\gamma}_{\mu \nu}\equiv \frac{1}{4}T^{\gamma \mu \nu}T_{\gamma \mu \nu}+\frac{1}{2}T^{\gamma \mu \nu}T_{\nu \mu \gamma}-T^{\gamma}_{\gamma \mu}T^{\nu \mu}_{\nu},
\end{equation}
where
\begin{equation}\label{17}
S^{\mu \nu}_{\gamma}=\frac{1}{2}(K^{\mu \nu}_{\gamma}+\delta^{\mu}_{\gamma}T^{\alpha \nu}_{\alpha}-\delta^{\nu}_{\gamma}T^{\alpha \mu}_{\alpha}),
\end{equation}
represents the superpotential tensor. The distinction between the Levi-Civita connection and the Weitzenböck connection lies in the contortion tensor, which is defined as
\begin{equation}\label{18}
K^{\mu \nu}_{\gamma}=-\frac{1}{2}(T^{\mu \nu}_{\gamma}-T^{\nu \mu}_{\gamma}-T^{\mu \nu}_{\gamma}).
\end{equation}
The action for Teleparallel Equivalent to GR (TEGR) reads
\begin{equation}\label{e92}
S= \frac{1}{2\kappa}\int f(T) \sqrt{-g} d^4x + \int  L_m\sqrt{-g} d^4x \text{.} 
\end{equation}
where $e=\sqrt{-g}$. Note that, the flat and teleparallel connections are employed in the transition from Einstein-Hilbert action to TEGR action. Other than this, there in no change in the matter action.
\subsection{Symmetric Teleparallel Equivalent to GR}
Symmetric Teleparallel Equivalent to GR (STEGR) is formulated by considering flat, vanishing torsion in the general connection, and the non-metricity tensors. In this subsection, we shall discuss the non-metricity tensors.

The non-metricity tensor and its traces are such that
\begin{equation}
\label{20}
Q_{\gamma\mu\nu}=\nabla_{\gamma}g_{\mu\nu}\,,
\end{equation}
\begin{equation}
\label{21}
Q_{\gamma}={{Q_{\gamma}}^{\mu}}_{\mu}\,, \qquad \widetilde{Q}_{\gamma}={Q^{\mu}}_{\gamma\mu}\,.
\end{equation}
Moreover, the superpotential as a function of non-metricity tensor is given by
\begin{equation}
\label{22}
4{P^{\gamma}}_{\mu\nu}=-{Q^{\gamma}}_{\mu\nu}+2Q_{({\mu^{^{\gamma}}}{\nu})}-Q^{\gamma}g_{\mu\nu}-\widetilde{Q}^{\gamma}g_{\mu\nu}-\delta^{\gamma}_{{(\gamma^{^{Q}}}\nu)}\,,
\end{equation}
where the trace of non-metricity tensor \cite{R51a} reads
\begin{equation}
\label{23}
Q=-Q_{\gamma\mu\nu}P^{\gamma\mu\nu}\,.
\end{equation}
The action for STEGR is the following
\begin{equation}\label{24}
S=-\frac{1}{2\kappa}\int Q  \sqrt{-g}d^4x + \int  L_m \sqrt{-g}d^4x \text{.}
\end{equation}
Note that the flat and non-metricity tensors are employed in the transition from Einstein-Hilbert action to STEGR action with no change in the matter action.

\section{The standard cosmological model}
In the previous section, we describe the physics of GR and its mathematical foundations. Now, we will see how these formulations can describe the various cosmological observations.

\subsection{Comoving distances}
The cosmological principle presents the Universe as a cosmic fluid, wherein galaxies act as fundamental particles. A fluid element represents a volume that encompasses numerous galaxies but remains very small in comparison to the entire Universe. Therefore, the movement of a cosmic fluid element reflects the smear motion of the galaxies within it. This motivates us to use a special coordinate system known as the comoving coordinate system, which evolves along with the expansion of the Universe. The relation between comoving distance $x$ and the actual physical distance $r$ is given by
\begin{equation}\label{e40}
 r = a(t) x \text{,}  
\end{equation}
where $a(t)$ is the cosmic expansion factor known as the scale factor. Now, consider a galaxy with recessional velocity $v$, then 
\begin{equation}\label{e41}
v= \frac{dr}{dt}  = \dot{r} = \dot{a} x = \frac{\dot{a}}{a} a x = \frac{\dot{a}}{a} r \text{.}
\end{equation}
According to the Hubble's law of expansion, we have $v= H r$, hence on comparing we obtain the following,
\begin{equation}\label{e42}
 H = \frac{\dot{a}}{a} \text{.}   
\end{equation}
This expression we call the Hubble parameter. 

\subsection{ Friedmann–Lemaître–Robertson–Walker metric Universe}
The terms open, closed, and flat are traditionally used to differentiate between the three possible isotropic and homogeneous spatial geometries. In the case of flat space, the spatial curvature is zero at every point. The closed scenario is characterized by a positive constant spatial curvature, whereas the open Universe exhibits a constant negative spatial curvature. The following line element represents these three different geometries of an isotropic and homogeneous spatial background
\begin{equation}\label{e43}
ds^2=-c^2dt^2+a^2(t)\left[d\xi^2+V(\xi)^2(d\theta^2+sin^2\theta d\phi^2)\right] \text{,}
\end{equation}
where 
\begin{equation}\label{e44}
V(\xi)=
\begin{dcases}
sin\xi ; &  closed \\
\xi ;&  flat\\
sinh\xi ; &  open 
\end{dcases} \text{.}
\end{equation}
One can rewrite the above metric in a more unified form as follows,
\begin{equation}\label{e45}
ds^2=-dt^2+a^2(t)\left[\frac{dr^2}{1-k r^2}+r^2(d\theta^2+sin^2\theta d\phi^2)\right]\text{.}
\end{equation}
The above line element characterizes the spatially isotropic and homogeneous evolution of the Universe, as the scale factor $a(t)$ varies. This line element is known as the Friedmann-Lemaitre-Robertson-Walker metric (FLRW ) \cite{R27a,R28a}. Note that from now on, we will work in units for which $c=1$. The quantity $k$ represents the spatial curvature, such as the case $k = -1$ represents the open Universe, the case $k = +1$ represents the closed Universe, while the case $k = 0$ represents the spatially flat Universe.

\subsection{Energy conditions} 
Energy conditions (ECs) fulfill three key roles and enforce coordinate invariant constraints on the stress-energy-momentum tensor associated with matter.
First, since Einstein's equations rely solely on the stress-energy tensor and not on other properties of matter, the ECs enable the study of gravitating systems without requiring a detailed understanding of matter's behavior. This approach was pivotal for Penrose and Hawking in proving their singularity theorems \cite{Penrose/1965,Hawking/1966} by bypassing a complicated and comprehensive analysis using this method, as it allowed them to avoid complex and exhaustive analyses. Second, ECs provide a generalized notion of normal matter that applies to a wide range of matter types. Third, ECs offer conceptual simplicity. For instance, the positivity of energy density can be linked to the stability of a system, at least in the classical sense, where stability is associated with energy being bounded from below. ECs are highly relevant in classical GR, addressing spacetime singularity issues and explaining the behavior of spacelike, timelike, or lightlike geodesics \cite{Visser/2000}. These conditions can be derived from the well-known Raychaudhuri equations of the form \cite{Poisson/2004,Poisson/2004a,Hawking/1973}.
\begin{eqnarray}
\label{2.19}
\frac{d\theta}{d\tau} &=& -\frac{1}{3}\theta^2-\sigma_{\mu\nu}\sigma^{\mu\nu}+\omega_{\mu\nu}\omega^{\mu\nu}-R_{\mu\nu}u^{\mu}u^{\nu}\,, \\
\label{2.20}
\frac{d\theta}{d\tau} &=& -\frac{1}{2}\theta^2-\sigma_{\mu\nu}\sigma^{\mu\nu}+\omega_{\mu\nu}\omega^{\mu\nu}-R_{\mu\nu}n^{\mu}n^{\nu}\,,
\end{eqnarray}
where $\theta$, $\sigma^{\mu\nu}$ and $\omega_{\mu\nu}$ are the expansion factor, shear and rotation associated with the geodesic congruence defined by the vector field $u^{\mu}$ and the null vector $n^{\mu}$. 

Consequently, there are various ECs, each with unique strengths and limitations in terms of their validity, significance, and interpretations. These ECs are constructed by forming scalars from the stress-energy tensor $\mathcal{T}_{\mu \nu}$, typically achieved by contracting it with arbitrary timelike vectors $t^{\mu}$ or null vectors $n^{\mu}$ \cite{Kontou/2020,Capozziello/2014}.

The Weak Energy Condition (WEC) is perhaps the most intuitive of the ECs. It states that the energy density measured by any observer moving along a timelike trajectory must be non-negative. Mathematically it can be written as 
\begin{equation}
\mathcal{T}_{\mu \nu} t^{\mu} t^{\nu} \geq 0,
\end{equation}
for any timelike vector $t^{\mu}$. For a perfect fluid, the weak energy condition suggests $\rho+ p \geq 0$ and also $\rho \geq 0$. 
The Null Energy Condition (NEC) is a variation of the WEC, where the timelike $n^{\mu}$ is replaced by the null vector. Mathematically it can be written as 
\begin{equation}
\mathcal{T}_{\mu \nu} n^{\mu} n^{\nu} \geq 0.
\end{equation}
In the context of a perfect fluid, the NEC states that $\rho+p \geq 0$. The NEC is particularly significant as it plays a key role in determining whether the Universe will undergo inflation, evolve into a singularity, or exhibit a bounce solution. Although the NEC is weaker than the WEC, it remains a fundamental constraint in gravitational theories and cosmology. 

The WEC can be extended to the Dominant Energy Condition (DEC). The DEC includes the requirements of the WEC but adds the constraint that $\mathcal{T}_{\mu \nu} t^{\mu}$ must be arbitrary or null for any arbitrary vector. In the context of a perfect fluid, this translates into conditions $\rho \geq |p|$. 

The Strong Energy Condition (SEC) imposes a bound
\begin{equation}
\left(\mathcal{T}_{\mu \nu} - \frac{1}{2}g_{\mu \nu} \mathcal{T}\right) t^{\mu} t^{\nu} \geq 0
\end{equation}
for every timelike vector $t^{\mu}$. The SEC transforms into $\rho + p \geq 0$ and $\rho + 3p \geq 0$ for a perfect fluid. According to the Einstein equation, SEC is strictly geometric, so $R_{\mu \nu} t^{\mu} t^{\nu} \geq 0$. This condition is widely employed and is one of the most important conditions of the Hawking and Penrose singularity theorems.

After establishing the foundational principles of a cosmological model, several quantities, often referred to as cosmological parameters, may remain unspecified. It is standard practice to define cosmological models using a small set of parameters, which are then measured or constrained through observations. These parameters help identify which version of the model most accurately describes the Universe. In the next section, we will explore the cosmological parameters that are typically considered in such analyses. 

\subsection{Cosmological parameters}
This section focuses on the key cosmological parameters relevant to observational cosmology. To understand the evolution of the Universe, we consider the Taylor expansion of the scale factor about the present time $t_{0}$. The general form of $a(t)$ is 
\begin{equation}
a(t) = a(t_{0}) + \dot{a(t_{0})} (t-t_{0}) + \frac{1}{2} \ddot{a(t_{0})} (t-t_{0})^{2} + \frac{1}{3!} \dddot{a(t_{0})} (t-t_{0})^{3} + \frac{1}{4!} \ddddot{a(t_{0})} (t-t_{0})^{4}+......
\end{equation}
Dividing by $a(t_{0})$, one gets
\begin{equation}
\frac{a(t)}{a(t_{0})} = 1 + H_{0}(t-t_{0}) - \frac{q_{0}}{2}H_{0}^{2} (t-t_{0})^{2} + \frac{1}{3!} j_{0} H_{0}^{3}(t-t_{0})^{3} + \frac{1}{4!} s_{0} H_{0}^{4} (t-t_{0})^{4}+......
\end{equation}
where $t_{0}$ is the present time. Here, the coefficients are named as Hubble, deceleration, jerk, snap, and lerk parameters, respectively. That are 
\begin{eqnarray}
H = \frac{\dot{a}}{a}, \quad q = -\frac{\ddot{a}}{a H^{2}}, \\
j = \frac{\dddot{a}}{a H^{3}}, \quad  s = \frac{1}{a H^{4}} \frac{d^{4}a}{dt^{4}}, \quad l = \frac{a^{(5)}}{a H^5}
\end{eqnarray}
The deceleration parameter $q$ provides insight into the nature of the expansion of the Universe. Specifically, $q<0$ indicates that the expansion is accelerating, while $q>0$ signifies deceleration. Furthermore, the jerk parameter $j$, which describes the rate of acceleration change, plays a significant role. A change in the sign of $j$ within an expanding model signals whether acceleration increases or decreases with time. These parameters are essential for understanding the dynamics of cosmic expansion and its evolution.

Now, we turn our attention to the simplest model of the Universe, which we will explore in detail. This model serves as a foundational framework for understanding the large-scale structure and evolution of the cosmos.

\subsection{History of $\Lambda$}
Now, on using the line element \eqref{e45}, we obtain the components of the Einstein's equation \eqref{9} as follows
\begin{equation}\label{e46}
\left(\frac{\dot{a}}{a}\right)^2 + \frac{k}{a^2}=\frac{8\pi G}{3}\rho 
\end{equation}

\begin{equation}\label{e47}
2\frac{\ddot{a}}{a} + \left(\frac{\dot{a}}{a}\right)^2 + \frac{k}{a^2} = - 8\pi G p \text{.}
\end{equation}
On combining above two equations, we obtain the following acceleration equation,
\begin{equation}\label{e48}
\frac{\ddot{a}}{a} = - \frac{4\pi G}{3}(\rho + 3p) \text{.} 
\end{equation}
At the time of the formulation of GR, the scientific community strongly believed in the static Universe i.e. the scale factor $a(t)$ must be constant. As a consequence of constant scale factor $a(t)$ and the above set of field equations, one can obtain the following results,
\begin{equation}\label{e49}
\rho=-3p= \frac{3k}{8\pi G a^2} \text{.}   
\end{equation}
Note that the positivity of energy density $\rho$ implies that the pressure component exhibits negative value, or if we assume $p=0$, then we obtain $\rho=0$. In both cases, the results obtained are absurd. To bypass this issue, later Einstein modified his field equation by adding a constant term $\Lambda$ called the cosmological constant. Thus, the new set of equations that are compatible with the static Universe read as follows,
\begin{equation}\label{e50}
\left(\frac{\dot{a}}{a}\right)^2 + \frac{k}{a^2}=\frac{8\pi G}{3}\rho + \frac{\Lambda}{3}
\end{equation}

\begin{equation}\label{e51}
2\frac{\ddot{a}}{a} + \left(\frac{\dot{a}}{a}\right)^2 + \frac{k}{a^2} = - 8\pi G p + \Lambda \text{.}
\end{equation}
Note that, in the year 1917, Einstein incorporated this cosmological constant $\Lambda$ into his field equation to achieve a static Universe model \cite{R29a}. However, following the discovery of cosmic expansion, he removed $\Lambda$ in 1931 \cite{R30a}. Later, in 1967, Zel'dovich revived the idea of the cosmological constant $\Lambda$, considering vacuum fluctuations \cite{R31a}. In 1987, Weinberg depicted a tiny non-vanishing cosmological constant $\Lambda$ \cite{R32a}. Finally, in the year 1998, the discovery of the accelerating expansion of the Universe brought back the cosmological constant $\Lambda$ into focus as a potential candidate for DE, which is driving this accelerated expansion.

\subsection{The $\Lambda$CDM model}
Recall that, an isotropic and homogeneous Universe having cosmic matter as perfect fluid characterize by the following energy-momentum tensor,
\begin{equation}\label{e52}
\mathcal{T}_{\mu\nu}=(\rho+p)u_{\mu}u_{\nu}+ p g_{\mu\nu} \text{,}
\end{equation}
where $u_\mu = (1,0,0,0)$ represents the four velocity vector of the cosmic fluid in a comoving coordinate. Note that, the vanishing covariant divergence of the energy-momentum tensor i.e. $\nabla^\mu \mathcal{T}_{\mu\nu} = 0 $ implies
\begin{equation}\label{e53}
\dot{\rho}+3 \frac{\dot{a}}{a}(\rho+p)=0 \text{.}  
\end{equation}
On employing the barotropic  EoS $p=\omega \rho$ in the above equation, we can have
\begin{equation}\label{e54}
\rho \propto a^{-3(1+\omega)} \text{.} 
\end{equation}
Note that, for different values of EoS parameter $\omega$, one can obtained the various expressions for the energy densities that corresponds to different cosmological epochs, as follows
\begin{itemize}
\item $\omega=1/3 \implies \rho_r = \rho_{r_0}a^{-4}$ represents the radiation phase
\item $\omega=0 \implies \rho_m = \rho_{m_0}a^{-3}$ represents the matter phase
\item $\omega=-1 \implies \rho_\Lambda = constant$ representing cosmological constant case as DE.
\end{itemize}
Now on using the equation \eqref{e50}, along with the assumption $\rho = \rho_m + \rho_r$ and $ \rho_\Lambda = \frac{\Lambda}{8 \pi G}$, we obtain the following expression,
\begin{equation}\label{e55}
 H^2 = \frac{8 \pi G}{3} \left[ \rho_{m_0}a^{-3} + \rho_{r_0}a^{-4}  + \rho_{\Lambda} \right] - \frac{k}{a^2} \text{.} 
\end{equation}
Here, $\rho_m = \rho_b + \rho_{cdm}$, where $\rho_b$ denotes the density of ordinary (baryonic) matter and $\rho_{cdm}$ represents the cold (non-relativistic) DM density. We define the dimensionless density parameter for the various component as follows
\begin{equation}\label{e56}
\Omega_{m_0} = \frac{\rho_{m_0}}{\rho_{crit_0}}, \:\: \Omega_{r_0} = \frac{\rho_{r_0}}{\rho_{crit_0}},  \:\: \Omega_{\Lambda} =\Omega_{\Lambda_0} = \frac{\rho_{\Lambda}}{\rho_{crit_0}}=\frac{\Lambda}{3H_0^2}, \:\: \text{and} \:\: \Omega_{k_0} = -\frac{k}{H_0^2} \text{,} 
\end{equation}
where $a_0=1$ (conventional assumption) and $\rho_{crit_0}=\frac{3H_0^2}{8 \pi G}$. Now, on utilizing the above setting along with the scale factor-redshift relation $a^{-1} = 1+z$, the equation \eqref{e55} becomes,
\begin{equation}\label{e57}
H^2 = H_0^2  \left[ \Omega_{m_0}(1+z)^{3} + \Omega_{r_0}(1+z)^{4}  + \Omega_{\Lambda_0} + \Omega_{k_0}(1+z)^2 \right] \text{.} 
\end{equation}
The model discussed above is widely known as the standard cosmological model or the $\Lambda$CDM model. The Planck 2018 results \cite{R2a} estimated the free parameter constraints of this standard model and predicted the values $H_0 = 67.4 \pm 0.5 \: km/s/Mpc$, $\Omega_{m_0}=0.315 \pm 0.007$, and $\Omega_{k_0}= 0.001 \pm 0.002$ that favor the spatial curvature of the Universe to be flat.

\subsection{Drawbacks of the $\Lambda$CDM model}

The standard cosmological model has been remarkably successful in describing the Universe and is largely consistent with observational data. It provides a comprehensive framework for understanding the evolution of the cosmos, the formation of large-scale structures through gravitational interactions, and the abundance of light elements, which can only be explained by primordial nucleosynthesis. This model incorporates key components such as inflation, DE, and DM, which are essential for explaining the observed dynamics and composition of the Universe. However, the $\Lambda$CDM approach faces certain shortcomings.

There is a significant discrepancy between the theoretical and observed values of the cosmological constant. The theoretical value, derived from quantum-mechanical processes within the framework of the standard model, is approximately $10^{-60}\, M_{Pl}^{4}$, where $M_{Pl}$ is the Planck mass. In contrast, the observed value is around $10^{-120}\, M_{Pl}^{4}$. This enormous difference, which spans nearly $60$ orders of magnitude, is known as the \textbf{ cosmological constant problem} \cite{Joyce/2015}. 

\textbf{The Horizon problem} \cite{Tsujikawa/2013} is a cosmological fine-tuning issue within the Big Bang theory, as supported by the $\Lambda$CDM model. It is also known as the homogeneity problem and arises from the challenge of explaining why regions of the Universe that are causally disconnected (i.e., too far apart to have interacted) exhibit such a high degree of homogeneity. Without a mechanism to establish uniform initial conditions across these regions, their observed similarity is difficult to justify. The most widely accepted solution to this problem is cosmic inflation, a period of exponential expansion in the early Universe. Inflation posits that these regions were once in close contact before rapidly expanding, thereby explaining their uniformity.

Another is the \textbf{cosmological coincidence} issue \cite{Velten/2014}, which highlights the observation that according to the $\Lambda$CDM model, we appear to be living in a transitional epoch between the matter-dominated era and the late-time acceleration era. This is puzzling because the current values of the cosmological constant and matter densities are of comparable magnitude, despite their vastly different evolutionary histories. This coincidence raises questions about why these two components, which scale differently over time, have similar energy densities in the present era. Resolving this problem remains a key challenge in modern cosmology.

 \textbf{DM} \cite{Astesiano/2021,Katsuragawa/2017,Zaregonbadi/2016} is a nonbaryonic form of matter that interacts with other components of the Universe solely through gravitational forces. Without assuming the existence of DM with these properties, it would be impossible to explain phenomena such as the rotational curves of galaxies, the formation of large-scale structures, and their distribution in the Universe. Although $\Lambda$CDM model incorporates DM as a fundamental component, neither ground-based nor space-based experiments have yet directly detected DM particles. This absence of direct detection remains one of the most significant challenges in modern physics and cosmology.

There have been notable discrepancies between high-redshift and low-redshift measurements of the present Hubble constant local measurements, which tend to yield higher values for $H_{0}$, while measurements based on the CMB and assuming the  $\Lambda$CDM model typically result in lower values. This tension, often referred to as the Hubble tension \cite{Valentino/2021a}, highlights a potential inconsistency in our understanding of the expansion rate of Universe and raises questions about the validity of the $\Lambda$CDM model or the presence of unknown systematic errors in the measurement and corresponds to roughly $4.4\sigma$ tension. The Planck data, measurements of weak lensing, and redshift surveys produce a second significant source of tension known as \textbf{$\sigma_{8}$ tension} \cite{Valentino/2021b}. This corresponds to the matter density ($\Omega_{m}$) and the amplitude or growth rate of the structure $(\sigma_{8},f\sigma_{8})$. Based on $\Lambda$CDM, the Planck collaboration estimates $S_{8}= \sigma_{8}\sqrt{\Omega_{m}/0.3} = 0.834 \pm 0.016$, while the KIDS-450 collaboration estimates $S_{8}= 0.745 \pm 0.039$ \cite{Joudaki/2017}. This results in approximately $2\sigma$ tension.

These challenges and inconsistencies have led to the exploration of alternative explanations beyond $\Lambda$CDM. In the following sections, we will review and discuss various theoretical frameworks that extend or modify $\Lambda$CDM aiming to address its limitations and provide a more comprehensive understanding of the Universe.

\section{Modified gravity}
The key motivation for the modified gravity scenario is to find suitable alternatives for the three main missing pieces of the standard model of cosmology, particularly inflation, DM, and DE. There are various methods to modify the GR, which can be categorized into two ways. One approach involves introducing new fields, while the other focuses on modification of the geometric framework. In this investigation, we will focus on the geometrical modification of GR, and hence, from now on, by the term modified gravity we mean the geometrical modification. Notably, some modified gravity models have shown promise in addressing the $H_0$ tension, especially through late-time solutions. Modified gravity models are effective in studying late-time epochs, particularly in producing late-time acceleration without the need for a DE component \cite{R34a}.

The Einstein-Hilbert action can be naturally extended within the framework of Riemannian geometry by replacing the Ricci scalar $R$ with any arbitrary function $f(R)$. This leads to  $f(R)$ modified theory of gravity.
There are two primary approaches to $f(R)$ gravity: the metric formulation and the palatini formulation. In the the metric formulation, the metric tensor is treated as the only dynamical variable, the connection is derived from it, and in the Palatini formulation, the metric tensor and the connection are considered independent fundamental variables. These formulations provide different perspectives on the modification of gravity and offer potential solutions to issues not fully addressed by GR. Detailed discussions of $f(R)$ gravity can be found in \cite{Sotiriou/2006,Capo/2008a,Motohashi/2019,Stachowski/2017}. The most glaring disadvantage of the $f(R)$ gravity theory is that the scalar field in the Palatini formulation is not dynamic. This implies that no additional degrees of freedom can be introduced, resulting in the existence of physically impossible infinite tidal forces.

\subsection{$f(R)$ gravity and its extensions}

One can generalize the Einstein-Hilbert action by replacing the Ricci scalar $R$ with a more general function  of $f(R)$ as 
\begin{equation}
\label{R}
S = \frac{1}{2\kappa}\int f(R) \, \sqrt{-g} d^{4}x  + \int  L_{m}\,  \, \sqrt{-g} d^{4}x.
\end{equation}
GR is also immediately recovered at $f(R)=R$.  To obtain the field equations of $f(R)$ gravity, we take into account the variation of action in equation \eqref{R} with respect to the metric tensor, which yields 
\begin{equation}
f'(R) R_{\mu \nu} - \frac{f(R)}{2} g_{\mu \nu} - \left(\nabla_{\mu} \nabla_{\nu} - g_{\mu \nu} \square  \right)f'(R) = 8 \pi G\, \mathcal{T}_{\mu \nu},
\end{equation}
where $\square = \nabla_{\mu}\nabla^{\mu}$ is the D'Alembert operator, $f'(R) =\frac{df(R)}{dR}$, and $\mathcal{T}_{\mu \nu}$ is the stress-energy-momentum tensor. 

DE models based on $f(R)$ theories have been extensively investigated as one of the simplest modified gravity frameworks to explain the late-time accelerated expansion of the Universe.  A notable example is the model with $f(R) = R + \beta R^{2}$ ($\beta>0$) which can drive an accelerated expansion due to the inclusion of the $R^{2}$ term. This specific model was first proposed by Starobinsky in 1980 as an early inflationary scenario demonstrating how modifications to the Einstein-Hilbert action can lead to significant cosmological effects, such as rapid expansion in the early Universe or late-time acceleration \cite{Staro/1980}. Another model with $f(R)= R - \frac{\beta}{R^{n}}$, ($\beta>0, n>0$) was proposed as a potential explanation for DE within the metric formulation of $f(R)$ gravity \cite{Nojiri/2003a,Carroll/2004}. However, this model was found to suffer from several significant issues. These include matter instability, where small perturbations in the matter density can grow uncontrollably \cite{Dolgov/2003,Faraoni/2006a}, and difficulties in satisfying local gravity constraints \cite{Faraoni/2006b,Erick/2006}, which are essential for consistency with solar system tests. Additionally, the model lacks a conventional matter-dominated epoch because of a strong coupling between DE and DM. These challenges highlight the complexity of constructing viable and reliable DE models within the framework of modified gravity theories.

Henceforth, the function $f(R)$ in modified gravity theories must satisfy the following key requirements to ensure a viable and physically consistent model \cite{Sotiriou/2010,Felice/2010}.
\begin{itemize}
\item To avoid ghost states, which are unphysical instabilities in theory, the function$f(R)$ must satisfy $f(R)>0$, for $R \geq R_{0}$, where $R_{0}$ represents the present value of the Ricci scalar. This ensures that the theory remains free of negative energy modes and maintains physical consistency. Additionally, this condition helps guarantee that the gravitational interaction remains attractive and avoids pathological behavior in the curvature regime relevant to cosmological observations.

\item To prevent the existence of a scalar degree of freedom with negative mass, known as tachyons, the second derivative of the function $f(R)$ with respect to $R$ must satisfy $f_{RR}>0$, for $R \geq R_{0}$. This condition ensures that the scalar mode associated with $f(R)$  gravity has a positive squared mass, avoiding tachyonic instabilities and maintaining the stability of the theory. It also guarantees that the theory remains free from unphysical behavior in the curvature regime relevant to cosmological observations.
\item $f(R)\rightarrow R-2\Lambda$, for $R \geq R_{0}$ is essential to ensure the presence of a matter-dominated era and to maintain agreement with local gravity constraints.
\item The condition for stability and late de Sitter limit of the Universe is given by $0< \frac{R f_{RR}}{f_{R}} <1$. 
\end{itemize}

An intriguing extension of gravitational theories involves incorporating a non-minimal coupling between geometry and matter into the action. This is achieved by introducing an arbitrary function that depends both on the scalar curvature $R$ and the Lagrangian density of matter leading to $f(R,L_{m})$ gravity or $f(R,\mathcal{T})$ gravity, where $\mathcal{T}$ is the trace of the energy-momentum tensor. These theories provide a richer framework for exploring the interplay between spacetime geometry and matter.
A particularly interesting feature of $f(R,\mathcal{T})$  gravity is that its field equations reduce to those of $f(R)$ when the energy-momentum tensor is traceless or $\mathcal{T}=0$.

The action in $f(R,\mathcal{T})$ gravity is given by \cite{Harko/2011}
\begin{equation}
\label{RT}
S = \frac{1}{2\kappa} \int f(R,\mathcal{T})\, d^{4}x \, \sqrt{-g}  + \int  L_{m}\, d^{4}x \, \sqrt{-g}.
\end{equation}
By varying the action in equation \eqref{RT} with respect to metric tensor, the gravitational field equations of $f(R,\mathcal{T})$ gravity is given as 
\begin{equation}
\label{FRT}
f_{R}(R,\mathcal{T}) R_{\mu \nu} -\frac{1}{2} g_{\mu \nu} f(R,\mathcal{T}) + \left(g_{\mu \nu} \square - \nabla_{\mu}\nabla_{\nu} \right) f_{R}(R,\mathcal{T}) = 8\pi G\, \mathcal{T}_{\mu \nu} -f_{\mathcal{T}}(R,\mathcal{T}) \mathcal{T}_{\mu \nu} - f_{\mathcal{T}}(R,T) \Theta_{\mu \nu},
\end{equation}
where $\Theta_{\mu \nu}= g^{\sigma \lambda} \frac{\delta \mathcal{T}_{\sigma \lambda}}{\delta g^{\mu \nu}}$, $f_{R}(R,\mathcal{T}) = \frac{df(R,\mathcal{T})}{dR}$, and $f_{\mathcal{T}}(R,\mathcal{T}) = \frac{df(R,\mathcal{T})}{d \mathcal{T}}$. 
In addition, $f(R,\mathcal{T})$ cosmology has been extensively studied, revealing a wide range of interesting cosmological behaviors and applications. For instance, in the case of a dust fluid, this theory can reproduce the $\Lambda$CDM model, which describes a Universe dominated by DE and cold dark matter. Additively $f(R,\mathcal{T})$ gravity can model phantom and non-phantom eras, where the Universe undergoes accelerated expansion with different equations of state. It can also mimic the behavior of exotic fluids like the Chaplygin gas, which interpolates between DM and DE, and it can be used for scalar field reconstruction, where the dynamics of a scalar field are effectively described by the modified gravity framework \cite{Jamil/2012,Alvarenga/2013,Fisher/2019}.
One can also write the covariant derivative of equation \eqref{RT} as 
\begin{equation}
\nabla^{\mu} \mathcal{T}_{\mu \nu} = \frac{f_{\mathcal{T}}(R,\mathcal{T})}{8\pi G - f_{\mathcal{T}(R,T)}} \left[(\mathcal{T}_{\mu \nu} + \Theta_{\mu \nu}) \nabla^{\mu} ln\,f_{\mathcal{T}}(R,\mathcal{T}) + \nabla^{\mu} \Theta_{\mu \nu} - \frac{1}{2} g_{\mu \nu} \nabla^{\mu} \mathcal{T} \right].
\end{equation} 

This highlights a key feature of $f(R,\mathcal{T})$: the energy-momentum tensor is not conserved, as its covariant divergence does not vanish. This nonconservation arises because of the coupling between matter and geometry in the theory. As a result, an additional force or acceleration is introduced, causing massive test particles to deviate from geodesic motion. Instead, they follow non-geodesic trajectories, reflecting the influence of the matter-energy coupling inherent in $f(R,\mathcal{T})$ gravity. This property distinguishes it from GR, where the energy-momentum tensor is conserved, and test particles move along geodesics \cite{Harko/2014b}.

\subsection{Geometric interpretation of torsion}
It is well known that an affine connection $\hat{\Gamma}$ and a metric structure $g$ are the building blocks of a spacetime manifold. This manifold is represented by $(M,g, \hat{\Gamma})$. The metric component describes distances, inner products and the mappings between contravariant and covariant tensor fields, whereas an affine connection characterizes the parallel transport of tensor fields with the help of covariant differentiation, enabling the comparison of vectors located in different vector spaces. In a general affine connection $\hat{\Gamma}$, both the non-metricity and torsion components contribute, and hence it is neither metric-compatible nor symmetric. This spacetime manifold structure equipped with a metric and a non-metric and non-symmetric affine connection is called non-Riemannian geometry \cite{R35a}.

Consider two curves $C$ and $\bar{C}$ described by $x^\mu(\tau)$ and $\bar{x}^\mu(\tau)$, along with the tangent vectors $u^\mu = \frac{dx^\mu}{d\tau}$ and $\bar{u}^\mu = \frac{d\bar{x}^\mu}{d\tau}$ respectively. Let us assume that in a parallel transport of the vector $u^\sigma$ along the curve $\bar{C}$ with the displacement $d\bar{x}^\mu$, we obtain the components of the displaced vector $u'^\sigma$ (up to first order) as follows
\begin{equation}\label{e71}
 u'^\sigma = u^\sigma + (\partial_\mu u^\sigma)d\bar{x}^\mu \text{.} 
\end{equation}
Since $u^\sigma$ is parallel transported along $\bar{C}$, we have
\begin{equation}\label{e72}
 \frac{d\bar{x}^\mu}{d\tau} (\hat{\nabla}_\mu u^\sigma) = 0 \implies \frac{d\bar{x}^\mu}{d\tau} (\partial_\mu u^\sigma + \hat{\Gamma}^\sigma_{\nu\mu} u^\nu) = 0 \implies (\partial_\mu u^\sigma)d\bar{x}^\mu = - \hat{\Gamma}^\sigma_{\nu\mu} u^\nu \bar{u}^\mu d\lambda \text{.} 
\end{equation}
From equation \eqref{e71} and equation  \eqref{e72}, we have
\begin{equation}\label{e73}
u'^\sigma = u^\sigma - \hat{\Gamma}^\sigma_{\nu\mu} u^\nu \bar{u}^\mu d\lambda \text{.} 
\end{equation}
Similarly, in the parallel transport of the vector $u'^\sigma$ along the curve $C$ with the displacement $dx^\mu$, we obtain
\begin{equation}\label{e74}
\bar{u}'^\sigma = \bar{u}^\sigma - \hat{\Gamma}^\sigma_{\mu\nu} \bar{u}^\mu u^\nu d\lambda \text{.} 
\end{equation}
On subtracting equation \eqref{e74} from equation \eqref{e73}, we obtain
\begin{equation}\label{e75}
 (u'^\sigma + \bar{u}^\sigma) - (u^\sigma + \bar{u}'^\sigma)  =  - ( \hat{\Gamma}^\sigma_{\nu\mu} - \hat{\Gamma}^\sigma_{\mu\nu} ) u^\nu \bar{u}^\mu d\lambda = - \hat{T}^\sigma_{\mu\nu} u^\nu \bar{u}^\mu d\lambda \text{.} 
\end{equation}
Observe that, in the case of the Levi-Civita connection of GR, i.e., $\hat{\Gamma} = \Gamma$, we obtain the right hand side of the above equation to be zero, and thus an infinitesimal parallelogram exists, since this connection is symmetric. But for the general affine connection $\hat{\Gamma}$, this is not the case; this parallelogram has been cracked into a pentagon due to the presence of torsion $\hat{T}^\sigma_{\mu\nu}$ in the spacetime geometry. One can define the vector $V^\sigma =  - \hat{T}^\sigma_{\mu\nu} u^\nu \bar{u}^\mu$ that measures this deviation of the cracked parallelogram \cite{R38a}.

\subsection{Geometric interpretation of non-metricity}
Consider a curve $C$ described by $x^\mu(\tau)$. Let us assume that two vectors $a^\mu$ and $b^\mu$ are parallely transported along this curve $C$ and having the inner product $a.b = a^\mu b^\nu g_{\mu\nu}$, then the total covariant derivative, denoted by $\hat{D}$, of the inner product $a.b$ along the curve $C$ is given by
\begin{equation}\label{e76}
\frac{\hat{D}}{d\tau}(a.b) = \frac{\hat{D}}{d\tau}(a^\mu b^\nu g_{\mu\nu}) = \frac{dx^\sigma}{d\tau} (\hat{\nabla}_\sigma a^\mu) b_\mu + \frac{dx^\sigma}{d\tau} (\hat{\nabla}_\sigma b^\nu) a_\nu  + \frac{dx^\sigma}{d\tau} (\hat{\nabla}_\sigma g_{\mu\nu} ) a^\mu b^\nu \text{.} 
\end{equation}
As the vectors $a^\mu$ and $b^\mu$ are parallely transported along this curve $C$, we have
\begin{equation}\label{e77}
 \frac{dx^\sigma}{d\tau} (\hat{\nabla}_\sigma a^\mu) = 0 \:\: \text{and}  \:\:  \frac{dx^\sigma}{d\tau} (\hat{\nabla}_\sigma b^\nu) = 0 \text{.} 
\end{equation}
On utilizing the above fact and the relation $\hat{Q}_{\sigma\mu\nu}=\hat{\nabla}_\sigma g_{\mu\nu} $, the equation \eqref{e76} becomes,
\begin{equation}\label{e78}
\frac{\hat{D}}{d\tau}(a.b) = \hat{Q}_{\sigma\mu\nu}  \frac{dx^\sigma}{d\tau} a^\mu b^\nu \text{.} 
\end{equation}
On taking $b^\mu=a^\mu$, we can have
\begin{equation}\label{e79}
\frac{\hat{D}}{d\tau}(|a|^2) =  \hat{Q}_{\sigma\mu\nu}  \frac{dx^\sigma}{d\tau} a^\mu a^\nu \text{.} 
\end{equation}
From the equation \eqref{e78} and equation \eqref{e79}, one can easily observe that both the inner product and the length of a vector change in a parallel transport along a given curve. Thus, in the case of the Levi-Civita connection of GR i.e. $\hat{\Gamma} = \Gamma$, we obtain the right-hand side of equation \eqref{e78} and equation \eqref{e79} to be zero, and hence the length of the vector is invariant under a parallel transport. But for the general affine connection $\hat{\Gamma}$, this is not the case; the length of a vector changes due to the non-metricity condition (metric incompatibility) or the presence of the non-metricity tensor $\hat{Q}_{\sigma\mu\nu}$ \cite{R38g}.

\subsection{$f(T)$ gravity and its extensions}
In the context of curvature, one of the simplest and most direct modifications is the theory known as $f(R)$ gravity \cite{R42a,Staro/1980}. This approach generalizes the Einstein-Hilbert action by extending it to an arbitrary function of the Ricci scalar, thereby providing a broader range of phenomena to explore. A decade ago, following the same spirit, the $f(T)$ theory was introduced \cite{R44a}, which is a generalization of the TEGR case involving nonlinear functions of the torsion scalar. Note that the teleparallel equivalent to the GR formulation is equivalent to GR at the level of field equations; however, the modification of these theories, i.e. $f(R)$ and $f(T)$ is not equivalent. The action of the $f(T)$ theory reads as follows,
\begin{equation}\label{e92}
S= \frac{1}{2\kappa}\int f(T) \sqrt{-g} d^4x + \int  L_m\sqrt{-g} d^4x \text{.} 
\end{equation}
On varying the above gravity action for the tetrad, we obtain the following field equation for the $f(T)$ theory,
\begin{equation}\label{e93}
 e^{-1}\partial_{\mu}(ee^{\gamma}_i S_{\gamma}{}^{\mu
\nu})f_T-f_Te^{\gamma}_i T^{\gamma}_{\mu\lambda}S_{\gamma}{}^{\lambda
\mu} +e^{\gamma}_i S_{\gamma}{}^{\mu\nu}\partial_{\mu}(T)f_{TT}+
\frac{1}{4}e^{\nu}_i f(T)= \frac{e^{\gamma}_iT^\nu_\gamma \text{,}}{2} 
\end{equation}
where $f_T=df(T)/dT$ and $f_{TT}=d^2f(T)/dT^2$. Note that the term $R=-T+B$ is a unique Lorentz scalar, but the single torsion scalar $T$ or the boundary term $B$ are not Lorentz scalars. Thus, GR and its modified gravity by curvature extension $f(R)$ are both locally Lorentz invariant, but this standard formulation of $f(T)$ is not locally Lorentz invariant. Therefore, the selection of tetrads is essential in $f(T)$ cosmological models, as various tetrads result in different field equations, which subsequently lead to distinct solutions. A tetrad can be considered a good tetrad if it does not place any limitations on the functional form $f(T)$ \cite{R45a}. However, this issue was eradicated in another invariant formulation of the theory $f(T)$, which is widely known as the covariant formulation of the $f(T)$ gravity \cite{R46a}. For a detailed view of $f(T)$ gravity and its cosmological implications, one can check the references \cite{R47a,R47aaa,R47abb,R47acc}.

In addition to $f(T)$ gravity, which modifies GR by introducing an arbitrary function of the torsion scalar $T$, the theory can be further extended by incorporating a dependence on the trace of the energy-momentum tensor $\mathcal{T}$.
This leads to $f(T,\mathcal{T})$ gravity, where the action includes an arbitrary function of both the torsion scalar and the energy momentum tensor. This gravity framework allows for new dynamics in the interplay between geometry (through torsion) and matter, offering potential solutions to cosmological and gravitational phenomena beyond those addressed by $f(T)$ or $f(R)$.
The action for $f(T,\mathcal{T})$ gravity is given by \cite{Harko/2014a}
\begin{equation}\label{e94}
S = \frac{1}{2 \kappa} \int \left[T+f(T,\mathcal{T})\right] \sqrt{-g}\, d^{4}x + \int L_{m} \sqrt{-g}\, d^{4}x.
\end{equation}
Varying the action with respect to tetrad yields the field equations given by 
\begin{multline}\label{e95}
(1+f_{T})\left[ e^{-1} \partial_{\mu} \left(e\,e^{\lambda}_{a} S_{\lambda}^{\,\,\,\, \alpha \mu}\right) - e^{\lambda}_{\,\,a} T^{\mu}_{\,\,\nu \lambda} S_{\mu}^{\,\,\,\, \nu \alpha} \right] + \left( f_{TT} \partial_{\mu}T + f_{T\mathcal{T}} \partial_{\mu}\mathcal{T} \right)e^{\lambda}_{\,\,a} S_{\lambda}^{\,\,\,\, \alpha \mu} + e^{\alpha}_{\,\,a} \left(\frac{f+T}{4} \right)\\
 - f_{\mathcal{T}} \left( \frac{e^{\lambda}_{\,\,a} \mathcal{T}_{\lambda}^{\,\, \alpha} + p e^{\alpha}_{\,\,a}}{2} \right) =\frac{e^{\lambda}_{\,\,a} \mathcal{T}_{\lambda}^{\,\, \alpha}}{2} 
\end{multline}
where $f_{T} = \frac{\partial f}{\partial T}$ and $f_{T\mathcal{T}} = \frac{\partial^{2}f}{\partial T \, \partial \mathcal{T}}$. 

\subsection{$f(Q)$ gravity and its extensions}
In the year 2018, the $f(Q)$ theory was introduced \cite{R51a}, which is a generalization of the STEGR case involving non-linear functions of non-metricity scalar. Note that the STEGR formulation is equivalent to GR at the level of field equations; however, the modification of this theory, that is, modified symmetric teleparallel gravity or $f(Q)$ gravity, is neither equivalent to $f(R)$ gravity nor to $f(T)$ gravity. The action of the $f(Q)$ theory reads as follows
\begin{equation}\label{e105}
S= \frac{1}{2\kappa}\int f(Q) \sqrt{-g} d^4x + \int  L_m\sqrt{-g} d^4x \text{.}  
\end{equation}
On varying the generic action \eqref{e105} with respect to metric, we obtained the following metric field equation of the $f(Q)$ gravity as follows,
\begin{equation}\label{e106}
\frac{2}{\sqrt{-g}}\nabla_{\sigma}\left( \sqrt{-g}f_Q {P^{\sigma}}_{\mu\nu}\right)+\frac{1}{2}g_{\mu\nu}f
+f_Q\left(P_{\mu\sigma\rho}{Q_{\nu}}^{\sigma\rho}-2Q_{\sigma\rho\mu}{P^{\sigma\rho}}_{\nu} \right)=-\mathcal{T}_{\mu\nu} \text{,} 
\end{equation} 
where $f_Q=\frac{df}{dQ}$. Again, on varying the generic action in equation \eqref{e105} with respect to symmetyric teleparallel connection, we obtained the following connection field equation of the $f(Q)$ gravity as follows
\begin{equation}\label{e107}
\nabla_{\mu}\nabla_\nu \left( \sqrt{-g}f_Q {P^{\mu\nu}}_{\sigma} \right) = 0 \text{.} 
\end{equation}

Another specific modified gravity that yields a general class of non-linear gravity model having the action as \cite{Xu/2019}
\begin{equation}
S = \frac{1}{2\kappa} \int f(Q,\mathcal{T}) \sqrt{-g}\, d^{4}x + \int L_{m} \sqrt{-g}\, d^{4}x,
\end{equation}
where $f(Q,\mathcal{T})$ is a general function of $Q$ and the trace of energy-momentum tensor $\mathcal{T}$. In the presence of geometry-matter coupling, the general field equation describing gravitational phenomena is obtained by varying the action with respect to the metric tensor
\begin{equation}
\label{QTF}
\frac{2}{\sqrt{-g}} \nabla_{\alpha}\left(\sqrt{-g}f_{Q} P^{\alpha}_{\,\,\mu \nu}\right) + \frac{1}{2}g_{\mu \nu} f -f_{\mathcal{T}} \left( \mathcal{T}_{\mu \nu} + \Theta_{\mu \nu} \right) + f_{Q}\left(P_{\mu \alpha \beta} Q_{\nu}^{\,\, \alpha \beta} - 2 Q_{\alpha \beta \mu} P^{\alpha \beta}_{\,\,\,\,\nu}\right) = - \mathcal{T}_{\mu \nu} .
\end{equation}

Moreover, another extension of non-metricity based modified theories has been proposed in \cite{PRR} as $f(Q,T_{\mu\nu}T^{\mu \nu})$ theory, also known as Energy Momentum Squared Symmetric Teleparallel Gravity (EMSSTG). The generic action of the $f(Q,T_{\mu\nu}T^{\mu \nu})$ theory is given by \cite{PRR}
\begin{equation}\label{actionv}
S = \frac{1}{2\kappa} \int d^4x \, \sqrt{-g} \, f\left(Q, \mathcal{T}_{\mu\nu} \mathcal{T}^{\mu\nu} \right) + \int d^4x \, \sqrt{-g} \, L_m \, .
\end{equation}

Here, $g = \det(g_{\mu\nu})$  and $\mathcal{T}_{\mu\nu} \mathcal{T}^{\mu \nu}$ where $\mathcal{T}_{\mu\nu}$ is the matter energy-momentum tensor.

We obtain the field equations of $f(Q,\mathcal{T}^2)$ as follows

\begin{equation}
\frac{2}{\sqrt{-g}}\nabla_{\alpha}(f_Q\sqrt{-g}P^{\alpha}_{\mu \nu}) + \frac{1}{2}f(Q, \mathcal{T}^2)g_{\mu \nu} -f_{\mathcal{T}^2}\theta_{\mu \nu} + f_{Q}(P_{\mu \alpha \beta}Q^{\alpha \beta}_{\nu} + 2Q^{\alpha \beta}_{\mu}P_{\alpha \beta \nu}) = \mathcal{T}_{\mu \nu}
\end{equation}

Here, ${\mathcal{T}}_{\mu\nu}$ is the stress-energy tensor defined as
\begin{equation}\label{2hv}
{\mathcal{T}}_{\mu\nu} = \frac{-2}{\sqrt{-g}} \frac{\delta(\sqrt{-g}L_m)}{\delta g^{\mu\nu}}
\end{equation}
and 
\begin{equation}
\theta_{\mu \nu} = \frac{(\delta \mathcal{T}_{\alpha \beta} \mathcal{T}^{\alpha \beta})}{\delta g^{\mu\nu}}  
\end{equation}
Furthermore, the connection field equation that results from varying equation \eqref{actionv} is as follows
\begin{equation}
\nabla_{\mu}\nabla_{\nu}(\sqrt{-g}f_QP^{\mu \nu}_{\alpha} + 4\pi H^{\mu \nu}_{\alpha}) = 0,
\end{equation}
where
\begin{equation}
H^{\alpha \beta}_{\rho} = \frac{\sqrt{-g}}{16 \pi}f_{\mathcal{T}{^2}}\frac{\delta T^2}{\delta \mathcal{T}^{\rho}_{\alpha \beta}} + \frac{\delta \sqrt{-g}L_{m}}{\delta \mathcal{T}^{\rho}_{\alpha \beta}}.
\end{equation}

\section{Parameter estimation method}

\subsection{$\chi^2$ minimization}
Consider a function $f_{model}(x,\theta)$ in the independent variable $x$ and the set of free parameters $\theta$. If the set $\{f_{k,obs}(x_{k,obs})\}_{k=1}^n$ represents $n$ independent observation along with the standard deviation $\sigma_{k,obs}$, then we define the $\chi^2$ as follows \cite{R52a}
\begin{equation}\label{e108}
\chi^2(\theta) = \sum_{k=1}^n  \frac{\left( f_{model}(x_{k,obs},\theta) - f_{k,obs} \right)^2}{\sigma_{k,obs}^2} \text{.} 
\end{equation}
Moreover, if these $n$ observations are not independent, then in this case a covariance matrix $C$ is utilized instead of the standard deviation. In such a case the $\chi^2$ function becomes
\begin{equation}\label{e109}
\chi^2(\theta) = \sum_{i,j=1}^n X_i C^{-1}_{ij} X_j \text{,} 
\end{equation}
where $X_i = f_{i,obs} - f_{model}(x_{i,obs},\theta) $. One can define another important statistical quantity called likelihood function as follows
\begin{equation}\label{e110}
\mathcal{L}(\theta) = P(\theta| D = Data) \text{}. 
\end{equation}
One can obtained the following relation between the above two statistical parameters as follows
\begin{equation}\label{e111}
\mathcal{L}(\theta) \propto exp\left( -\frac{1}{2} \chi^2(\theta) \right) \text{}.  
\end{equation}
Our ultimate aim is to find the best fit values of the parameter set $\theta$ so that the function $f_{model}(x,\theta)$ agrees with the observational data. In order to investigate this, one has to minimize the $\chi^2(\theta)$ function (which is equivalent to maximize the likelihood function), and the parameter corresponds to the $\chi^2_{min}$ value, say $\theta_{min}$ is the required best-fit parameter value. Several approaches exist in the literature to solve this optimization issue, such as the Gradient Descent algorithm, Newton's method, and the Random Walk algorithm \cite{R53a,R54a}. The disadvantage of these algorithms is that whenever the parameter space exhibits too many local minima, the best-fit value obtained by these methods is local rather than global. However, utilizing the features of these algorithms, a new efficient algorithm can be constructed. One such algorithm is widely adopted in the computational field of science known as the Markov Chain Monte Carlo (MCMC) algorithm. 

\subsection{The MCMC approach }
Over the past decade, probabilistic data analysis, particularly Bayesian statistical inference, has transformed scientific research. This approach relies on the use of either the posterior probability density function or the likelihood function. Although finding the optimal values of these functions can often be achieved using various algorithms, a more comprehensive understanding of the posterior PDF is frequently necessary. To address this, MCMC methods have been developed. These methods are specifically designed to efficiently sample from the posterior probability density function, even in high-dimensional parameter spaces, enabling robust and detailed exploration of complex models and their uncertainties. MCMC techniques have become the cornerstone of modern data analysis, allowing researchers to extract meaningful insights from intricate data sets \cite{R55a}. Mathematically, a Markov chain is a sequence of parameter values generated by a random process, where each step depends only on the current state and its immediate predecessor, embodying the memory less property. The Monte Carlo process complements this by exploring the entire parameter space, enabling efficient sampling from complex probability distributions. The core idea of MCMC is to construct a Markov chain that samples the parameter space of a model according to a specified probability distribution, such as the posterior distribution in Bayesian inference. The chain is built iteratively: at each step, a new parameter value is proposed based on a proposal distribution, which suggests a transition from the current state. The acceptance of this proposed value depends on its posterior probability, which incorporates both the likelihood of the observational data and the prior probability of the parameters. Once the chain converges to a stationary distribution, the posterior distribution of the parameters can be estimated by analyzing the frequency of parameter values in the chain. This posterior distribution provides insight into the values of the optimal parameters and their associated uncertainties, facilitating predictions for various observables. One of the most widely used MCMC algorithms is the Metropolis-Hastings algorithm, which employs a random walk approach to propose and accept new parameter values. This method has become a cornerstone of probabilistic data analysis, enabling researchers to explore high-dimensional parameter spaces and extract meaningful inferences from complex models. Another more advanced algorithm has recently been proposed by Mackey, called the EMCEE: The MCMC Hammer \cite{R56a}. The EMCEE algorithm performs better than traditional MCMC sampling techniques. When using an emcee, one can go with a large number of walkers, often hundreds. Theoretically, there is no downside to increasing the number of walkers, unless you encounter performance limitations. The detailed discussion on the different MCMC algorithms compared with the EMCEE sampling method is beyond the scope of this thesis.

\section{Conclusions}

In this chapter, we have discussed the modified gravity scenario as an alternative to the DE candidate. In the present thesis, we present the accelerating cosmological models in different non-Riemannian gravity formalisms such as modified TEGR and STEGR. This thesis aims to reproduce the DE effects originating from the modification of spacetime geometry, bypassing the need for a controversial cosmological constant. Furthermore, for the statistical assessment and model parameter estimation, the MCMC approach is utilized, along with the emcee sampler and Bayesian statistical inference. The following chapters enclose the detailed investigation and the corresponding outcomes.


\chapter{Squared torsion $f(T,\mathcal{T})$ gravity and its cosmological implications} 

\label{Chapter2} 

\lhead{Chapter 2. \emph{Squared torsion $f(T,\mathcal{T})$ gravity and its cosmological implications}} 

\vspace{10 cm}
* The work in this chapter is covered by the following publications: \\
 
\textit{Squared torsion $f(T,\mathcal{T})$ gravity and its cosmological implications}, Fortschritte der Physik {\bf 71}, 2200162 (2023).

\clearpage
\pagebreak

This chapter introduces the coupling of the torsional scalar $T$ and the trace of the energy moment tensor $\mathcal{T}$. Furthermore, consider the functional form $f(T,\mathcal{T})=\alpha\mathcal{T}+\beta T^2$. As an alternative to the cosmological constant, $f(T,\mathcal{T})$ theory can provide a theoretical explanation of slow acceleration. The most recent observational data on the model under consideration, particularly the limitations of the model parameters, are used in detail. Additionally, we analyze the cosmological behavior of delay, the effective EoS, and the overall equation of condition parameters. However, we can see that the deceleration parameters represent the transition from deceleration to acceleration, and that effective dark sectors exhibit quintessence like evolution.

\section{Introduction}\label{sec1x}
The thorough validation of late-time acceleration has sparked extensive research aimed at uncovering its underlying causes. It is widely recognized through observations of Type Ia SN \cite{R9a,R10a}, BAO \cite{Eisenstein/2005, Percival/2007}, CMB \cite{Komatsu/2011}, and $H(z)$ measurements \cite{Farooq/2017}.  DE, which attempts to explain late-time acceleration as a result of an energy associated with the cosmological constant, is one of the leading and widely accepted models. To explore alternatives beyond conventional DE models, one can extend the general theory of relativity by modifying the underlying geometry. Alternative theories such as $f(R)$ gravity \cite{Staro/2007,Capo/2008,Chiba/2007}, a coupling between matter and curvature through $f(R,\mathcal{T})$ gravity \cite{Harko/2011,Moraes/2017},  $f(R,G)$ \cite{Laurentis/2015,Gomez/2012} ($G$ is the Gauss-Bonnet) have all attempted to explain the DE phenomenon in the context of curvature.
Extending the action of modified gravity based on torsion gives rise to a distinct and fascinating class of modified gravity known as the teleparallel equivalent of GR or $f(T)$ gravity. However, numerous studies in $f(T)$ gravity have been conducted, such as cosmological solutions \cite{Paliathanasis/2016}, late-time acceleration \cite{Myrzakulov/2011, Bamba/2011}, thermodynamics \cite{Salako/2013}, cosmological perturbations \cite{Chen/2011}, cosmography \cite{Capozziello/2011} as documented in the literature.  For a complete analysis of $f(T)$ gravity, one can refer to \cite{Cai/2016}.\\
Another novel approach in modified gravity involves incorporating the coupling between torsion and the trace of the energy-momentum tensor, $f(T,\mathcal{T})$ theory, analogous to $f(R,T)$ gravity. This theory was introduced in \cite{Harko/2014a}, but its compatibility with cosmological data and the essential physical conditions for a consistent cosmological framework still require verification. The interaction between torsion and matter broadens the potential explanations for the nature of DE, or more specifically, the driving mechanism behind the observed acceleration. This theory has been explored in the context of reconstruction and stability \cite{Junior/2016,Momeni/2014}, late-time acceleration and inflationary phases \cite{Harko/2014a}, growth factor of subhorizon modes \cite{Farrugia/2016}, quark stars \cite{Pace/2017}. \\
In this chapter, we investigate a squared-torsion $f(T,\mathcal{T})$ model that raises a question about the viability of such a theory as a candidate to account for late-time acceleration. Further, the parameters are constrained using the set of observational data sets and in particular, we check the late-time accelerating behavior holds true for $f(T,\mathcal{T})$ using the cosmological parameters. \\
The plan of the work is the following: starting from the motion equations in the flat FLRW background in the framework of $f(T,\mathcal{T})$ gravity in section \eqref{sec2x}. Section \eqref{sec3x} is devoted to the cosmological framework and the solutions to the field equations. Specifically, in section \eqref{sec4x}, we deal with the observational data and methodology used to constrain the parameters involved. 
The late-time accelerated phase is examined in section \eqref{sec5x} through cosmological evolution. Finally, a conclusion is given in section \eqref{sec6x}.

\section{Field equations} \label{sec2x}

We incorporate the following flat FLRW metric to obtain modified Friedmann equations
\begin{equation}
\label{9x}
 ds^{2}=-dt^{2}+a(t)^{2} \delta_{ij}dx^{i} dx^{j}, 
\end{equation}
where $a(t)$ is the scale factor. Further, equation \eqref{e95} give rise to modified Friedmann equations
\begin{equation}
\label{10x}
H^2 =\frac{8\pi G}{3}\rho_m - \frac{1}{6}\left(f+12H^2f_T \right)+f_\mathcal{T}\left(\frac{\rho_m+p_m}{3} \right),
\end{equation}
\begin{multline} 
\label{11x}
\dot{H}= -4\pi G(\rho_m+p_m)-\dot{H}(f_T-12H^2 f_{T \mathcal{T}})-H(\dot{\rho_m}-3\dot{p_m}) f_{T \mathcal{T }} - f_\mathcal{T}\left(\frac{\rho_m+p_m}{2} \right).
\end{multline}
Here, $\mathcal{T}=\rho_m-3p_m$ in the above equation is true for the fluid of perfect matter.

Comparing the modified Friedmann equation \eqref{10x} and equation \eqref{11x} to GR equations 
\begin{eqnarray}
\label{12x}
H^2 &=& \frac{8 \pi G}{3}\left(\rho_m + \rho_{eff}\right),\\
\label{13x}
\dot{H} &=& - 4\pi G \left(\rho_m + p_m + \rho_{eff}+p_{eff}\right).
\end{eqnarray}
we obtain 
\begin{equation}\label{14x}
\rho_{eff} =\frac{1}{16\pi G}[f+12f_T H^2-2f_\mathcal{T}(\rho_m +p_m)],
\end{equation}
\begin{multline}\label{15x}
p_{eff} = \frac{1}{16\pi G}[f+12f_T H^2-2f_\mathcal{T}(\rho_m +p_m)]+ \\ (\rho_m+p_m)\left[\frac{(1+\frac{f_T}{8\pi G})} {1+f_T 12H^2 f_{TT}+H(\frac{d\rho_m}{dH})(1-3{c_{s}}^2)f_{T \mathcal{T}}} -1\right].
\end{multline}
The effective and total equation-of-state parameter is defined as follows
\begin{eqnarray}
\label{16x}
\omega_{eff} &=& \frac{p_{eff}}{\rho_{eff}},\\
\label{17x}
\omega &= & \frac{p_{eff} +p_m}{\rho_{eff}+\rho_m}.
\end{eqnarray}
We consider $p_m =0$ for the dust Universe which implies $ \omega =\frac{\omega_{eff}}{1+\frac{\rho_m}{\rho_{eff}}}$. The conservation equation involving the effective energy and pressure reads
\begin{equation}\label{18x}
\dot{\rho}_{eff}+\dot{\rho_m}+3H(\rho_m+\rho_{eff}+p_m+p_{eff})=0. 
\end{equation}
  
\section{Cosmology} \label{sec3x}
 This section examines the cosmological impacts of $f(T,\mathcal{T})$ gravity while emphasizing on a specific model. We consider the functional form $f(T,\mathcal{T)}=\alpha\mathcal{T}+\beta T^2=\alpha \rho_m +\beta T^2 =\alpha \rho_m+ \gamma H^4 $, where $\alpha $ and $\gamma=36\beta$  are constants \cite{Harko/2014a}. For simplicity, we use $8 \pi G= c=1$. The model defines a straightforward deviation from GR within the framework of $f(T,\mathcal{T)}$. In case $\alpha=0$, the model behaves as a power-law cosmology in the $f(T)$ theory \cite{Capozziello/2011}. In this case, we obtain $f_T=\frac{\gamma T}{18}$, $f_{TT} =\frac{\gamma}{18}$ ,$f_\mathcal{T}=\alpha$,  $f_{T \mathcal{T}}=0$.\\
 Hence, using the above expressions, equation  \eqref{10x} \& equation  \eqref{11x}, we have the following
\begin{eqnarray}
\label{19x}
\rho_{m} &=& \frac{3\left(1-\frac{\gamma H^2}{2} \right)}{1+\frac{\alpha}{2}} H^2, \\ 
\label{20x}
\dot{H} &=& -\frac{3(1+\alpha) \left(1-\frac{\gamma H^2}{2}\right)}{(\alpha +2) (1-\gamma H^2)} H^2,\\
\label{21x}
 q &=&\frac{3(1+\alpha)\left(1-\frac{\gamma H^2}{2}\right)}{(\alpha+2)(1-\gamma H^2)}-1.
\end{eqnarray}

Moreover, the effective DE density and pressure from equation \eqref{14x} and equation \eqref{15x} can be obtained as
\begin{eqnarray}
\label{22x}
    \rho_{eff} &=&\frac{3H^2(\alpha+\gamma H^2)}{\alpha+2},\\
    \label{23x}
    p_{eff}&=&-\frac{3H^2(\alpha+\gamma H^2)}{(\alpha+2)(\gamma H^2-1)},
    \end{eqnarray}
which gives $ \omega_{eff}=\frac{1}{1-\gamma H^2}$.\\
Now, we replace the term $d/dt$ by $d/dlna$ via the expression $d/dt = H \frac{d}{dlna}$, ($a=\frac{1}{1+z}$) such that solution of equation \eqref{20x} is
\begin{equation}
\label{24x}
    H(z)= H_0 \sqrt{\frac{\sqrt{1-2^{-2 A} (2 z+2)^{2 A} \left(2 \gamma  H_{0}^2-\gamma ^2 H_{0}^4\right)}+1}{\gamma H_{0}^2}},
\end{equation}
where $A=\frac{3 (\alpha +1)}{\alpha +2}$.

\section{Observational constraints and methodology} \label{sec4x}
In this section, we perform a statistical analysis using the MCMC approach, comparing model predictions with observational data to evaluate its viability. Specifically, we used SNeIa data, BAO data, and H(z).

\subsection{SNeIa data}
Since Type Ia supernovae serve as standard candles, they enable the estimation of cosmic distances and are widely utilized to place constraints on the DE sector. In this study, we specifically employ the Pantheon compilation, which consists of 1,048 data points covering a wide red-shift range $0.01<z<2.26$ \cite{R18a}. The $\chi^{2}$ function is given as 
\begin{equation}
\chi^2 _{SN}= \Delta\mu C^{-1}_{SN} \Delta\mu^{T},
\end{equation}
where $\Delta \mu= \mu_{i}-\mu_{th}$ is the difference between the observational and theoretical distance modulus, and $C^{-1}_{SN}$ corresponds to the inverse covariance matrix of the data. Furthermore, we define $\mu= m_{B}-M_{B}$, where $m_{B}$ is the apparent magnitude observed at a given red-shift, while $M_{B}$ is the absolute
magnitude (retrieving nuisance parameters according to the new approach called BEAMS with Bias Correction (BBC) \cite{Kessler/2017}). The theoretical value is computed as 
\begin{eqnarray}
\mu_{th}&=& 5 log_{10}\left[\frac{d_{L}}{1 Mpc}\right]+25,\\
d_{L}&=& c(1+z) \int_{0}^{z} \frac{dy}{H(y,\theta)}.
\end{eqnarray}
where $\theta$ is the parameter space.

\subsection{Hubble data}
We make use of measurements of Hubble parameters derived from the differential age method. Here, we consider 31 points compiled in \cite{Moresco/2015}.
The $\chi^{2}$ function is given as 
\begin{equation}
\chi^2_{Hz}= \sum_{i=1}^{31} \frac{\left[H(z_{i})-H_{obs}(z_{i})\right]^2}{\sigma(z_{i})^{2}},
\end{equation}
where $H_{obs}$ is the observed value, $\sigma(z_{i})$ is the observational error.

\subsection{Baryon acoustic oscillations}
Baryon Acoustic Oscillations (BAO) are pressure waves resulting from cosmological disturbances in the baryon-photon plasma during the recombination period, which are observed as distinct peaks on large angular scales. In our study, we integrate BAO data from the Six Degree Field Galaxy Survey (6dFGS), the Sloan Digital Sky Survey (SDSS), and the LOWZ samples of the Baryon Oscillation Spectroscopic Survey (BOSS) to improve the precision and robustness of our findings \cite{Blake/2011,Percival/2010}. The expressions used for BAO data are
\begin{eqnarray}
d_{A}(z) &=& c \int_{0}^{z} \frac{dz'}{H(z')},\\
D_{v}(z) &=& \left[\frac{d_{A}(z)^2 c z }{H(z)}\right]^{1/3},\\
\chi_{BAO}^2 &=& X^{T} C^{-1} X. 
\end{eqnarray}
Here, $d_{A}(z)$ is the distance between comoving angular diameters, $D_{v}(z)$ is the dilation scale, and $C$ is the covariance matrix \cite{Giostri/2012}.

\subsection{Results}
\begin{figure}[]
\centering
\includegraphics[scale=0.9]{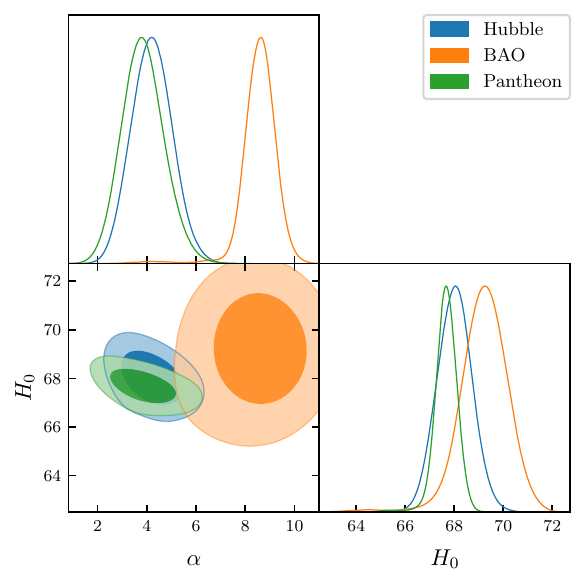}
\caption{One-dimensional and two-dimensional marginalized confidence regions (68\% CL and 95\% CL) for $\alpha$, $H_0$ obtained from the Hubble, BAO and Pantheon data for the f(T,$\mathcal{T}$) gravity model.}
\label{figure 1x}
\end{figure}

\begin{figure}[]
\centering
\includegraphics[scale=0.9]{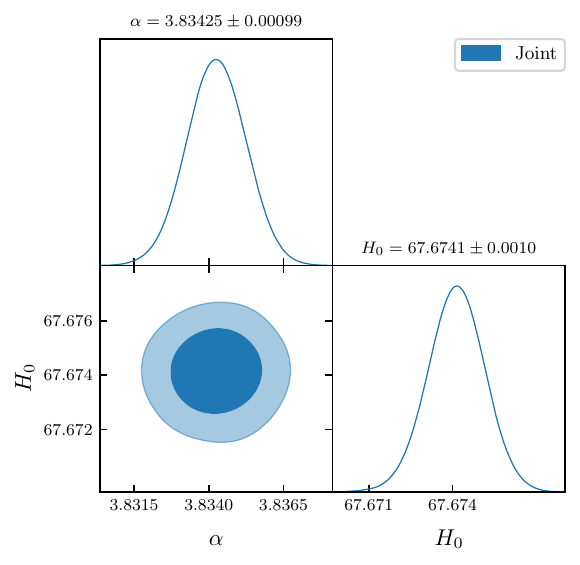}
\caption{One-dimensional and two-dimensional marginalized confidence regions (68\% CL and 95\% CL) for $\alpha$, $H_0$ obtained from the Hubble+BAO+Pantheon data for the f(T,$\mathcal{T}$) gravity model.}
\label{figure 2x}
\end{figure}

The statistical results for the model are presented as contour plots in 
figure  \eqref{figure 1x} and figure \eqref{figure 2x}. In addition, Table \eqref{table1x} summarizes the parameter space values obtained from the combination of different data sets. We noticed weaker constraints for the case of BAO, whereas stronger constraints were observed for SNeIa and joint analysis (Hz + BAO + SNeIa). We assumed $\gamma=0.0006$ \cite{Harko/2014a} so that $\frac{\gamma H^2}{2}<1$. We can observe that the BAO data is anticorrelated with other data sets. We also get a constraint $H_{0}$ consistent with the Planck results \cite{Planck/2018} favored by $\Lambda$CDM. 

\begin{figure}[H]
\centering
\includegraphics[scale=0.5]{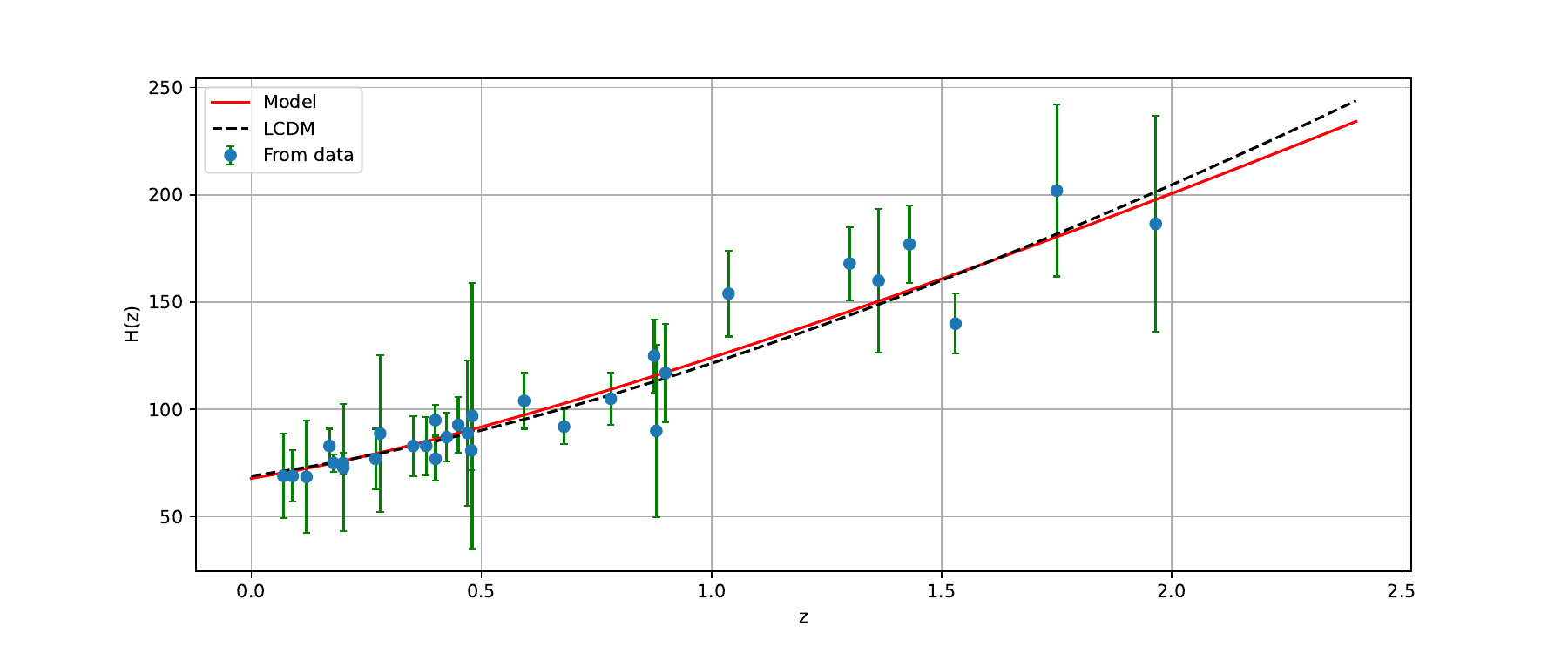}
\caption{Plot shows the expansion rate of the Universe with theoretical predictions (red curve) and $\Lambda$CDM (black curve with $\Omega_{\Lambda_0}=0.7$ and $\Omega_{m_{0}}=0.3$). The blue dots represent 31 Hubble points with the corresponding error bars.}
\label{figure errorx}
\end{figure}

\begin{figure}[H]
\centering
\includegraphics[scale=0.5]{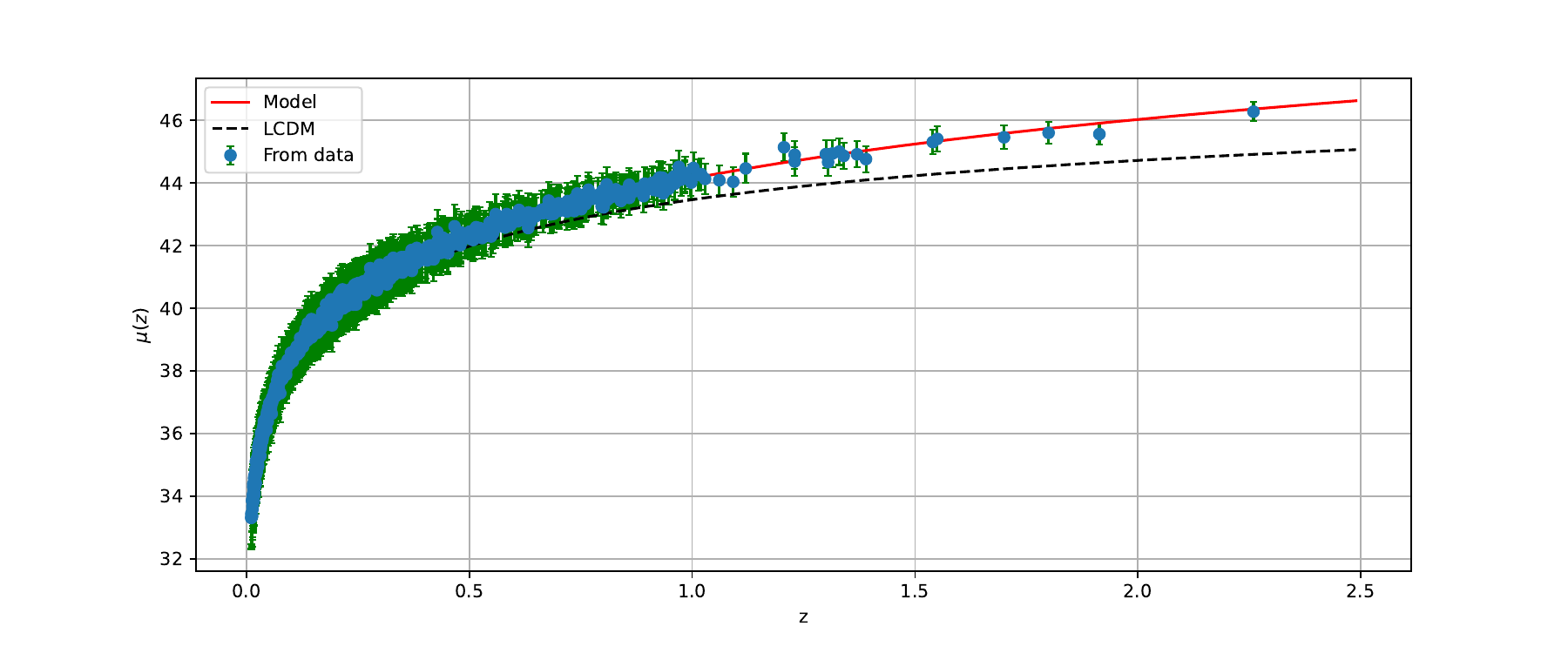}
\caption{Plot shows the $\mu$-red-shift relation of Pantheon SN sample  with theoretical predictions (red curve) and $\Lambda$CDM (black curve). with $\Omega_{\Lambda_0}=0.7$ and $\Omega_{m_{0}}=0.3$). The blue dots represent 1048 Pantheon points with the corresponding error bars.}
\label{figure mux}
\end{figure}

\begin{table}[H]
\begin{center}
 \caption{Best-fit values of model parameters obtained from observational data sets}
    \label{table1x}
\begin{tabular}{|l|c|c|c|c|c|}
\hline 
data sets              & $\alpha$ & $H_{0}$ & $z_{t}$ & $q_{0}$ & $\omega_{0}$\\
\hline

$BAO$           & $8.55^{+0.63}_{-0.53}$  & $69.20^{+1.0}_{-1.0}$ &  $0.36^{+0.04}_{-0.04}$ & $-0.36^{+0.03}_{-0.04}$ & $-0.57^{+0.024}_{-0.024}$\\
\hline
$Hz$     & $4.21^{+0.82}_{-0.82}$  & $68.01^{+0.71}_{-0.71}$  & $0.60^{+0.141}_{-0.101}$ & $-0.45^{+0.04}_{-0.04}$ &  $-0.63^{+0.024}_{-0.026}$ \\
\hline
$SNeIa$     & $3.85^{+0.79}_{-0.91}$  & $67.66^{+0.48}_{-0.48}$  & $0.65^{+0.198}_{-0.092}$ &  $-0.46^{+0.02}_{-0.04}$& $-0.64^{+0.019}_{-0.024}$\\
\hline
$Hz+BAO+SNeIa$             & $3.83^{+0.0009}_{-0.0009}$  & $67.67^{+0.0010}_{-0.0010}$ & $0.65^{+0.0004}_{-0.0017}$ &  $-0.46^{+0.0001}_{-0.937}$ & $-0.64^{+0.00003}_{-0.00003}$\\ 
\hline
\end{tabular}
\end{center}
\end{table}

\section{Cosmological evolution} \label{sec5x}
\begin{figure}[]
\centering
\includegraphics[scale=0.5]{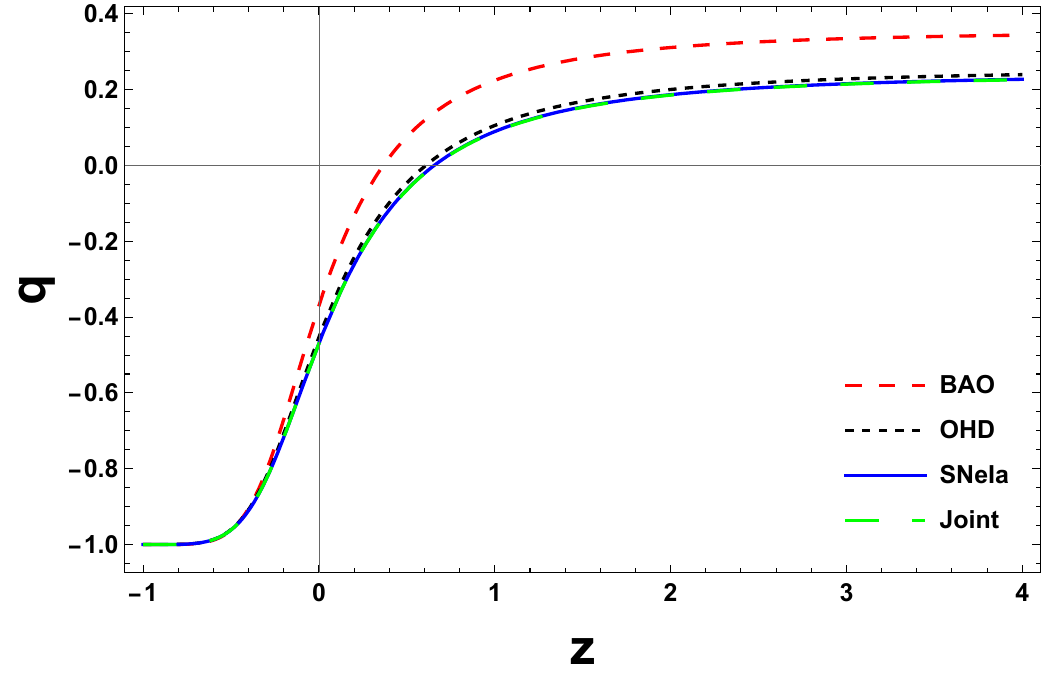}
\caption{Variation of the deceleration parameter $q$ as a function of the red-shift $z$ for different data sets.}
\label{figure 4x}
\end{figure}

The plots in this section illustrate how the Universe can exhibit fascinating dynamical behaviors, depending on the chosen values of the model parameters. 
The Hubble function, as shown in figure \eqref{figure errorx}, exhibits a monotonically increasing behavior with red-shift throughout the entire evolutionary history of the Universe.\\
The figure \eqref{figure 4x} shows that the Universe begins its history from deceleration ($q>0$) and shows the accelerating phase ($q<0$) after a red-shift transition $z_{t}$. The deceleration parameter is defined as $q= -\frac{\dot{H}}{H^2}-1$. This evolution aligns with the observed behavior of the Universe, which has undergone three distinct phases: a decelerating matter-dominated phase, a transition into an accelerating expansion phase, and a late-time accelerated expansion. Notably, the Universe asymptotically approaches a de Sitter expansion at lower red-shifts. We find that the present value of the deceleration parameter ($q_{0}$) \cite{Almada/2019,Basilakos/2012}  and $z_{t}$ \cite{Garza/2019,Jesus/2020} is in good agreement with the data sets SNeIa and Hz+BAO+SNeIa.

\begin{figure}[]
\centering
\includegraphics[scale=0.5]{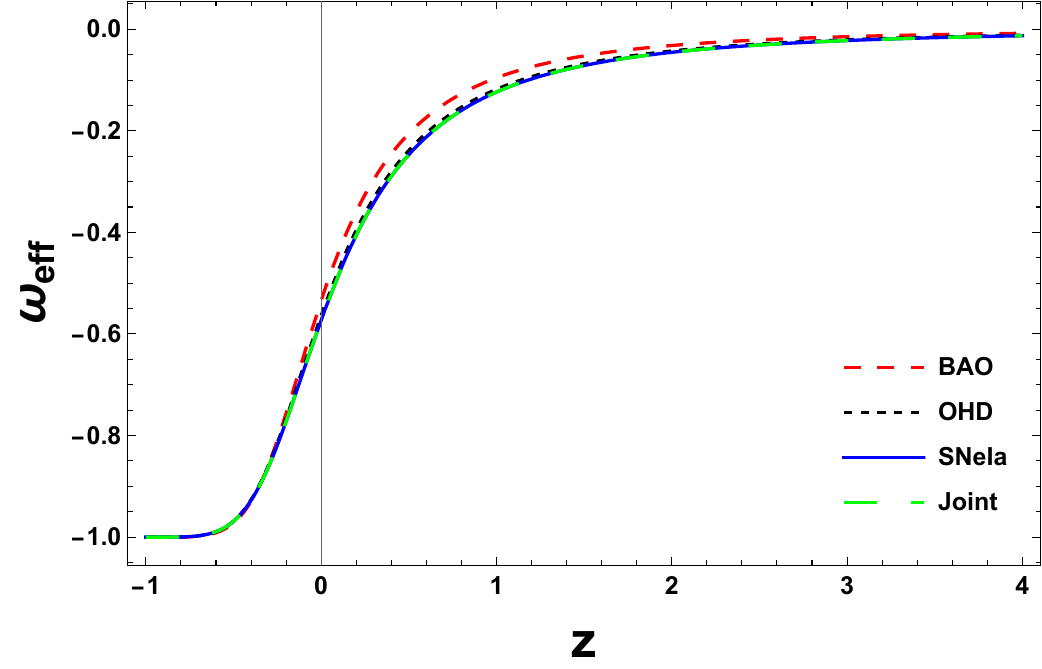}
\caption{Variation of $\omega_{eff}$ as a function of the red-shift $z$ for different data sets}
\label{figure 5x}
\end{figure}

\begin{figure}[]
\centering
\includegraphics[scale=0.5]{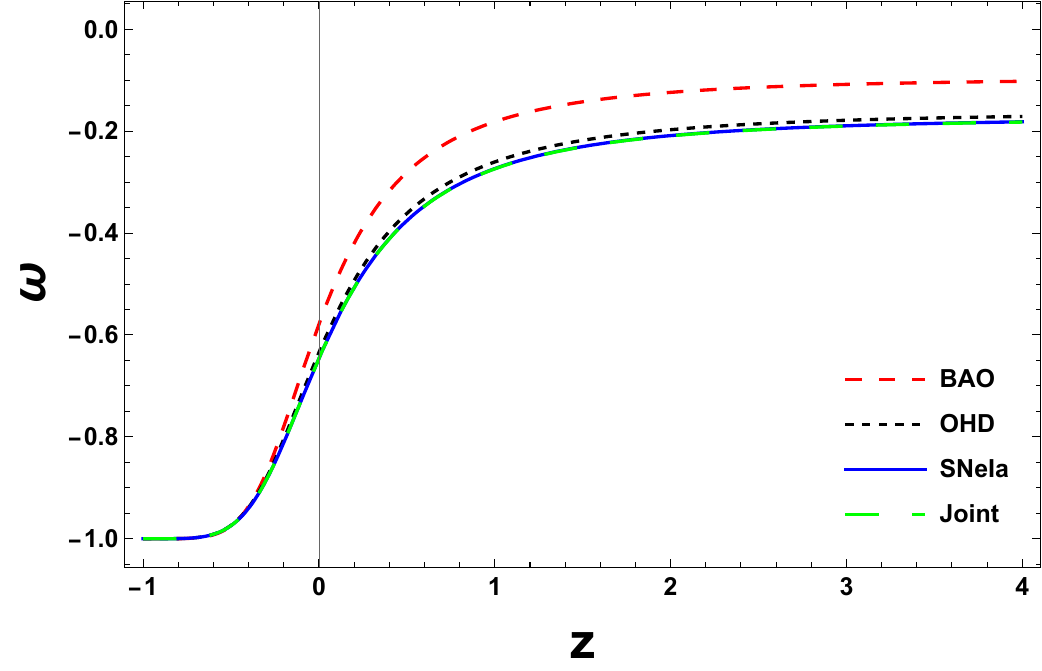}
\caption{Variation of $\omega$  as a function of the red-shift $z$ for different data sets}
\label{figure 6x}
\end{figure}

Determining the EoS value and its evolution is another attempt to comprehend the existence of DE. 
The EoS ($\omega_{eff}$) in figure \eqref{figure 5x} shows a similar evolution, moving toward negative at lower red shifts. Moreover, we show the total EoS parameter ($\omega$) in figure \eqref{figure 6x}. Hence, both equations of the state parameter lie in the quintessence regime ($-1<\omega<0$), approaching the cosmological constant ($\omega=-1$) at smaller red shifts. We find that the present value of $\omega_{0}$ is in good agreement with the data sets SNeIa and Hz+BAO+SNeIa.

\section{Conclusion} \label{sec6x}
Motivated by the teleparallel formulation of GR, we explored an extension of $f(T)$ by incorporating the coupling between the torsion scalar $T$ and the trace of the energy-momentum tensor $\mathcal{T}$. A key aspect of this approach is that both the $f(T)$ components, and the matter energy density and pressure contribute to the effective DE sector. The added flexibility in the Lagrangian imposed within the $f(T,\mathcal{T})$ framework enables a broad spectrum of conditions and dynamic behaviors.

In this study, we explore the cosmological implications of $f(T,\mathcal{T})$ theory by considering the squared-torsion model $f(T,\mathcal{T})= \alpha \mathcal{T}+ \beta T^2$, where $\alpha$ and $\beta$ are free parameters. We derived the solutions to the modified Friedmann equations, expressing the Hubble parameter as a function of the red shift $z$. Further, in section \eqref{sec3x}, we utilized recent observational data sets: Hubble, BAO, SNeIa and the joint analysis to constrain the model's free parameters. The best-fit values for these parameters were obtained and presented in figure  \eqref{figure 1x} and figure \eqref{figure 2x}. Compared with the $\Lambda$CDM model, the obtained $H(z)$ and the $\mu(z)$ of our model are confronted with the cosmic data in figure  \eqref{figure errorx} and figure \eqref{figure mux}, respectively. 

Based on the constrained model parameters, we discovered a diverse range of intriguing cosmological behaviors. Notably, our analysis revealed the evolution of the deceleration parameter, which explicitly transitions from a decelerating phase to an accelerating one, effectively accounting for the late-time expansion of the Universe. Additionally, the effective EoS ($\omega_{eff}$) and the total EoS ($\omega$) behave in a similar fashion, demonstrating that the cosmic fluid has the characteristics of quintessence DE. Moreover, we find that the
present values of $q_{0}$, $\omega_{0}$ and $z_{t}$ are in good agreement with SNeIa and Hz + BAO + SNeIa data sets.

Finally, it is important to emphasize that $f(T,\mathcal{T})$ when confronted with observational data provides a viable explanation for the late-time accelerating Universe and can be extended to various regimes to establish a robust gravitational framework. Moreover, perturbation analysis could be further developed to include vector and tensor modes, offering valuable insights into the inflationary scenario. We hope that this study inspires further exploration of torsional modified gravity as a promising candidate for describing the Universe.\\
Building on this foundation, the next chapter extends our investigation to the early Universe by exploring inflationary dynamics within the $f(T,\mathcal{T})$ gravity framework. Assuming the validity of slow roll conditions, we analyze key inflationary observables such as the tensor-to-scalar ratio $r$, scalar spectral index $n_s$, running of the spectral index $\alpha_s$, and tensor spectral index $n_t$. Using the Hubble slow-roll parameters for a specific $f(T,\mathcal{T})$ model, we assess the theoretical consistency of the inflationary scenario and examine whether the model can satisfy observational constraints through appropriate parameter choices.


\chapter{Slow-roll inflation in $f(T,\mathcal{T})$ modified gravity} 

\label{Chapter3} 

\lhead{Chapter 3. \emph{Slow-roll inflation in $f(T,\mathcal{T})$ modified gravity}} 


\vspace{10 cm}
* The work in this chapter is covered by the following publication:

\textit{Slow-roll inflation in $f(T,\mathcal{T})$ modified gravity}, Chinese Physics C \textbf{47}, 125104 (2023).
\clearpage
In this chapter, we investigate the phenomenon of cosmological inflation in the context of the $f (T,\mathcal{T})$ theory of gravity, where $f$ represents a generic function dependent on the torsion scalar $T$ and the trace $\mathcal{T}$ of the energy-momentum tensor. The analysis assumes the validity of the slow-roll inflation conditions within the framework of $f (T,\mathcal{T})$ gravity. To compute key inflationary observables—such as the tensor-to-scalar ratio $r$, the scalar spectral index $n_s$, the spectral index $ \alpha_s $, and the tensor spectral index $n_T$, we employ the Hubble slow-roll parameters for a specific $f (T,\mathcal{T})$ model. In addition, a numerical evaluation of the model parameters is performed to assess their viability. The results demonstrate that compatibility with observational constraints on the slow-roll parameters can be achieved by appropriately tuning the free parameters of the model.

\section{Introduction}\label{sec1y}
Two useful models for describing the early evolution of the Universe are the inflationary theory \cite{linde/2018,gorb/2011,baff/2021} and the bouncing cosmology \cite{db/2015,mv/2008}. Inflationary cosmology is a widely recognized framework that explains the early stages of the Universe, occurring shortly after the Big Bang. The initial inflationary models were designed to tackle problems related to the primordial singularity and naturally evolved from the particles and entropy present in the early Universe. However, these early models were partial adjustments to the Big Bang theory, as they presupposed that the Universe was in a state of thermal equilibrium and large-scale homogeneity before inflation began. The chaotic inflation proposal later addressed this limitation. On the other hand, the big bounce, or the Phoenix Universe, is a cyclical cosmological model that alternates between phases of expansion (Big Bang) and contraction (Big Crunch). This theory was popular until the inflationary model gained prominence as a solution to the horizon problem, which was supported by observations revealing the large-scale structure of the Universe. The researchers in \cite{s/2008} investigated the future development of $f(R)$ gravity models, ensuring that they align with local tests and provide a unified explanation for the expansion history of the Universe. In \cite{KB}, the authors analyzed how inflation can be realized within unimodular $f(T)$ gravity, a modified version of teleparallel gravity. Their findings suggested that unimodular $f(T)$ gravity is a viable framework to describe the early stages of the Universe.
 
The chapter aims to examine the generation of a viable inflationary era in $f (T,\mathcal{T})$ gravity, which is a torsion matter coupling gravity. The primary objective is to determine the viability of each inflationary scenario by analyzing key observational indices, such as the tensor-to-scalar ratio ($r$), the scalar spectral index ($n_s$), the running ($\alpha_s$) of the spectral index, and the tensor spectral ($n_T$). The observational indices will be assessed using conventional techniques, and we will examine whether the resulting cosmologies comply with the most recent Planck and BICEP2/Keck Array data \cite{PA/2002,PA/2016}. To carry out our research, we have examined a well-known model $f(T,\mathcal{T}) =\alpha \mathcal{T}+\beta T^2$  where $\alpha$  and $\beta$  are free parameters. The chapter is organized as follows: section \eqref{sec3y} focuses on exploring the possibility of slow roll inflation within the framework of $f (T,\mathcal{T})$ gravity. The results and insights of this investigation are summarized in the concluding remarks provided in section \eqref{sec4y}.

\section{Slow-roll inflation with  $f(T,\mathcal{T})$ gravity} \label{sec3y}
Slow-roll conditions are essential criteria that inflationary models in cosmology must meet to ensure a phase of rapid exponential expansion in the early Universe. These conditions play a significant role in evaluating and constraining alternative theories of gravity, including modified gravity models that aim to explain the observed accelerated expansion of the Universe without invoking dark energy. By analyzing data from the cosmic microwave background radiation and large-scale structure, cosmologists can impose limits on the parameters of these theories and eliminate those that fail to satisfy the slow-roll conditions. These conditions are generally defined in terms of the inflation field's potential energy, kinetic energy, and time derivatives.
The tensor-to-scalar ratio is essential to compute the values of key inflation-related observables, such as the tensor-to-scalar ratio $r$, the scalar spectral index $n_S$, the running of the spectral index $\alpha_s$, and the tensor spectral $n_T$. The tensor-to-scalar ratio is defined as the ratio of the amplitude of tensor perturbations, which correspond to primordial gravitational waves, to the amplitude. The scalar spectral index $n_s$ characterizes the variation in the clumpiness of matter across different scales immediately after cosmic inflation. It is a crucial parameter for understanding the properties of primordial density perturbations. In principle, determining these observables requires a thorough and complex perturbation analysis. However, this process can be simplified by transforming the given scenario into the Einstein frame, where all the essential information about inflation is encapsulated in the (effective) scalar potential $V(\phi)$, defining the slow-roll parameters $\epsilon$, $\eta$ and $\xi$ in terms of this
potential and its derivatives \cite{JAE,JAEa,JAEb,MA}.
\begin{equation}
    \epsilon \equiv \left( \frac{M}{2} \right)^2 \left( \frac{1}{V} \frac{dV}{d\phi} \right)^2,
\end{equation}
\begin{equation}
    \eta \equiv \frac{M^2}{p V} \left( \frac{d^2V}{d\phi^2} \right),
\end{equation}
\begin{equation}
    \xi^2 \equiv \frac{M^4}{pV^2} \left( \frac{dV}{d\phi} \right) \left( \frac{d^3V}{d\phi^3} \right).
\end{equation}

The slow-roll parameter $\epsilon$ must be much smaller than unity
$$\epsilon = \frac{1}{2}\left(\frac{V'(\phi)}{V(\phi)}\right)^2 \ll 1,$$
where $V'(\phi)$ is the derivative of the potential energy with respect to $\phi$.\\
The second slow-roll parameter $\eta$ must also satisfy the condition of being much smaller than unity
$$\eta = \frac{V''(\phi)}{V(\phi)} \ll 1,$$
where $V''(\phi)$ represents the second derivative of the potential energy with respect to the inflation field $\phi$.

In this section, we assume that the slow-roll inflation conditions are satisfied within the $f(T,\mathcal{T})$ gravity framework. These conditions are expressed in terms of the Hubble parameter $H$ as follows
\begin{eqnarray}
    \label{16y}
    \frac{\dot{H}}{H^2} \leq1 , \\ 
    \label{17y}
    \frac{\ddot{H}}{H \dot{H}} \leq 1.
\end{eqnarray}

To explore the potential for inflation within a theory of gravity that incorporates a coupling between torsion and the trace of the energy-momentum tensor, let us consider the following Lagrangian as an illustrative example.

We have a function$f(T,\mathcal{T)}=\alpha\mathcal{T}+\beta T^2=\alpha \rho_m +\beta T^2 =\alpha \rho_m+ \gamma H^4 $, where $\alpha $ and $\gamma=36\beta$ are constants. To simplify, we use $8\pi G = c = 1$. This model represents a departure from GR in the context of $f(T,\mathcal{T)}$. When $\alpha = 0$, the model reduces to a simpler form, aligning more closely with standard gravitational theories \cite{Capozziello/2011}. In this case, we get $f_T=\frac{\gamma T}{18}$, $f_{TT} =\frac{\gamma}{18}$ ,$f_\mathcal{T}=\alpha$,  $f_{T \mathcal{T}}=0$.\\
Using these equations along with equation \eqref{10x} \& equation \eqref{11x}, we can derive the following
\begin{eqnarray}
\label{18y}
\rho_{m} &=& \frac{3\left(1-\frac{\gamma H^2}{2} \right)}{1+\frac{\alpha}{2}} H^2, \\ 
\label{19y}
\dot{H} &=& -\frac{3(1+\alpha) \left(1-\frac{\gamma H^2}{2}\right)}{(\alpha +2) (1-\gamma H^2)} H^2.
\end{eqnarray}
The deceleration parameter, denoted $q$, is defined as $q= -\frac{\dot{H}}{H^2}-1$. This parameter reflects the evolution of the Universe, which has exhibited three distinct phases: an early decelerating phase, followed by a period of accelerating expansion, and finally a late-time acceleration phase. In our model, the deceleration parameter $q$ is expressed as
\begin{equation}
    \label{20y}
 q =\frac{3(1+\alpha)\left(1-\frac{\gamma H^2}{2}\right)}{(\alpha+2)(1-\gamma H^2)}-1.
 \end{equation}

Furthermore, the effective DE density and pressure can be derived from equation \eqref{14x} and equation \eqref{15x} as follows
\begin{eqnarray}
\label{21y}
    \rho_{eff} &=&\frac{3H^2(\alpha+\gamma H^2)}{\alpha+2},\\
    \label{22y}
    p_{eff}&=&-\frac{3H^2(\alpha+\gamma H^2)}{(\alpha+2)(\gamma H^2-1)},
    \end{eqnarray}
which gives
\begin{equation}
    \label{23y} 
    \omega_{eff}=\frac{1}{1-\gamma H^2}.
\end{equation}

Applying the slow-roll conditions and using the above equations in the equation \eqref{18x}, we have
\begin{eqnarray} 
\label{24y}
2\dot{H}\gamma+3k_1(\gamma*H^2-2)=0 ,\\  
\label{25y}
H(t)=p \text{tan}(q_1t) ,\\
\label{26y}
H(t)=p\text{tan}\mathcal{T}_1,
\end{eqnarray}
where 
\begin{eqnarray} \label{27y}
p=\frac{\sqrt{2}}{\sqrt{-\gamma}},\,\,\, k_1 = \frac{(\alpha+1)}{(\alpha+2)},\\
q_1=\sqrt{2}\sqrt{-\gamma}, \,\,\, q_1\,t=\mathcal{T}_1.
\end{eqnarray}
In the slow-roll regime, equation \eqref{23y}  yields
\begin{equation}
 \label{29y}
\omega_{eff}= \frac{1}{1-\gamma*p^2*\text{tan}^2q_1t}.\\
\end{equation}
The inflationary model provides a consistent and compelling explanation for the Universe's rapid expansion and the subsequent cosmological perturbations that lead to its observed anisotropy. We examined the behavior of the deceleration parameter $q(t)$  and observed that during the very early stages of evolution, it indicates a phase of rapid expansion. Later, it transitions to a de Sitter like expansion in the late-time evolution. 
According to the Planck results, the value of $n_s$ is estimated to be $0.968\pm0.006$ (with the confidence level $68\%$), the value of $r$ is less than 0.11 (with the confidence level $95\%$) and the value of $\alpha_s$ is estimated to be $-0.003\pm 0.007$ (with the confidence level $68\%$). These parameters are derived from the slow-roll parameters \cite{FR,KB}.
\begin{eqnarray}\label{30y}
 \epsilon_1 \equiv -\frac{\dot{H}}{H^2},\\
 \label{31y}
 \epsilon_2 \equiv \frac{\ddot{H}}{H\dot{H}} - \frac{2\dot{H}}{H^2},\\
 \label{32y}
 \epsilon_3 \equiv \left(H\ddot{H} - 2\dot{H}^2\right)^{-1},
 \end{eqnarray}
and rewritten as 
\begin{eqnarray}\label{33y}
 r\approx 16\epsilon_1,\\ 
 \label{34y}
 n_s \approx 1 - 2\epsilon_1 - 2\epsilon_2, \\
 \label{35y}
 n_s \approx  - 2\epsilon_1\epsilon_2 - \epsilon_2\epsilon_3, \\
 \label{36y}
 n_T \approx -2\epsilon_1.
\end{eqnarray}

Now, we can express the slow-roll parameters as follows
\begin{eqnarray}\label{37y}
 r=-\frac{16 q_1 \csc^2{\mathcal{T}_1}}{p}, \\
 \label{38y}
 n_s=1 + \frac{2q_1(-1+\cot^2{\mathcal{T}_1})}{p}, \\
 \label{39y}
\alpha_s=-\frac{2q_1^2(1-2pq_1+\cos{2\mathcal{T}_1})\cot^2{\mathcal{T}_1}\csc^4{\mathcal{T}_1}}{p^2(pq_1-\cot^2{\mathcal{T}_1})},\\
\label{40y}
 n_T= -\frac{2q_1}{p}\csc^2{\mathcal{T}_1}. 
\end{eqnarray}
We can estimate the number of e-folds that quantifies the amount of exponential expansion during inflation as \\
\begin{equation}\label{41y}
N = \ln \frac{a_f}{a_i} = \int_{t_i}^{t_f} H(t)dt,
\end{equation}
where $a_i = a(t=t_i)$ is the initial value of the scale factor $a$ at the beginning of inflation $t_i$, and $a_f = a(t=t_f)$ is its final value at the end of inflation $t_f$. This relationship highlights the exponential growth of the scale factor during the inflationary period
\begin{equation}
\label{42y}
N=\frac{p}{q_1} \ln \left[ \frac{\cos(q_1t_f)}{\cos(q_1t_i)} \right].
\end{equation}
Assuming that the Hubble parameter $H(t)$ can be expanded as a series around $\mathcal{T}_1= 0$, we can obtain the second-order approximation as 
\begin{equation}
   \label{43y}
    H(t) \approx \ \ p \mathcal{T}_1 + \mathcal{O}(\mathcal{T}^2_1).
\end{equation}
and the slow-roll parameters become
\begin{eqnarray}
\label{44y}
\varepsilon_1 \approx \ \ -\frac{1}{\mathcal{T}^2_1},\\
\label{45y}
\varepsilon_2\approx \ \ -2\frac{1}{p\mathcal{T}^2_1},\\
\label{46y}
\varepsilon_3\approx \ \ -2\frac{1}{p\mathcal{T}^2_1}.
\end{eqnarray}
Hence, in this scenario, we have\\
\begin{eqnarray}
\label{47y}
r\approx \ \ -16\frac{1}{p\mathcal{T}_1},\\
\label{48y}
n_s=1,\\
\label{49y}
\alpha_s \approx \ \ -\frac{8}{p^2 \mathcal{T}^2_1},\\
\label{50y}
n_T  \approx \ \ -2 \frac{1}{p\mathcal{T}^2_1}.
\end{eqnarray}
According to Hubble parameter $H(t)$, the e-folding number is  given by 
\begin{equation}\label{51y}
    N=\frac{pq_1(t_f^2-t_i^2)}{2}.
\end{equation}
These statements imply that when the Hubble parameter $H(t)$ can be expanded in a series around $\mathcal{T} = 0$, which corresponds to low values of $\mathcal{T}$, the inflationary parameters take on large values under the condition that the slow-roll requirements are satisfied. This occurs specifically when the slow-roll conditions are met.\\
To assess the viability of our model, we present the numerical results for various inflation-related parameters given by equation \eqref{37y}, equation \eqref{38y}, equation \eqref{39y} by comparing the theoretical results with the observational data from PLANK 2015 and the BICEP2 / Leck-Array data \cite{PA/2002,PA/2016}.
\begin{figure}
    \centering
\includegraphics[scale=0.6]{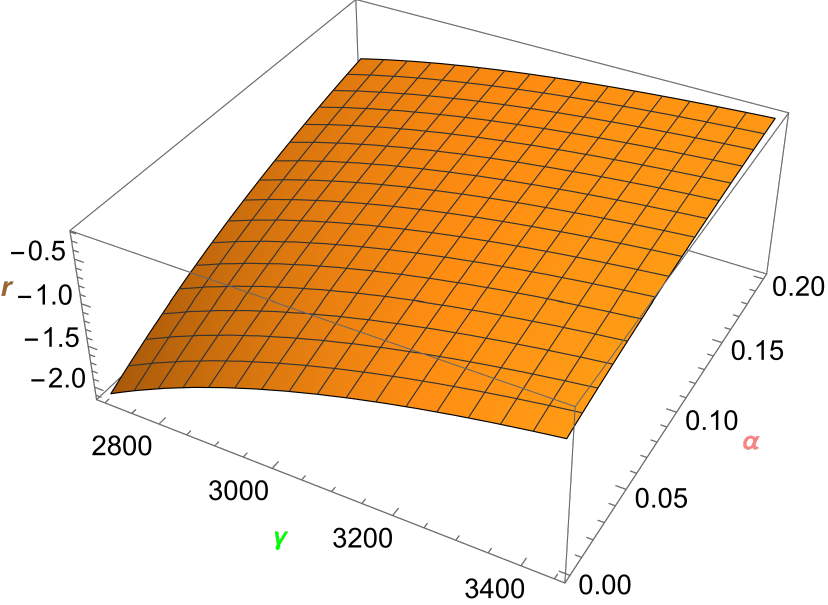}
\includegraphics[scale=0.7]{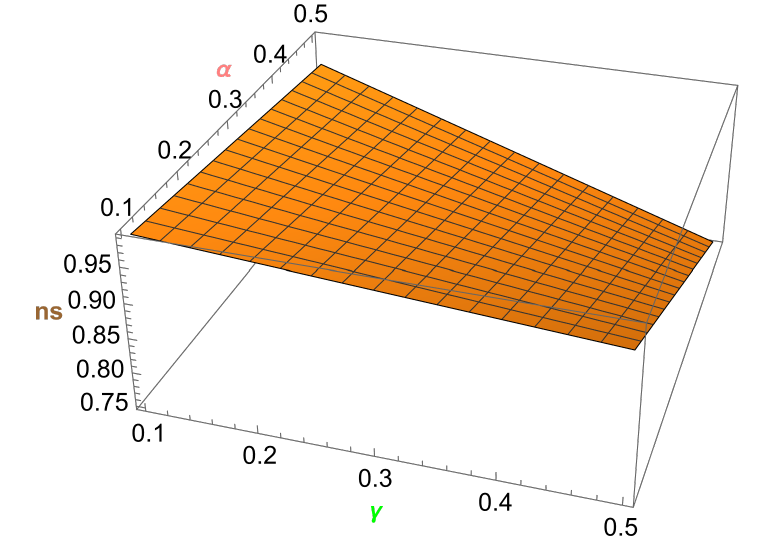}
\includegraphics[scale=0.68]{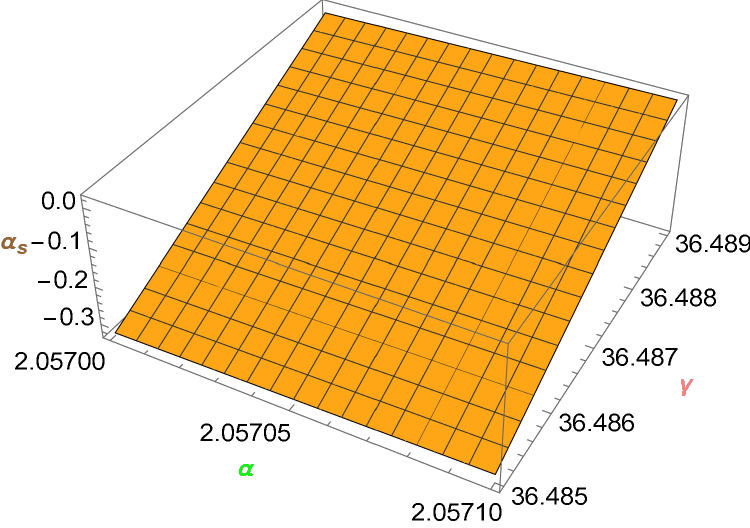}
\caption{Profiles of tensor-to-scalar ratio $(r)$, scalar spectral index $(n_s)$, and running spectral index $(\alpha_s)$ with varying model parameters $\alpha$ and $\gamma$.}
\end{figure}


The graphs provided represent the evolution of the inflation parameters of the model $f(T,\mathcal{T})$ at a specific time $t = 0.1$ seconds. The first graph illustrates the tensor-to-scalar ratio $r$ for values of $\alpha$  ranging from $0.0$ to $0.2$, while the second graph displays the scalar spectral index $n_s$ for values of $\alpha$ between $0.1$ and $0.5$. The final graph highlights the behavior of the spectral index spectral index
$\alpha_s$. Analysis of the curves in the parametric diagrams indicates that the values of the tensor-to-scalar ratio $(r)$, scalar spectral index $(n_s)$, and the running of the spectral index values $(\alpha_s)$ align well with the observed cosmological data.
\section{Conclusion} \label{sec4y}
This chapter investigates the inflationary scenario in torsion-trace coupling gravity, which involves a Lagrangian density derived from a function called $f(T,\mathcal{T})$, which makes use of the torsion $T$ and the trace of the energy-momentum tensor $\mathcal{T}$. We found an evolution of the deceleration parameter, explicitly experiencing a change from deceleration to acceleration, capable of explaining the late-time Universe. The study focuses on a Universe containing ordinary matter and dark energy, and a primary differential equation is derived in the Hubble parameter. Slow-roll conditions are applied for a specific $f(T,\mathcal{T})$ model, and various measurements related to inflationary phenomena are evaluated, including the ratio of tensor to scalar perturbations $(r)$, the spectral index of scalar perturbations $(n_s)$, the running of the spectral index $(\alpha_s)$, the spectral tensor $n_T$, and the number of e-folds parameter. The numerical findings are consistent with the observational data. So far, numerous studies have been conducted within the framework of torsion-based modified gravity theories, with a particular focus on explaining the current accelerated expansion of the Universe and its late-time acceleration phase \cite{Harko/2014a, AAQID/2023, dsg/2016}. However, in this chapter, we aim to study the early-inflationary scenario and successfully presented it. We have not only successfully presented the slow-roll inflationary scenario, but also presented a way to constrain the parameters of the $f(T,\mathcal{T})$ cosmological model. We also constrained the model parameters for the desirable results and discussed them in the last of section \eqref{sec3y}. We hope that this study will provide valuable insight and open up new avenues in the exploration of torsion-based modified gravity cosmology.

In the near future, researchers can investigate more generalized models to impose constraints on the parameters and extend this study to explore additional cosmological scenarios. Upcoming observational surveys are expected to provide a more accurate and detailed understanding of the Universe, which will help narrow down the range of viable models. This progress is expected to reduce the number of potential candidates and identify those that offer a more complete and robust explanation for the inflationary epoch.\\

 Building upon this foundation, the next chapter shifts focus from the early Universe to the dark sector, encompassing both DM and DE. We introduce a modified $f(Q)$ gravity model, where the dark energy component follows a power-law function, $f(Q)= \gamma \left(\frac{Q}{Q_0}\right)^n$, and the dark matter sector is modeled using the EBEC EoS. The interaction between these components is incorporated through an energy exchange term, $\mathcal{Q} = 3b^2H \rho$, which allows us to examine their coupled evolution.


\chapter{Extended Bose-Einstein Condensate dark matter in $f(Q)$ gravity} 

\label{Chapter4} 

\lhead{Chapter 4. \emph{Extended Bose-Einstein condensate dark matter in $f(Q)$ gravity}} 

\vspace{8 cm}
* The work in this chapter is covered by the following publications: \\
 
\textit{Extended Bose-Einstein condensate dark matter in $f(Q)$ gravity}, General Relativity and Gravitation \textbf{56}, 63 (2024).

\clearpage

In this chapter, we aim to investigate the dark sector of the Universe, which encompasses both DM and DE. The DE component is associated with a modified $f(Q)$ Lagrangian, particularly a power-law function of the form $f(Q)= \gamma \left(\frac{Q}{Q_0}\right)^n$. On the other hand, the DM component is characterized by the EBEC EoS, specifically expressed as $p = \alpha \rho + \beta \rho^2$. This approach allows us to explore the interplay between these two components and their roles in shaping the dynamics of the Universe. We derive the corresponding Friedmann-like equations and the continuity equation for both DM and DE, incorporating an interaction term $\mathcal{Q} = 3b^2H \rho$, which represents the energy exchange between the dark sectors of the Universe.
Additionally, we obtain the analytical expression for the Hubble function and determine the best-fit values of the free parameters using Bayesian analysis to estimate the posterior probability, along with the MCMC sampling technique, applied to the CC+Pantheon+SH0ES data sets. To assess the robustness of our MCMC analysis, we performed a statistical evaluation using the Akaike Information Criterion (AIC) and the Bayesian Information Criterion (BIC). Furthermore, from the evolutionary profiles of the deceleration parameter and the energy density, we observe a transition from a decelerated epoch to an accelerated expansion phase, with the present value of the deceleration parameter value as $q(z=0)=q_0=-0.56^{+0.04}_{-0.03}$ ($68 \%$ confidence limit), which is quite consistent with cosmological observations. In addition, we observe the expected positive behavior of the effective energy density, which aligns with physical requirements. Finally, by analyzing the sound speed parameter, we conclude that the assumed theoretical $f(Q)$ model is thermodynamically stable, ensuring its consistency and viability as a framework for describing the dark sector of the Universe.

\section{Introduction}\label{sec1z}

As is widely recognized, the Universe is composed of both visible and dark components. The visible elements encompass all visible entities within the cosmos, while the dark components encompass DM and DE. Among these mysterious components, DM stands out as an undetectable entity within the electromagnetic radiation spectrum. The observational phenomena such as the cosmic microwave background and the gravitational lensing \cite{R1d,R2d,R3d,R4d} provide compelling evidence for the presence of DM, in order to elucidate the difference between the estimated mass of large celestial bodies and the mass derived from luminous matter such as stars, gas, and dust within them. This implies that empirical data indicate the emergence of DM through gravitational pull on ordinary matter. The DM encompasses both baryonic and non-baryonic forms. In the baryonic form, DM manifests itself as astronomical entities like massive and compact halos, primarily made up of ordinary baryonic matter yet emitting negligible electromagnetic radiation. Conversely, non-baryonic DM is characterized by hypothetical and actual particles, whereas the Weakly Interacting Massive Particles and axions are the hypothetical ones.  Additionally, the state of matter known as BEC arises in the non-baryonic realm, formed when particles called bosons undergo cooling to near absolute zero \cite{R5d,R6d,R7d,R8d,R9d,R10d,R11d}. As a consequence of the extremely low temperature, a phase transition takes place, causing the majority of boson gases to occupy the lowest quantum state, leading to the manifestation of macroscopic quantum phenomena. At this condition, cold bosons interact, giving rise to superparticles that exhibit microwave-like behavior. The assumption is made that DM exists in the form of a bosonic gas below a critical temperature, leading to the formation of BEC. For more on BEC, see the references \cite{ADD1d,ADD2d,ADD3d,ADD4d,ADD5d}. Using the generalized Gross-Pitaevskii equation, the EoS for DM is derived as that of a barotropic fluid. It should be noted that this particular EoS is termed conventional DM. Taking into account the DM halo that exists in a quantum ground state, the EoS was derived as $p \propto \rho^2$ \cite{R10d}.  It is important to highlight that the origins of both the usual DM and the quantum ground state can be attributed to the one-body and two-body interactions among bosonic particles. Note that EoS $p=0$, $p=\alpha \rho$, and $p=\beta \rho^2$ characterize the cold DM, normal DM, and DM halo, respectively. These observations prompt the introduction of the EBEC model, a comprehensive model that combines normal DM and the quantum ground state \cite{R13d}. The merit of this approach lies in its capacity to concurrently account for both one-body and two-body interactions, offering insights into the components of the Universe, particularly DM.

Further, the conventional theory of relativity, particularly GR, which interprets gravity as the curvature of spacetime, may not provide the ultimate solution to explaining DE. This encourages the exploration of alternative theoretical frameworks in cosmology that can effectively account for cosmic acceleration while aligning with observational data. GR and its curvature-based extensions have been adequately formulated and studied in the past \cite{R15d}. Recently, modified theories of gravitation within a flat spacetime geometry, dependent solely on non-metricity, have been established and extensively investigated \cite{R50a,R51a}. Various astrophysical and cosmological implications of $f(Q)$ gravity have been widely investigated \cite{R18d,R19d,R20d,R21d,R22d,R23d,R24d,R25d,R26d,R27d,R28d,R29d}. The manuscript is organized as follows: In section \eqref{sec2z}, we present the Friedmann equations in the $f(Q)$ gravity. In section \eqref{sec3z}, we employ EBEC dark matter EoS along with the power law $f(Q)$ Lagrangian to derive the analytical solution of the Friedmann-like equations. In section \eqref{sec4z}, we find the best-fit values of free parameters utilizing Bayesian analysis to estimate the posterior probability and MCMC sampling. In addition, we employ AIC and BIC tools to examine the robustness of our MCMC analysis and then test the stability of the considered cosmological $f(Q)$ model. Finally in section \eqref{sec5z}, we discuss our findings.

\section{Friedmann equations in $f(Q)$ theory}\label{sec2z}
\justifying
We assume the following homogeneous and isotropic flat FLRW line element
\begin{equation}\label{2nz}
ds^2 = -dt^2 + a^2(t) (dx^2+dy^2+dz^2).   
\end{equation}
We obtained the non-metricity scalar $Q=6H^2$ for the line element equation \eqref{2nz}. The Friedmann like equations for the generic $f(Q)$ functional corresponding to the line element given in equation \eqref{2nz} is obtained as follows \cite{R292d}
\begin{equation}\label{2oz}
3H^2=\frac{1}{2f_Q} \left( -\rho+\frac{f}{2}  \right),
\end{equation}
\begin{equation}\label{2pz}
\dot{H}+3H^2+ \frac{\dot{f_Q}}{f_Q}H = \frac{1}{2f_Q} \left( p+\frac{f}{2} \right).
\end{equation}
We can rewrite equations \eqref{2oz} and \eqref{2pz} as follows 
\begin{equation}\label{2qz}
3H^2= \rho + \rho_{de},
\end{equation}
\begin{equation}\label{2rz}
 \dot{H}=- \frac{1}{2}\left[  \rho + \rho_{de} +p + p_{de} \right].
\end{equation}
where $\rho_{de}$ and $p_{de}$ are energy density and pressure of the DE fluid part arising due to non-metricity component, and can be expressed as follows
\begin{equation}\label{2sz}
\rho_{de}= \frac{1}{2} (Q-f) + Q f_Q,
\end{equation}

\begin{equation}\label{2tz}
 p_{de}=-\rho_{de} - 2\dot{H} (1+f_Q+2Qf_{QQ}).
\end{equation}
Further, we write the continuity equation for both matter and DE component as
\begin{equation}\label{2uz}
\dot{\rho} + 3H(\rho + p) = \mathcal{Q},  
\end{equation}
\begin{equation}\label{2vz}
\dot{\rho}_{de} + 3H(\rho_{de} + p_{de}) = - \mathcal{Q},  
\end{equation}
where $\mathcal{Q}$ is defined as an interaction term arising due to the energy transfer between dark components of the Universe. It is evident that the parameter $\mathcal{Q}$ must possess a positive value, indicating the occurrence of energy transfer from DE to DM, ensuring the second law of thermodynamics. In this context, considering $\mathcal{Q}$ as the product of the energy density and the Hubble parameter is a natural choice, given that it represents the inverse of cosmic time. Therefore, we adopt the specific expression $\mathcal{Q} = 3b^2H \rho$, where $b$ is the intensity of energy transfer \cite{R13d}.\\

\section{Dark matter as a EBEC}\label{sec3z}
\justifying
In this section, we examine normal DM as bosonic particles whose number density is determined by Bose-Einstein statistics, signifying the formation of these particles through the decoupling of the residual plasma during the early Universe. However, the energy density of DM is expressed as the product of the particle number density and the mass of DM. The pressure of DM, according to Bose-Einstein statistics, is defined within a sphere characterized by the radius of the momentum of the particles \cite{R30,R31}. Consequently, one can express the normal DM pressure as a linear relationship in terms of the energy density as follows
\begin{equation}\label{3az}
p=\alpha \rho.  
\end{equation}
Now, we examine the BEC dark matter as non-relativistic bosons engaged in a two-particle interaction within a quantum system. As discussed earlier, BEC recognized as a state of matter, emerges when a dilute Bose gas undergoes significant cooling to attain extremely low temperatures.
According to experimental observations of Einstein and Bose, as the temperature approaches absolute zero, the waves associated with the particles eventually overlap. This phenomenon results in the merging of elementary particles into a single quantum state, termed Bose-Einstein condensation. The generalized Gross-Pitaevskii equation characterizes the physical behavior of the BEC \cite{R32d}. When considering BEC within a gravitational context, the corresponding EoS for BEC dark matter is formulated in the subsequent manner
\begin{equation}\label{3bz}
p=\beta \rho^2,   
\end{equation}
where $\beta$ is defined as the coefficient related to the DM mass and its scattering length \cite{R33,R34}.
Now in order to gain a more profound comprehension of the Universe, we assume an extended form of the EoS for DM known as the EBEC for DM EoS as \cite{R13d}
\begin{equation}\label{3cz}
p=\alpha \rho + \beta \rho^2.
\end{equation}
Here, $\alpha$ represents the single-body interaction arising from conventional DM, while $\beta$ is introduced to signify the two-body interaction originating from the DM halo. In particular, $\alpha=\beta=0$ reduces to the cold DM case, whereas case $\beta=0$ reduces to the normal matter scenario. Furthermore, the case $\alpha=0$ represents the DM halo, while $\alpha \neq 0$ and $\beta \neq 0$ represent the contribution of both the DM halo and the normal matter.

We consider the following dynamically tested power-law $f(Q)$ function that can efficiently describes an evolution of the Universe from a matter dominated phase to the de Sitter era \cite{R341}
\begin{equation}\label{3dz}
f(Q)= \gamma \bigg(\frac{Q}{Q_0}\bigg)^n
\end{equation}
where $Q_0=6H_0^2$ and $\gamma$ and $n$ are free parameters. Then by using equation \eqref{3dz} in the equation \eqref{2qz}, we obtained
\begin{equation}\label{3ez}
\rho = \frac{(1-2n)}{2}\gamma \left( \frac{H}{H_0} \right)^{2n}.
\end{equation}
On evaluating the equation \eqref{3ez} at present redshift $z = 0$, we have
\begin{equation}\label{3fz}
\rho_0 = \frac{(1-2n)}{2}\gamma     
\end{equation}
and therefore, we have
\begin{equation}\label{3gz}
\rho = \rho_0 \left( \frac{H}{H_0} \right)^{2n} .
\end{equation}
Now, on integrating the continuity equation \eqref{2uz} for the matter component, we acquired
\begin{equation}\label{3hz}
 \rho = \rho_0 \left( \frac{c \eta - \beta}{c \eta (1+z)^{3 \eta} - \beta}  \right).
\end{equation}
Here, $c$ is the constant of integration, and $\eta=\alpha+1-b^2$. We obtained the expression of the Hubble parameter, by utilizing equation \eqref{3gz} and equation \eqref{3hz}, as follows
\begin{equation}\label{3iz}
H(z)=H_0 \left( \frac{c \eta - \beta}{c \eta (1+z)^{-3 \eta} - \beta}  \right)^{\frac{1}{2n}}.
\end{equation}

\section{Best fit value of parameters}\label{sec4z}
\justifying
In this section, a statistical analysis is performed to compare the predictions of the theoretical model with the observational data, with the goal of constraining the free parameters of the model. The analysis incorporates a dataset of 31 CC measurements and the Pantheon+SH0ES sample, which consists of 1701 data points. Bayesian statistical methods are applied to estimate the posterior probability utilizing the likelihood function and the MCMC random sampling technique. This approach allows for a robust determination of the parameter constraints and provides insight into the compatibility of the model with observational evidence.
\subsection{Pantheon+SH0ES}
\justifying
The Pantheon+SH0ES samples encompass a wide range of redshifts, ranging from 0.001 to 2.3, and represent an advancement over previous data sets by incorporating the most recent observational data. SNIa, known for their uniform luminosity, are highly reliable standard candles for measuring relative distances using the distance modulus method. This extensive data set enhances our ability to constrain cosmological parameters and test theoretical models with greater precision. In the last two decades, several compilations of Type Ia supernova data have been introduced, such as Union \cite{R15a}, Union2 \cite{R16a}, Union2.1 \cite{R17a}, JLA \cite{R40d}, Pantheon \cite{R18a}, and the most recent addition, Pantheon+SH0ES \cite{R42d}. The corresponding $\chi^2$ function is expressed as, 
\begin{equation}\label{4bz}
\chi^2_{SN}= D^T C^{-1}_{SN} D,
\end{equation}
Here, $C_{SN}$ \cite{R42d} represents the covariance matrix associated with the Pantheon+SH0ES samples, encompassing both statistical and systematic uncertainties. Moreover, the vector $D$ is defined as $D=m_{Bi}-M-\mu^{th}(z_i)$, where $m_{Bi}$ and $M$ are the apparent magnitude and the absolute magnitude, respectively. In addition, the $\mu^{th}(z_i)$ represents the distance modulus of the assumed theoretical model, and it can be expressed as
\begin{equation}\label{4cz}
\mu^{th}(z_i)= 5log_{10} \left[ \frac{D_{L}(z_i)}{1 Mpc}  \right]+25, 
\end{equation}
where,  luminosity distance $D_{L}(z)$ for the assumed theoretical model can be expressed as,
\begin{equation}\label{4dz}
D_{L}(z)= c(1+z) \int_{0}^{z} \frac{ dx}{H(x,\theta)}
\end{equation}
where, $\theta$ is the parameter space of the assumed model.

Unlike the Pantheon dataset, the Pantheon+SH0ES compilation successfully resolves the degeneracy between the parameters $H_0$ and $M$ by redefining the vector $D$ as
\begin{equation}\label{4ez}
\bar{D} = \begin{cases}
     m_{Bi}-M-\mu_i^{Ceph} & i \in \text{Cepheid hosts} \\
     m_{Bi}-M-\mu^{th}(z_i) & \text{otherwise}
    \end{cases}   
\end{equation}
Here, $\mu_i^{Ceph}$ independently estimated using Cepheid calibrators. Hence, equation \eqref{4bz} is obtained as $\chi^2_{SN}=  \bar{D}^T C^{-1}_{SN} \bar{D} $.\\

\begin{figure}[H]
\includegraphics[scale=0.75]{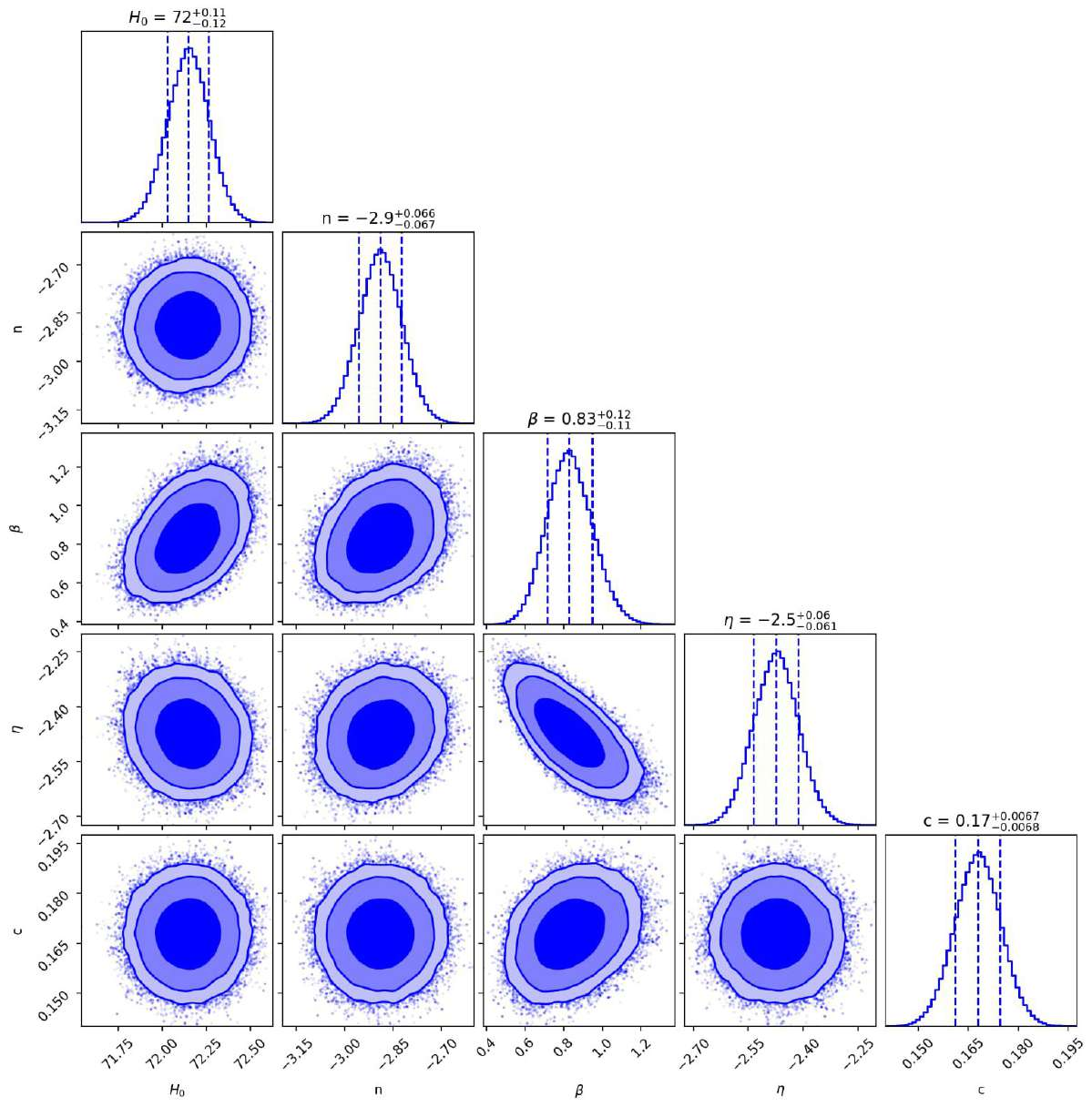}
\caption{The contour plot for the given model corresponding to the free parameter space $(H_0, n, \beta, \eta, c)$ within the $1\sigma-3\sigma$ confidence interval using CC+Pantheon+SH0ES samples.}\label{f1z}

\end{figure}

We derive constraints on the free parameter space for the combined CC+Pantheon+SH0ES data sets by employing Gaussian priors such as $[50,100]$ for $H_0$, $[-5,0]$ for $n$, $[0,5]$ for $\beta$, $[-5,0]$ for $\eta$ and $[0,1]$ for $c$. In order to obtain the best fit value of parameters, we minimize the total $\chi^2_{total}$ function that is defined as follows
\begin{equation}\label{4fz}
\chi^2_{total}= \chi^2_{CC}+\chi^2_{SN}. 
\end{equation}
The corresponding contour plot illustrating the correlation between different model parameters within the confidence intervals $1\sigma-3\sigma$ is shown in figure \eqref{f1z}.

We obtained constraints on the free parameter space with the $68 \%$ confidence limit as $H_0=72^{+0.11}_{-0.12}$, $n=-2.9^{+0.066}_{-0.067}$, $\beta=0.83^{+0.12}_{-0.11}$, $\eta=-2.5^{+0.06}_{-0.061}$ and $c=0.17^{+0067}_{-0.0068}$. In addition, we obtained the minimum value of $\chi^2_{total}$ as $\chi^2_{min}=1668.55$.

\subsection{Model comparison}
\justifying
To assess the robustness of our MCMC analysis, it is essential to perform a statistical evaluation using the AIC and the BIC \cite{R43d}. The initial criterion, AIC, can be expressed as follows
\begin{equation}\label{4gz}
AIC = \chi^2_{min} + 2d   .  
\end{equation}
Here, $d$ represents the number of parameters within the given model. For comparison of the model with the established $\Lambda$CDM model, we introduce $\Delta AIC = |AIC_{Model} - AIC_{\Lambda CDM}|$. A value of $\Delta AIC$ less than 2 suggests strong evidence in favor of the assumed theoretical model, while in the range of $4 < \Delta AIC \leq 7$, there is moderate support. Moreover, if the $\Delta AIC$ value exceeds 10, there is no evidence supporting the assumed model. The second criterion, BIC, can be expressed as follows
\begin{equation}\label{4hz}
BIC = \chi^2_{min} + d ln(N) .   
\end{equation}
Here, $N$ represents the number of data samples used in the MCMC analysis. Similarly, $\Delta BIC$ is less than 2 suggesting strong evidence in favor of the assumed theoretical model, while in the range of $2 < \Delta BIC \leq 6$, there is moderate support. Using the aforementioned $\chi^2_{min}$ minimum value, we obtained $AIC_{Model}=1678.85$ and $BIC_{Model}=1679.8$ and hence obtained $\Delta AIC=0.95$ and $\Delta BIC=15.5$, where the $\Lambda$CDM value is taken to be $AIC_{\Lambda CDM}=1679.4$ and $BIC_{\Lambda CDM}=1664.3$. Thus, it is evident from the $\Delta AIC$ value that there is strong evidence in favor of the assumed theoretical $f(Q)$ model. However, it is well known that a large number of parameters compensate for the high $\Delta BIC $ value.

\subsection{Evolutionary parameters}
\justifying
The deceleration parameter is an essential tool to quantify the evolutionary phase of expansion of the Universe.

\begin{figure}[H]
\centering
\includegraphics[scale=0.475]{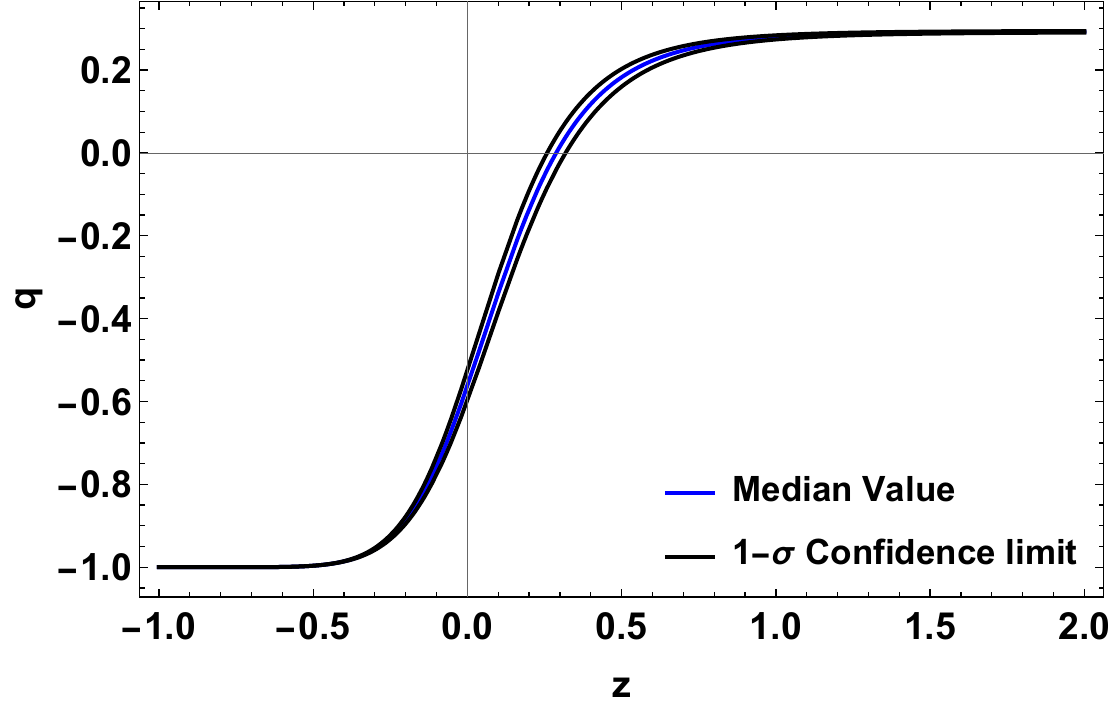}
\caption{Profile of the deceleration parameter vs redshift corresponding to obtained parameter constraints with $68 \%$ confidence limit. }
\label{f2z}
\end{figure}

\begin{figure}[H]
\centering
\includegraphics[scale=0.45]{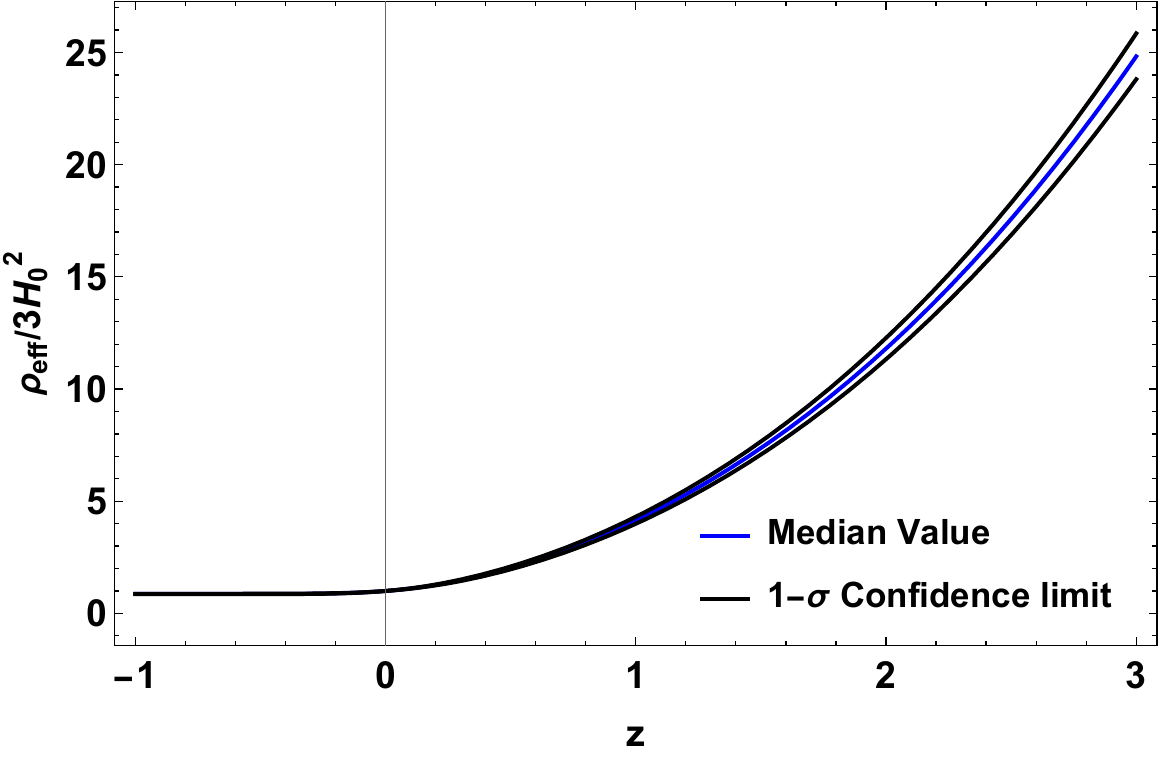}
\caption{Profile of the effective energy density vs redshift corresponding to obtained parameter constraints with $68 \%$ confidence limit.}
\label{f3z}
\end{figure}
From figure \eqref{f2z}, it is evident that the assumed model shows a transition from a decelerated epoch to the de Sitter type accelerated expansion phase, with the redshift transition $z_t=0.288^{+0.031}_{-0.029}$. The present value of the deceleration parameter is obtained as $q(z=0)=q_0=-0.56^{+0.04}_{-0.03}$ ($68 \%$ CL), which is quite consistent with the observed values. From figure \eqref{f3z}, we observed the anticipated positive trend in the effective energy density, which diminishes as the Universe expands.

\subsection{Thermodynamical stability}
\justifying
To conduct a thorough assessment, we investigate the thermodynamical stability of the assumed theoretical model by examining the sound speed parameter. In this analysis, we assume that the Universe operates as an adiabatic system, where there is no transfer of heat or mass from within the Universe to its external environment, resulting in a zero entropy perturbation. Under these conditions, the variation of pressure in relation to energy density becomes the primary focus, leading us to introduce the sound speed parameter, denoted as $c_s^2$, in the subsequent expression,
\begin{equation}\label{4jz}
c_s^2=\frac{\partial p}{\partial \rho} =  \frac{\partial_z p}{\partial_z \rho }
\end{equation}
where $\partial_z= \frac{\partial}{\partial z}$. Here, it should be noted that the condition $c_s^2 > 0$ indicates stability, while $c_s^2 < 0$ indicates instability.
\begin{figure}[H]
\centering
\includegraphics[scale=0.51]{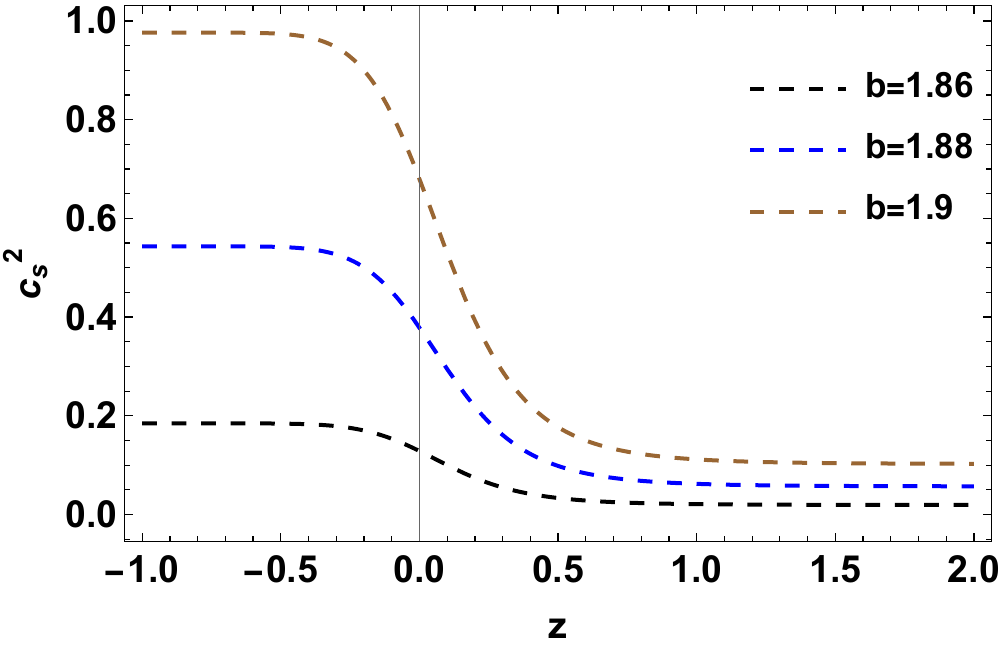}
\caption{Profile of the sound speed parameter vs redshift corresponding to the value $b=1.86$, $b=1.88$, and $b=1.9$. }
\label{f4z}
\end{figure}
From figure \eqref{f4z}, it is evident that the assumed theoretical $f(Q)$ model shows the evolution of the Universe from a decelerated to an accelerated epoch in a stable way. Thus, the considered model can efficiently address the late-time expansion phase with the observed transition epoch. 

\section{Conclusion}\label{sec5z}
\justifying
In this chapter, our objective was to investigate the dark sector of the Universe, with a particular emphasis on DM and DE. To achieve this, we used an extended version of the EoS for DM, known as the EBEC EoS (as presented in equation \eqref{3cz}), combined with a modified $f(Q)$ Lagrangian. The state of matter known as BEC arises in the non-baryonic realm, when particles called bosons undergo cooling to near absolute zero \cite{R5d}. The assumption is made that DM exists in the form of a bosonic gas below a critical temperature, leading to the formation of BEC. Using the generalized Gross-Pitaevskii equation, the EoS for DM is derived as that of a barotropic fluid. Furthermore, the DM halo exists in a quantum ground state, and the EoS was derived as $p \propto \rho^2$ \cite{R10d}. These observations motivate us to consider the EBEC model, a comprehensive model that combines normal DM and the quantum ground state \cite{R13d}.

Now to describe another prominent dark component i.e. undetected DE, we consider the modified theories of gravitation within a flat spacetime geometry, dependent solely on non-metricity, particularly, we consider the power law $f(Q)$ Lagrangian $f(Q)= \gamma \left(\frac{Q}{Q_0}\right)^n$, where $\gamma$ and $n$ are free parameters \cite{R44d}. We derive the corresponding Friedmann-like equations and the continuity equation for both DM and DE, incorporating an interaction term to account for energy exchange between the two components. The interaction term is directly proportional to the product of the Hubble parameter and the energy density of DM. In other words, it signifies the energy exchange between the dark sector of the Universe. We obtained the analytical solution of the corresponding equations, i.e. the Hubble function in terms of redshift,  presented in the equation \eqref{3iz}. Further, to find the best-fit values of parameters of the assumed theoretical model, we utilize the Bayesian analysis to estimate the posterior probability through the utilization of the likelihood function and the MCMC sampling technique. The corresponding contour plot describing the correlation of model parameters within the $1\sigma-3\sigma$ confidence interval utilizing CC+Pantheon+SH0ES samples is presented in figure \eqref{f1z}. The obtained constraints on the free parameter space with $68 \%$ confidence limit are $H_0=72^{+0.11}_{-0.12}$, $n=-2.9^{+0.066}_{-0.067}$, $\beta=0.83^{+0.12}_{-0.11}$, $\eta=-2.5^{+0.06}_{-0.061}$, and $c=0.17^{+0067}_{-0.0068}$. Furthermore, to evaluate the reliability of our MCMC analysis, we conducted a statistical evaluation using the AIC and the BIC. We obtained $\Delta AIC=0.95$ and $\Delta BIC=15.5$, and hence it is evident from the $\Delta AIC$ value that there is strong evidence in favor of the assumed theoretical $f(Q)$ model. However, it is well known that a large number of parameters compensate for a high $\Delta BIC $ value. We presented the evolutionary profile of the deceleration parameter and the energy density, respectively in figure \eqref{f2z} and figure \eqref{f3z}. We found that the assumed model shows a transition from the decelerated epoch to the de Sitter type accelerated expansion phase, with the transition redshift $z_t=0.288^{+0.031}_{-0.029}$. Moreover, the present value of the deceleration parameter obtained as $q(z=0)=q_0=-0.56^{+0.04}_{-0.03}$ ($68 \%$ confidence limit), which is quite consistent with cosmological observations. Further, we found the expected positive behavior of the effective energy density. More on energy density, we obtained $\Omega_0 = \frac{(1-2n)\gamma}{Q_0} $ utilizing the relation \eqref{3fz}. For the STEGR case i.e. $n=1$ and $\frac{\gamma}{Q_0}=-1$, we obtained expected value $\Omega_0=1$ that represents matter dominated phase. Further, for the parameter value $n=-2.9$ (obtained in figure \eqref{f1z}), we acquired $\Omega_0 = \frac{6.8 \gamma}{Q_0} $ which aligns with the observed value $\Omega_0 \in [0.25, 0.35] $ for the viable parameter range $\frac{\gamma}{Q_0} \in [19, 27] $. Lastly, we investigated the thermodynamical stability of the assumed theoretical model by examining the sound speed parameter (presented in figure \eqref{f4z}). We observed that one can analyze the contribution of the normal DM and the DM halo existing in a quantum ground state separately by estimating the value of parameter $\alpha$ using the specific value of intensity parameter $b^2$ and the obtained constrained value of $\eta$, for instance, if we choose $b=1.9$ (obtained in figure \eqref{f4z}) and $\eta=-2.5$ (obtained in figure \eqref{f1z}), we obtained $\alpha=0.11$ utilizing the relation $\eta=\alpha+1-b^2$. As we already have $\beta=0.83$ (obtained in figure \eqref{f1z}), we found that the DM halo existing in a quantum ground state contributes nearly $7.5$ times more than that of the normal DM. We found that the considered theoretical $f(Q)$ model can efficiently address the late-time expansion phase of the Universe with the observed transition epoch in a stable way. \\
Building upon this foundation, the next chapter expands our exploration of modified gravity theories by introducing the $f(Q,\mathcal{T_{\mu\nu}} \mathcal{T^{\mu\nu}})$ gravity framework. This extension generalizes the existing $f(Q)$ and $f(Q,\mathcal{T})$ theories by incorporating a function dependent on both non-metricity $Q$ and the square of the energy-momentum tensor, $\mathcal{T}^2=\mathcal{T_{\mu\nu}} \mathcal{T^{\mu\nu}}$. We derive an analytical solution for a barotropic fluid with EoS $p=\omega \rho$ under the specific model $f(Q, \mathcal{T_{\mu\nu}} \mathcal{T^{\mu\nu}}) = Q + \eta(\mathcal{T_{\mu\nu}} \mathcal{T^{\mu\nu}})$. The free parameters of the model are constrained using MCMC sampling techniques combined with Bayesian statistical analysis, employing the latest CC, BAO and Pantheon + SH0ES data sets.


\chapter{Cosmology in energy-momentum squared symmetric teleparallel gravity} 

\label{Chapter5} 

\lhead{Chapter 5. \emph{Cosmology in energy-momentum squared symmetric teleparallel gravity}} 
\vspace{10 cm}
* The work presented in this chapter is covered by the following publication: \\
 
\textit{Cosmology in energy-momentum squared symmetric teleparallel gravity}, Physics Letters B \textbf{858}, 139068 (2024).

\clearpage
In this chapter, we investigate the $f(Q,\mathcal{T_{\mu\nu}} \mathcal{T^{\mu\nu}} )$ gravity theory, extending the frameworks established by the $f(Q)$ and $f(Q,T)$ gravity theories. Here, $\mathcal{T_{\mu\nu}}$ stands for the energy-momentum tensor. The proposed action incorporates an arbitrary function of both non-metricity $Q$ and the square of the energy-momentum tensor, specifically $\mathcal{T}^{2}=\mathcal{T_{\mu\nu}} \mathcal{T^{\mu\nu}}$. We derive an analytical solution for the barotropic fluid case $p=\omega \rho$ for the specific model $f(Q, \mathcal{T_{\mu\nu}} \mathcal{T^{\mu\nu}})  = Q + \eta(\mathcal{T_{\mu\nu}} \mathcal{T^{\mu\nu}}) $. Using the MCMC sampling technique combined with Bayesian statistical analysis, we constrain the parameters of the solution $H(z)$ employing CC, BAO, and the latest Pantheon+SH0ES data sets. Additionally, through the Om diagnostic test, we determine that the assumed cosmological model aligns with the quintessence regime.

\section{Introduction}\label{sec1v}
The advent of GR in 1916, courtesy of Albert Einstein, marked a paradigm shift in our understanding of gravity \cite{EIN}.
Over the ensuing century, GR has withstood rigorous empirical scrutiny, vindicated by its accurate predictions: from Mercury's perihelion precession to the bending of starlight by the Sun, gravitational redshift, and even the monumental detection of gravitational waves \cite{BP} from cosmic cataclysms. However, among these triumphs, challenges have emerged,  highlighted by the cosmic acceleration discovered in the early twentieth century \cite{R9a}. This revelation hinted at the limitations of GR on cosmological scales and underscored the pressing need for a more encompassing theory. This motivation sparked huge theoretical investigations into understanding our Universe over the past century \cite{SN,SD,SC}. As quantum mechanics matured throughout this period, scientists, including Einstein himself, embarked on the quest to formulate a coherent theory of quantum gravity. This pursuit gave rise to various contenders such as string theory, and loop quantum gravity theory, among others. However, despite their promise, none have yet achieved the status of being truly comprehensive.
To address the challenge of reconciling GR's predictions with observations on cosmological scales, mainstream research has diverged into two distinct paths.\\
The initial avenue explores the concept of DE, which theorizes that the material composition of the Universe can be depicted by an unusual fluid exerting negative pressure, consequently driving the observed acceleration of cosmic expansion. Numerous comprehensive reviews on DE can be found in the literature. In an alternative path, scientists explore modified gravity theories. Katirci and Kavuk proposed $f(R,\mathcal{T}^{2})$ in \cite{NK}, where $\mathcal{T}^{2} = \mathcal{T_{\mu\nu}} \mathcal{T^{\mu\nu}}$ and $\mathcal{T_{\mu\nu}}$ represent the matter energy-momentum tensor. Roshan and Shojai further delved into the theory, exploring the properties of the form $R + \mathcal{T}^{2}$, termed energy-momentum squared gravity (EMSG) \cite{MR}. In this chapter, our aim is to delve into a further extension of the symmetric teleparallel theory. Specifically, we are motivated to progress from $f(Q,\mathcal{T})$ to $f(Q,\mathcal{T}^{2})$ gravity, where $\mathcal{T}^{2}$ is defined as $ \mathcal{T_{\mu\nu}} \mathcal{T^{\mu\nu}}$, inspired by the extension of $f(R,\mathcal{T})$ gravity to $f(R,\mathcal{T}^{2})$. We shall refer to this extension as Energy-Momentum Squared Symmetric Teleparallel Gravity (EMSSTG), denoted by $f(Q, \mathcal{T}^{2})$. The gravitational action will be governed by an arbitrary function $f(Q, \mathcal{T}^{2})$ of $Q$ and $\mathcal{T}^{2}$. Subsequently, by varying the action with respect to the metric, we can derive the field equations within a metric-affine formalism. These equations will serve as the foundation for exploring the cosmological evolution of the theory in depth. Investigating a specific toy model may yield valuable insights into the effective understanding of the dynamics of the theory.

\section{Cosmology in EMSSTG} \label{sec3v}
The Friedmann like equations for the line element \eqref{2nz} is given by
\begin{equation}
6f_QH^2 - \frac{1}{2}f(Q, \mathcal{T}^{2}) = 8\pi \rho + f_{\mathcal{T}^{2}}(\rho + 4p\rho +3p^2)
\end{equation}
\begin{equation}
6f_QH^2 - \frac{1}{2}f(Q, \mathcal{T}^{2}) -2(\dot{f}_QH + f_Q\dot{H}) = - p 
\end{equation}\\
where dot represents the derivative with respect to time.

We consider the following $f(Q,  \mathcal{T}^{2})$ model based on the specific coupling nature between $Q$ and $ \mathcal{T}^{2}$ as follows
\begin{equation}
f(Q, \mathcal{T_{\mu\nu}} \mathcal{T^{\mu\nu}}) = f(Q, \mathcal{T_{\mu\nu}} \mathcal{T^{\mu\nu}}) = Q + \eta(\mathcal{T_{\mu\nu}} \mathcal{T^{\mu\nu}}) = Q + \eta( \mathcal{T}^{2}) .
\end{equation}

We assume the following barotropic EoS for a fluid that typically relates pressure $(p)$ to density $(\rho)$ in a way that depends only on the local density
\begin{equation}
p=\omega\rho.
\end{equation}

The corresponding Friedmann equation becomes
\begin{equation}
3H^2 = 8\pi \rho +\eta [\frac{3}{2}(1+3\omega^{2})+4\omega]\rho^{2}
\end{equation}
\begin{equation}
3H^2 - 2\dot{H} = \frac{\eta}{2}({\rho}^2 + 3p^2) - 8\pi p,
\end{equation}
where $\omega$ is the EoS parameter.
    
The continuity equation for the assumed $f(Q, T^2)$ model is given as \cite{PRR} 
\begin{equation}  
\dot{\rho}+3 H(\rho+p) =\frac{3 H\left[18 H^{2}-4 \dot{H}+\rho(48 \omega \pi+\eta(1+\omega(14+\omega(19+18 \omega))) \rho)\right]}{16 \pi+2 \eta(3+\omega(8+9 \omega)) \rho}.
\end{equation}

Using the above equations, we solve for $H(z)$ and get
\begin{small}
\begin{equation}
H^2(z)=H_0^2(1+z)^{3(1+\omega)}[1+ \eta  \Omega_0 ^{2} \{ \frac{3}{2}(1+3\omega^2)+4\omega \}
\{(1+z)^{3(1+\omega)}-1 \}].    
\end{equation}    
\end{small}

\section{Observational constraints and methodology}
In this section, we will utilize a statistical approach based on the MCMC technique. Our goal is to determine the efficacy of a model by comparing its predictions with observational data from the cosmos. In particular, we will test the model's validity by analyzing its alignment with BAO measurements and observational Hubble data. The MCMC method plays a crucial role in cosmological studies, as it is widely used to explore parameter spaces and obtain the corresponding probability distributions \cite{Almada/2019}. At its core, the MCMC method involves constructing a Markov chain that systematically explores the parameter space by sampling a specified probability distribution. The chain evolves as a sequence of parameter values, where each new value is generated from the previous one based on transition rules defined by a proposal distribution. This proposal distribution suggests potential new parameter values, and their acceptance is determined by their posterior probability, which combines observational data with prior probability functions. Once the chain reaches convergence, the posterior distribution of the parameters can be estimated by analyzing the frequency of parameter values within the chain. This posterior distribution then enables the determination of the best-fit parameter values and their corresponding uncertainties, ultimately supporting predictions for various observable quantities.

\subsection{Baryon Acoustic Oscillations}
BAO serve as a vital tool in cosmology, enabling us to probe the vast structure of the Universe on a grand scale. These fluctuations originate from acoustic waves that propagated through the early Universe, causing the compression of baryonic matter and radiation within the photon-baryon fluid. This compression creates a unique peak in the correlation function of galaxies or quasars, providing a consistent ruler for measuring cosmic distances. The characteristic size of the BAO peak is determined by the sound horizon at the time of recombination, which depends on factors such as the density of baryons and the temperature of the cosmic microwave background. On large angular scales, BAO occur as separate peaks and are thought to be pressure waves caused by cosmic perturbations in the baryon-photon plasma during the recombination era \cite{Blake/2011,Percival/2010}. The expressions utilized for non-correlated BAO data are as follows

\begin{equation}
\chi^2_{BAO/non cov }= \sum_{i=1}^{26} \frac{\left[H_{th}(z_{i,\vartheta})-H_{obs}^{BAO}(z_{i})\right]^2}{\sigma^{2}(z_{i})}.
\end{equation}
In these expressions, $H_{\text{th}}$ denotes the theoretical values of the Hubble parameter for a particular model characterized by model parameters $\vartheta$. Conversely, $H_{\text{obs}}^{\text{BAO}}$ corresponds to the observed Hubble parameter acquired through the BAO method, while ${\sigma_H}$ represents the error associated with the observed values of $H^{\text{BAO}}$. For the correlated BAO samples, the following expressions are utilized,

\begin{eqnarray}
d_{A}(z) &=& c \int_{0}^{z} \frac{dz'}{H(z')},\\
D_{v}(z) &=& \left[\frac{d_{A}(z)^2 c z }{H(z)}\right]^{1/3},\\
\chi_{BAO/cov}^2 &=& X^{T} C^{-1} X. 
\end{eqnarray}
The comoving angular diameter distance is denoted by $d_{A}(z)$, the dilation scale by $D_{v}(z)$, and the covariance matrix by $C$ \cite{Giostri/2012}.
Hence the total chi-square function for BAO samples is defined as 

\begin{equation}\label{4fv}
\chi^2_{BAO}= \chi^2_{BAO/non cov}+\chi^2_{BAO/cov} 
\end{equation}

\begin{figure}[H]
\includegraphics[scale=0.8]{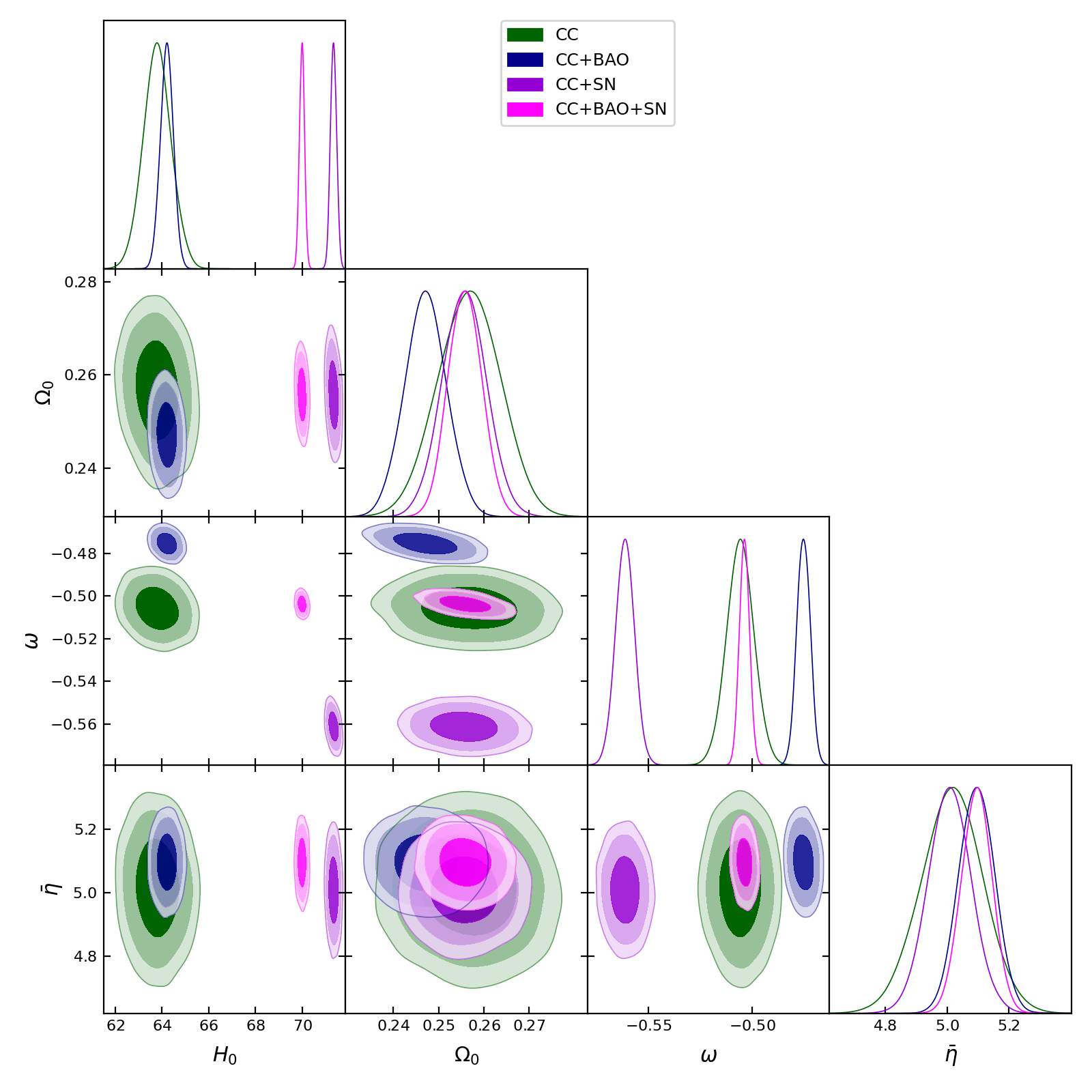}
\caption{The contour plot for the assumed model corresponding to free parameters within the $1\sigma-3\sigma$ confidence interval using CC, CC+BAO, CC+SN, and CC+BAO+SN samples.}\label{f1v}
\end{figure}

The contour plot for the assumed model corresponding to the free parameters within the $1\sigma-3\sigma$ confidence interval using CC, CC+BAO, CC+SN, and CC+BAO+SN samples presented in the figure \eqref{f1v}. The constraints of the parameters obtained are listed in Table \eqref{table1v}.

\centering

\begin{table*}
\begin{center}
 \caption{Best-fit values of model parameters determined  from observational datasets}
 
    \label{table1v}
\begin{tabular}{|l|c|c|c|c|}
\hline 
Datasets              & $CC$ & $CC+BAO$ & $CC+SN$ & $CC+SN+BAO$ \\\hline

$H_0$           & $64^{+0.59}_{-0.59}$  & $64^{0.27}_{-0.28}$ &  $71^{+0.13}_{-0.13}$ & $70^{+0.11}_{-0.11}$ \\
\hline

$\Omega_0$           & $0.26^{+0.0009}_{-0.0071}$  & $0.25^{0.0046}_{-0.0045}$ &  $0.26^{+0.0048}_{-0.0049}$ & $0.26^{+0.00038}_{-0.0037}$\\
\hline

$\omega$           & $-0.51^{+0.0065}_{-0.0065}$  & $-0.48^{0.0032}_{-0.0032}$ &  $-0.56^{+0.0045}_{-0.0045}$ & $-0.5^{+0.0024}_{-0.0024}$\\
\hline

$\eta$           & $5^{+0.098}_{-0.1}$  & $5.1^{0.058}_{-0.058}$ &  $5^{+0.068}_{0.069}$ & $5.1^{+0.049}_{-0.05}$\\
\hline

$q_0$ & $-0.108$  & $-0.066$ &  $-0.19$ & $-0.088$\\ 

\hline

$z_t$ & $0.646$  & $0.354$ &  $1.65$ & $0.491$\\ 
\hline

$\omega_0$ & $-0.405$  & $-0.377$ &  $-0.46$ & $-0.392$\\
\hline

$\chi^2 _{min}(model) $           & $19.448$  & $42.933$ &  $1621.46$ & $1644.631$\\
\hline

$\chi^2 _{min}(\Lambda CDM) $           & $26.597$  & $55.926$ &  $1640.198$ & $1669.527$\\

\hline

$AIC(model) $           & $30.597$  & $59.866$ &  $1644.198$ & $1674.257$\\
\hline

$AIC (\Lambda CDM) $           & $27.448$  & $50.933$ &  $1629.146$ & $1653.631$\\

\hline

$\Delta AIC $           & $3.149$  & $8.933$ &  $15.052$ & $20.626$\\
\hline

\end{tabular}
\end{center}
\end{table*}

\section{Evolutionary behavior of cosmological parameters}
\begin{figure}[H]
\centering
\includegraphics[scale=0.4]{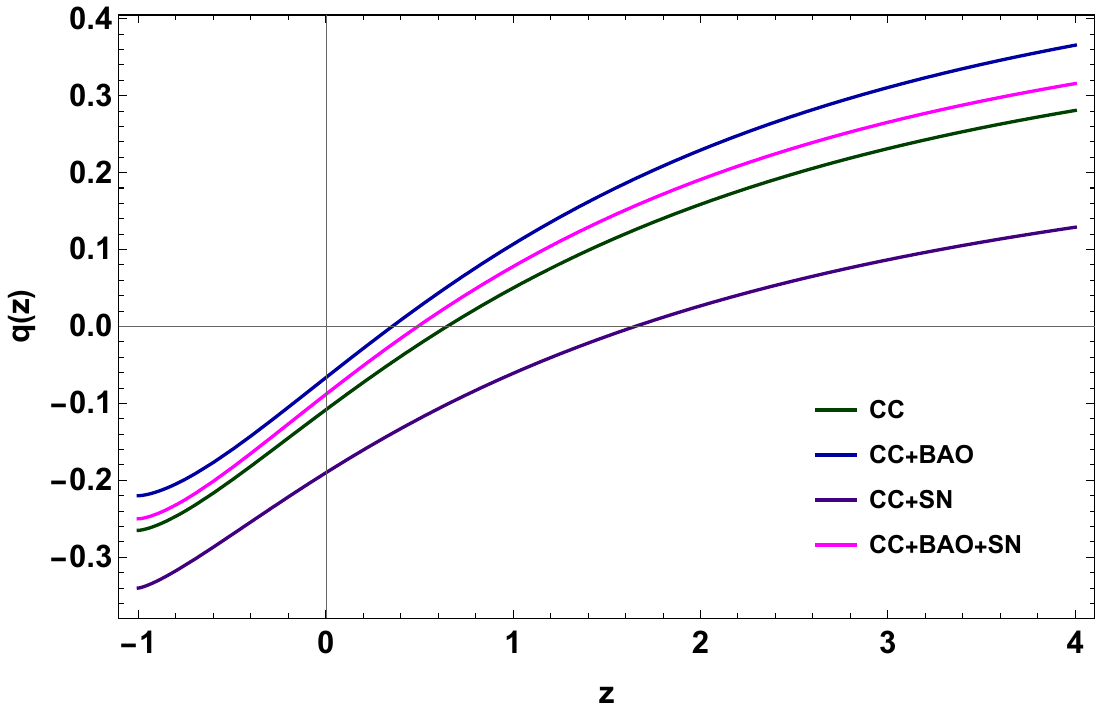}
\caption{Variation of the deceleration parameter $q$ as a function of the redshift $z$ for different datasets.}
\label{f2v}
\end{figure}

\begin{figure}[H]
\centering
\includegraphics[scale=0.52]{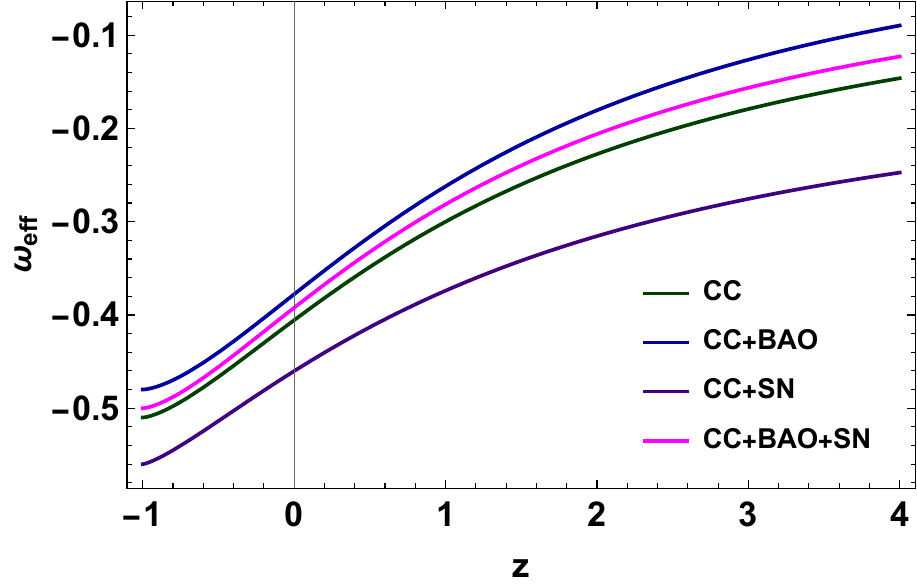}
\caption{Variation of the effective EoS parameter $\omega_{eff}$ as a function of the redshift $z$ for different datasets.}
\label{f2av}
\end{figure}

\begin{figure}[H]
\centering
\includegraphics[scale=0.6]{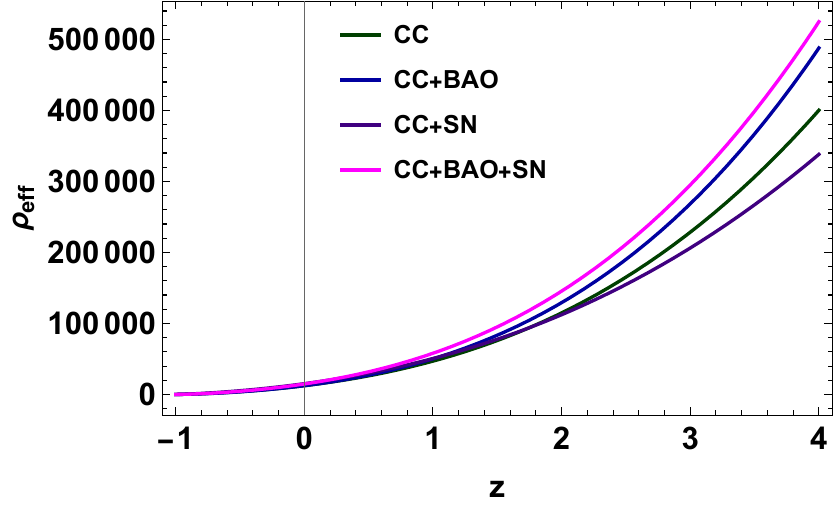}
\caption{Variation of the effective energy density $\rho_{eff}$ as a function of the redshift $z$ for different datasets.}
\label{f2bv}
\end{figure}

\justifying
The visualizations shown below vividly illustrate how the dynamics of the Universe can exhibit remarkable intricacies, depending upon the specific values of the parameters involved. Further, figure \eqref{f2v} illustrates the trajectory of the Universe, commencing with a decelerating phase ($q>0$) before transitioning to an accelerating phase ($q<0$) following a redshift transition denoted $z_{t}$. The deceleration parameter, denoted $q$, is calculated using the expression $q= -\frac{\dot{H}}{H^2}-1$. This evolutionary pattern is consistent with the current understanding of the dynamics of the Universe, which is characterized by three distinct phases: an initial decelerating phase, followed by a transition to an accelerating expansion, and culminating in a late-time acceleration phase. This behavior aligns with observational evidence and theoretical models that describe the evolution of the Universe over cosmic time. Remarkably, our results show that the current value of the deceleration parameter ($q_{0}$) depicts the acceleration phase \cite{Almada/2019,Basilakos/2012} and the transition redshift ($z_{t}$) \cite{Garza/2019,Jesus/2020} align well with the observations of the data set taken, listed in Table \eqref{table1v}. Furthermore, the same result is reflected in the behavior of the effective EoS parameter defined by $\omega_{eff}= -\frac{\dot{2H}}{3H^2}-1$, presented in figure \eqref{f2av}. Moreover, the effective matter-energy density shows expected positive behavior in the entire redshift domain, presented in figure \eqref{f2bv}. Note that we observe that the trajectories of the cosmological evolutionary parameters corresponding to the CC + SN samples are much deviated in comparison to other dataset combinations. The underlying root cause of this deviation is the nature of datasets, such as BAO and CC datasets are more sensitive to the early Universe and can describe the proper transition epoch, whereas the SN datasets is concentrated at lower redshifts, mostly less than one; therefore, it is more focused on the present cosmic acceleration rather than the full history of expansion. Therefore, the presence of SN data points in CC+SN samples favors a high transition value, whereas the CC and BAO combination provides a true transition value. For example, one must check the reference \cite{CPS,R29d,R15a} to see how the SN data points prefer a high transition redshift, generally greater than one, whereas the observations of BAO and CC prefer a low transition redshift, generally less than one, and thus this will lead to a discrepancy in the $H_0$, $\Omega_0$, as well as $z_t$ value due to the nature of the underlying datasets and its measurement techniques.\\
A simple test technique that uses only the first-order derivative of the cosmic scale factor is the Om diagnostic. Its equation for a spatially flat Universe is as follows

\begin{equation}\label{5bv}
Om(z)= \frac{\big(\frac{H(z)}{H_0}\big)^2-1}{(1+z)^3-1}.
\end{equation}

\begin{figure}[H]
\centering
\includegraphics[scale=0.4]{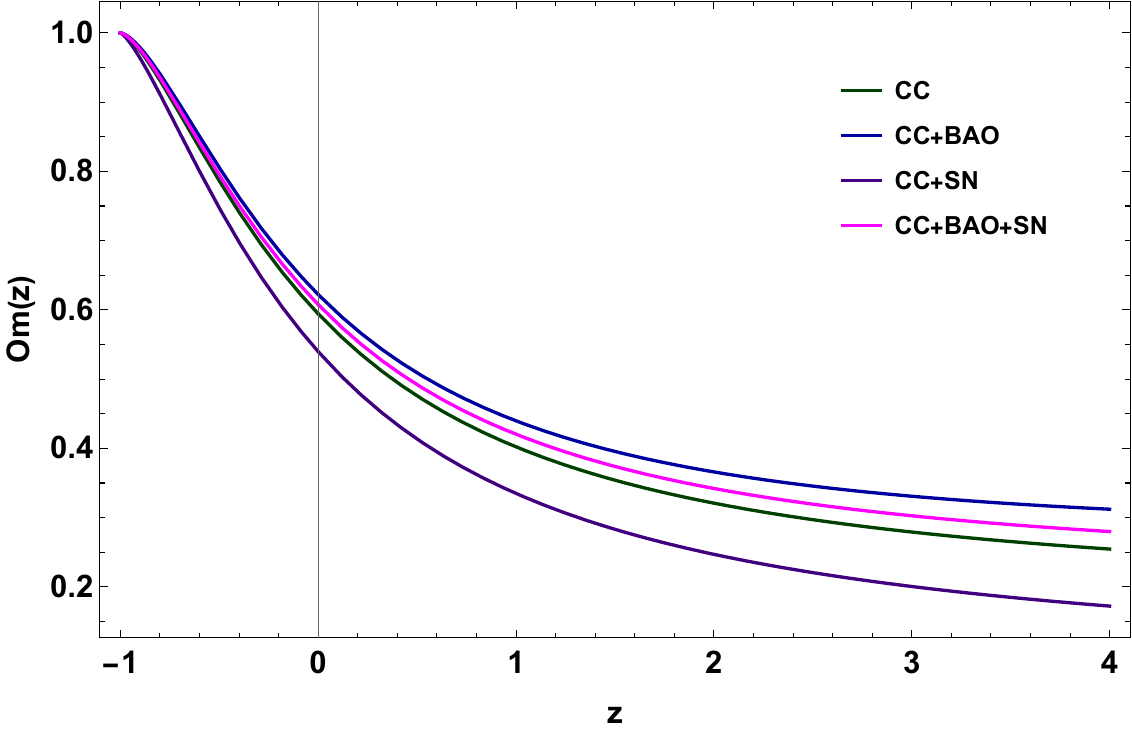}
\caption{Variation of Om diagnostic parameter as a function of the redshift $z$ for different datasets.}
\label{f3v}
\end{figure}

The descending slope of the $Om(z)$ curve indicates quintessence-like behavior, while an ascending slope corresponds to phantom behavior. In contrast, a constant $Om(z)$ signifies the characteristics of the $\Lambda$CDM model. From the behavior of the Om diagnostic parameter presented in figure \eqref{f3v}, it can be inferred that our cosmological framework exhibits quintessence-like behavior.

\section{Conclusion}\label{sec4}
In this chapter, we presented an extended formulation of symmetric teleparallel gravity by generalizing the gravitational Lagrangian to include an arbitrary function of the form $f(Q,\mathcal{T_{\mu\nu}} \mathcal{T^{\mu\nu}})$. We derived the FLRW equations for a flat, homogeneous, and isotropic spacetime. To deepen our understanding of the cosmological framework within this theory, we found the analytical solution for the barotropic fluid case $p=\omega \rho$ for the model $f(Q, \mathcal{T_{\mu\nu}} \mathcal{T^{\mu\nu}})  = Q + \eta(\mathcal{T_{\mu\nu}} \mathcal{T^{\mu\nu}})$. Furthermore, we constrained the parameters of the derived solution $H(z)$ by incorporating CC data, BAO measurements, and the latest SN samples. This was achieved using the MCMC sampling technique combined with Bayesian statistical analysis. The resulting constraints on the parameters of the considered cosmological model are summarized in Table \eqref{table1v}, along with the corresponding contour plots depicting the parameter correlation, in figure \eqref{f1v}. In addition, we analyze the behavior of the deceleration parameter in figure \eqref{f2v} depicting the observed accelerating phenomenon with the transition epoch. The present value of the deceleration parameter along with the redshift of the transition is listed in Table \eqref{table1v}. Furthermore, the same result is reflected in the behavior of the effective equation of the state parameter defined by $\omega_{eff}= -\frac{\dot{2H}}{3H^2}-1$, presented in figure \eqref{f2av}. Moreover, the effective matter-energy density shows expected positive behavior in the entire redshift domain, presented in figure \eqref{f2bv}.  In addition, we used the Om diagnostic test to assess the behavior of the supporting DE. We found that the behavior of Om diagnostic parameter presented in figure \eqref{f3v} favors the quintessence-type dark energy model. Thus, our investigation successfully describes the late-time expansion phase of the Universe. However, we would like to note that since square gravity is inherently dependent on the choice of Lagrangian density ($\mathcal{L}$), it cannot accommodate scalar fields like inflation fields as $P_{\phi}=\frac{1}{2}\dot{\phi}^2-V(\phi)$, which is very different from $\mathcal{L}_{fluid}=P$. This, incorporating the scalar field into $T$ and $T^2$ gravity, comes with more complex issues, as discussed the same in the case of $f(R,\mathcal{T})$ gravity \cite{Harko/2011}. 

Building upon this study, the next chapter focuses on the investigation of the dark sector of the Universe, specifically DE and DM, within an alternative gravity framework based entirely on non-metricity. We explore the $f(Q)$ gravity model with quadratic functional form, $f(Q) = \gamma Q^2$, where $\gamma$ is a free parameter, and describe DM using the EBEC EoS. The interaction between DM and DE is incorporated through an energy exchange term $\mathcal{Q} = 3bH \rho$, with $b > 0$ representing the transfer strength. To further analyze the system, we introduce dimensionless phase-space variables, transforming the equations into an autonomous system. A detailed stability analysis of the dynamical system reveals that the cosmological model effectively captures the evolution of the Universe from a decelerated matter-dominated phase to a stable accelerated expansion era.

\chapter{Stability analysis of a $f(Q)$ cosmological model along with EBEC dark matter equation of state} 

\label{Chapter6} 

\lhead{Chapter 6. \emph{Stability analysis of a $f(Q)$ cosmological model along with EBEC dark matter equation of state}} 
\vspace{10 cm}
* The work presented in this chapter is covered by the following publication: \\
 
\textit{Stability analysis of a $f(Q)$ cosmological model along with EBEC dark matter equation of state}, Physics of the Dark Universe \textbf{47}, 101772 (2025).

\clearpage

This chapter aims to investigate the mysterious components of the Universe, specifically DE and DM. To characterize the origin of DE, we investigate alternative gravity theory to general relativity exhibiting a flat background framework that is based entirely on non-metricity. Specifically, we consider the $f(Q)$ functional from as $f(Q) = \gamma Q^2$, where $\gamma$ serves as a free parameter in the model, whereas to describe DM, we assume the relation $p = \alpha \rho + \beta \rho^2$ known as the EBEC dark matter EoS. We obtain the motion equations and the continuity relation that incorporate both the DM and the DE fluid, along with an interaction term, particularly $\mathcal{Q} = 3bH \rho$, where $b > 0$ represents the strength of the energy exchange from DE to DM. Further, we invoke a set of dimensionless phase-space variables that enables us to transform the dynamics of the cosmological system into an autonomous system. Then we perform a detail stability analysis of the considered cosmological model, and we conclude that the cosmological model under consideration effectively captures the dynamics of the Universe from a decelerated matter-dominated era to a stable phase of accelerated expansion.

\section{Introduction}\label{sec1q}
\justifying

It is commonly acknowledged that, when exploring cosmological models, the use of auxiliary variables enables the transformation of cosmological equations into an autonomous dynamical system \cite{COPE}. This leads to a system of the form $X' =f(X)$, where $X$ is a column vector composed of the auxiliary variables and $f(X)$ represents the vector field. Analyzing the stability of such an autonomous system generally follows a structured process. First, the critical points (or equilibrium points) $X_c$ are determined by setting $X'=0$.  Next, linear perturbations around the critical point $X_c$ are introduced by expressing $X=X_c+P$, where $P$ represents the perturbed auxiliary variables. This leads to the linearized system (up to first order) $P'=AP$, where $A$ 
is the coefficient matrix derived from the perturbed equations. Ultimately, the stability properties of each hyperbolic critical point are determined by examining the eigenvalues of the matrix $A$. A critical point $X_c$ is classified as stable if all the corresponding eigenvalues have negative real parts, unstable if all the eigenvalues have positive real parts, or as a saddle point if the eigenvalues have real parts with mixed signs (i.e., some positive and some negative). Numerous intriguing results in the study of modified gravity, employing the dynamical system approach, have been documented in various references \cite{R341,WOM,Mishra-1,Mishra-2,HAMID}.

The work in this chapter is presented as follows: In section \eqref{sec5q}, we investigate the stability of the considered non-linear $f(Q)$ model along with EBEC EoS utilizing dynamical system analysis. In section \eqref{sec6q}, we highlight our findings.

\section{Stability analysis}\label{sec5q}
\justifying
We assume the following $f(Q)$ functional form that effectively models the progression of Universe from a matter-dominated phase to the de Sitter era \cite{RA},
\begin{equation}\label{3dq}
f(Q)= \gamma Q^2,
\end{equation}
where $\gamma$ is free model parameter.

We start by introducing the following dimensionless variables that capture the complete evolution of the system, called phase-space variables. These variables enable us to reformulate the dynamics of the considered physical system into the framework of an autonomous system, simplifying the analysis of the system's behavior. The dimensionless phase space variables under consideration are as follows
\begin{equation}\label{4aq}
x=\Omega_m=\frac{\rho_m}{3H^2}, \:\: y=\frac{1}{1+\frac{H_0}{H}}, \:\: \text{and} \:\: z=\Omega_{de}=\frac{\rho_{de}}{3H^2}.
\end{equation}
Then the equation \eqref{2qz} becomes
\begin{equation}\label{4bq}
x+z=1 .
\end{equation}
Hence, we have $z=1-x$ and $0 \leq x \leq 1$ as well as $0 \leq z \leq 1$. Note that $H \rightarrow \infty \implies y \rightarrow 1$ represents the early phase of the Universe, whereas the limit $H \rightarrow 0 \implies y \rightarrow 0$ denotes the late-time phenomenon of the Universe. In addition, at present epoch $H=H_0 \implies y=\frac{1}{2}$.

\begin{figure}[H]
\centering
\includegraphics[scale=0.8]{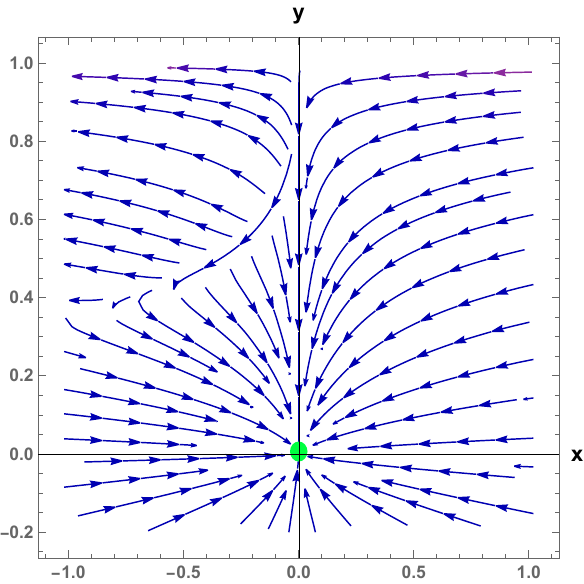}
\caption{Phase space portrait for the given autonomous system with the parameter constraints  $\alpha=-0.95$, $b=0.001 $, and $\beta=-1$, where the green dot representing an attractor. }\label{f1q}
\end{figure}

On utilizing equation \eqref{2qz}, equation \eqref{2sz}, equation \eqref{2tz}, and equation \eqref{3dq} in the equation \eqref{2rz}, we obtained the following expression
\begin{equation}\label{4cq}
\frac{\dot{H}}{H^2}=-\frac{3}{4} \left[1+\alpha+\bar{\beta}x \left( \frac{y}{1-y} \right)^2 \right],
\end{equation}
where $\bar{\beta}=3H_0^2 \beta$. 
The deceleration parameter and the effective EoS parameter play a vital role in describing the behavior of the expansion phase of the Universe.

\begin{equation}\label{4eqw}
\omega_{tot} = \frac{p+p_{de}}{\rho+\rho_{de}}= -1-\frac{2\dot{H}}{3H^2}.
\end{equation}
Utilizing the equation \eqref{4cq} in equation \eqref{4eqw}, we acquired the following expressions
\begin{equation}\label{4fqq}
q=-1+\frac{3}{4} \left[1+\alpha+\bar{\beta}x \left( \frac{y}{1-y} \right)^2 \right],
\end{equation}
\begin{equation}\label{4gq}
\omega_{tot} =-1+\frac{1}{2} \left[1+\alpha+\bar{\beta}x \left( \frac{y}{1-y} \right)^2 \right].
\end{equation}
On differentiating the expression $x$ and $y$ in the equation \eqref{4aq} with respect to $N=ln(a)$, we obtain
\begin{equation}\label{4eq}
x'= \frac{dx}{dN} = \frac{3x}{2} \left[  6b-1-\alpha - \bar{\beta}x \left( \frac{y}{1-y} \right)^2  \right] = f_1(x,y),
\end{equation}
\begin{equation}\label{4fq}
y'= \frac{dy}{dN} = \frac{3y(y-1)}{4} \left[ 1+\alpha + \bar{\beta}x \left( \frac{y}{1-y} \right)^2  \right]  = f_2(x,y).
\end{equation}
Clearly, the system of equations \eqref{4eq} and \eqref{4fq} is an autonomous non-linear system. Now, we obtained the critical point of this system by putting $x'=0$ and $y'=0$ as follows
\begin{equation}\label{4gq}
(x_c,y_c) = (0,0).
\end{equation}
We investigate the stability of this particular autonomous system around its critical points. To begin, we linearize the autonomous set of equations by considering small perturbations in the vicinity of critical points, expressed as $(x,y) \longrightarrow (x_c+\delta x, y_c+\delta y)$. In order to analyze whether these perturb quantities $(\delta x, \delta y)$ grows or decays, we construct the following linear system of autonomous differential equations by utilizing the Taylor series expansion with the variable $(\delta x, \delta y)$ around the critical point $(x_c,y_c)$ and neglecting the tiny quadratic expressions
\begin{equation}\label{4hq}
\left[
\begin{array}{c}
\delta x'  \\ 
\delta y'  \\ 
\end{array}
\right] \,=   \left[
\begin{array}{cc}
\left( \frac{\partial f_1}{\partial x} \right)_{0} & \left( \frac{\partial f_1}{\partial y}  \right)_{0}  \\ 
\left( \frac{\partial f_2}{\partial x}  \right)_{0} & \left( \frac{\partial f_2}{\partial x}  \right)_{0} \\ 
\end{array}
\right]  \left[
\begin{array}{c}
\delta x \\ 
\delta y  \\ 
\end{array}
\right] .
\end{equation}
The Jacobian matrix in the equation \eqref{4hq} becomes
\begin{equation}\label{4iq}
J= \left[
\begin{array}{cc}
\left( -\frac{3 (\alpha +1)}{2}+9 b-\frac{3 \beta  x y^2}{(y-1)^2}  \right)_{0} & \left( \frac{3 \beta  x^2 y}{(y-1)^3}  \right)_{0}  \\ 
\left( \frac{3 \beta  y^3}{4 (y-1)} \right)_{0} & \left( \frac{3}{4} \left(\frac{\beta  x (2 y-3) y^2}{(y-1)^2}+(\alpha +1) (2 y-1)\right) \right)_{0}  \\ 
\end{array}
\right],
\end{equation}    
where suffix 0 indicates the critical point $(x_c,y_c)$. Now, on evaluating the Jacobian matrix in equation \eqref{4iq} at the critical point obtained in equation \eqref{4gq}, we acquired the following eigenvalues of the above Jacobian matrix corresponding to the critical point in equation \eqref{4gq}
\begin{equation}\label{4jq}
\lambda_1 = -\frac{3}{4} (\alpha +1) \:\:\: \text{and} \:\:\: \lambda_2 = \frac{3}{2} (-\alpha +6 b-1).
\end{equation}
It is evident from the equation \eqref{4jq}, the critical point $(x_c,y_c)=(0,0)$ exhibits stable behavior for the parameter constraints $\alpha > -1$ and $b < \frac{\alpha+1}{6}$. For instance, we choose $\alpha=-0.95$, $b=0.001 < \frac{\alpha+1}{6} = 0.0083$, and $\beta=-1$. The phase space portrait corresponding to these parameter constraints is shown in figure \eqref{f1q}.

\begin{figure}[H]
\centering
\includegraphics[scale=0.475]{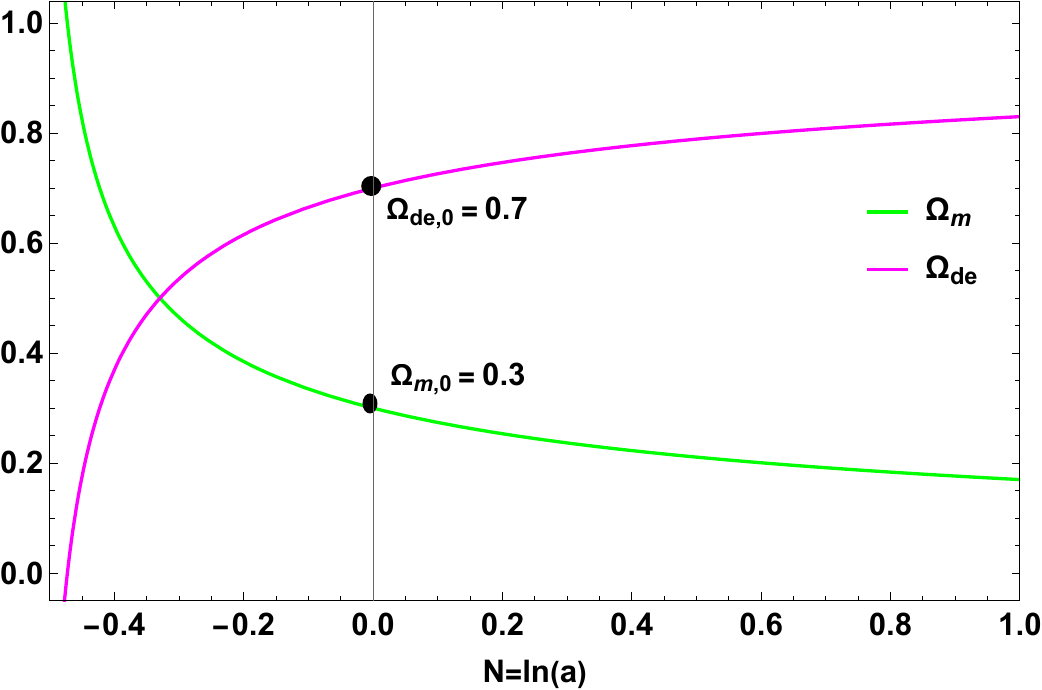}
\caption{Profile of the dimensionless matter and DE density parameter corresponding to the parameter constraints  $\alpha=-0.95$, $b=0.001 $, and $\beta=-1$.}
\label{f2q}
\end{figure}

\begin{figure}[H]
\centering
\includegraphics[scale=0.475]{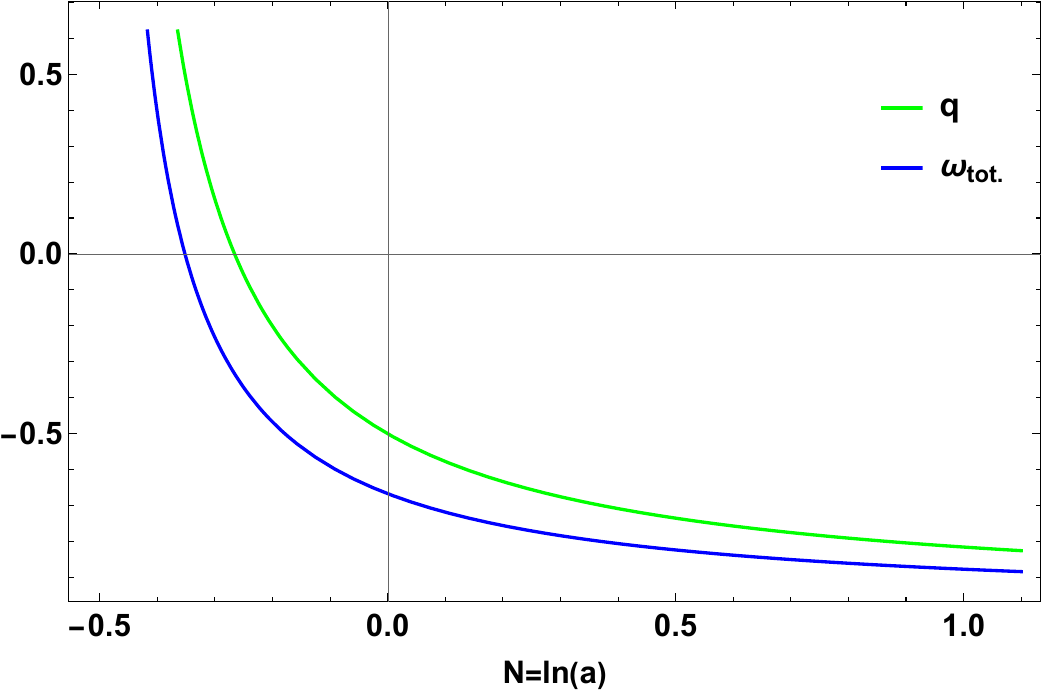}
\caption{Profile of the equation of state and the deceleration parameter corresponding to the parameter constraints  $\alpha=-0.95$, $b=0.001 $, and $\beta=-1$.}
\label{f3q}
\end{figure}

The evolutionary profile of the dimensionless density parameter is presented in figure \eqref{f2q}, while the profile of the effective EoS and the deceleration parameter are presented in figure \eqref{f3q}.

From equation \eqref{4jq}, we obtained the eigenvalues $\lambda_1=-0.0375$ and $\lambda_2=-0.066$ that correspond to the critical point $(x_c,y_c)=(0,0)$ for the parameter constraints $\alpha=-0.95$, $b=0.001 $, and $\beta=-1$. It is clear from figure \eqref{f1q} that the phase-space trajectories are converged to the critical point $(x_c,y_c)=(0,0)$. Moreover, at this point, we obtain $q=-0.9625$ and $\omega_{tot} =-0.975$. Thus, the critical points $(x_c,y_c)=(0,0)$ indicate a stable accelerated behavior of the given physical system. Note that at the critical point $(x,y)=(0,0)$, the equation \eqref{4fqq} becomes $q=-1+\frac{3}{4}(1+\alpha)$. So to make the critical point $(x,y)=(0,0)$ stable we choose the parameters in the regime $\alpha > -1$ and $b < \frac{\alpha+1}{6}$ and that stable critical point shows the de Sitter type accelerated expansion if we choose $\alpha$ in such a way that the value of $q$ becomes close to $-1$. Thus, the choice $\alpha=-0.95$ gives the value $q=-0.962$ which is much closer to the de Sitter limit. Also note that one cannot obtain the perfect limit $q=-1$, since the choice $\alpha=-1$ that gives the limit $q=-1$ makes the critical point unstable, as evident from the expression of the first eigenvalue $\lambda_1 = -\frac{3}{4}(\alpha+1)$. As a consequence, a model with saddle- or unstable-de Sitter-accelerated epoch is not physically viable. Thus, the parameter choice is uniquely made in order to get a stable critical point along with the de Sitter accelerated behavior. Also, note that a de Sitter type critical point is physically viable only if it is obtained for the late-time epoch. From the choice of the dynamical variable $ y=\frac{1}{1+\frac{H_0}{H}}$, it is evident that $y\longrightarrow 0$ can be obtained for the limit $H \rightarrow 0$. Thus, the critical point $(0,0)$ shows the late-time behavior of the Universe. In addition, from figure \eqref{f2q}, we found that the DE density dominates, whereas the matter energy density vanishes with expansion of the Universe. Moreover, the present value of dimensionless matter energy and DE density parameter are obtained as $\Omega_{m,0}=0.3$ and $\Omega_{de,0}=0.7$, which is consistent with cosmological observations. Further, from figure \eqref{f3q}, we conclude that the considered $f(Q)$ gravity model along with the extended BEC EoS successfully describes the evolutionary phase of the Universe from a decelerated matter era to the accelerated expansion epoch. In addition, the present value of the deceleration parameter $q_0 \approx -0.5$ is quite consistent with the observational constraints.

\section{Conclusion}\label{sec6q}
\justifying
In this chapter, we delved into the dark sector of the Universe, specifically focusing on DM and DE. We examined an extended version of the EoS for DM, commonly referred to as the EBEC equation of state for DM, given as $p=\alpha \rho + \beta \rho^2$, along with the modified $f(Q)$ Lagrangian. For a deeper understanding of the various parameterization schemes, one can check the references \cite{SSS1,SSS2,SSS3,SSS4,SSS5}. This assumed EoS has already been considered in the references \cite{R13d,RA}. However, the assumption made in the work \cite{R13d} is very complex due to the presence of the viscosity term as well as the boundary term in teleparallel gravity. Moreover, the work \cite{RA} investigates the observational aspects of this model, whereas the work presented in this manuscript investigates the dynamical aspects of the model. To describe another significant dark component, namely the elusive DE, we explore alternative gravity theory to general relativity exhibiting a flat background framework that is based entirely on non-metricity. Specifically, we focus on the power law $f(Q)$ Lagrangian given by $f(Q)= \gamma Q^2$, where $\gamma$ represents the free parameter of the model \cite{R44d}. We derived the corresponding motion equations and the continuity relation for both dark fluid component, incorporating an interaction term. The interacting term considered $\mathcal{Q} = 3bH \rho$ indicates an energy exchange from DE to DM, where $b > 0$ is the strength of the energy transfer.

We introduced a set of dimensionless phase space variables in equation \eqref{4aq}. These phase space variables allow us to reframe the dynamics of the given cosmological system into an autonomous system. Using these phase space variables in equation \eqref{4aq}, we obtained an autonomous two-dimensional nonlinear system presented in equation \eqref{4eq} and equation \eqref{4fq}. Furthermore, putting $x'=0$ and $y'=0$, we obtained a single critical point of the system of equation \eqref{4eq} and equation \eqref{4fq} as $(x_c,y_c)=(0,0)$. In order to probe the behavior of this critical point, we linearize the autonomous system by considering small perturbations near the critical points as $(x,y) \longrightarrow (x_c+\delta x, y_c+\delta y)$, and then investigate that these perturbations $(\delta x, \delta y)$ grow or decay, via a linear system of autonomous differential equations obtained by using the Taylor series expansion with the variable $(\delta x, \delta y)$ around the critical point $(x_c,y_c)$ and neglecting the tiny quadratic expressions.

We calculated the eigenvalues of the Jacobian matrix in equation \eqref{4hq} as $\lambda_1 = -\frac{3}{4} (\alpha +1) $ and $\lambda_2 = \frac{3}{2} (-\alpha +6 b-1) $. We found that the critical point $(x_c,y_c)=(0,0)$ exhibits a stable behavior for the parameter constraints $\alpha > -1$ and $b < \frac{\alpha+1}{6}$. We choose the parameter values as $\alpha=-0.95$, $b=0.001 < \frac{\alpha+1}{6} = 0.0083$, and $\beta=-1$, and then present the corresponding phase space portrait in figure \eqref{f1q}. It is clear from figure \eqref{f1q} that the phase-space trajectories are converge to the critical point $(x_c,y_c)=(0,0)$, having $q=-0.9625$ and $\omega_{tot} =-0.975$. Thus, the critical points $(x_c,y_c)=(0,0)$ indicate a stable behavior of the given physical system. Moreover, the behavior of dimensionless density parameters have been presented in figure \eqref{f2q}, along with the present value obtained as $\Omega_{m,0}=0.3$ and $\Omega_{de,0}=0.7$, which is consistent with cosmological observations. Further, we predict that the DE density dominates, whereas the matter energy density vanishes with expansion of the Universe in the far future. Lastly, from figure \eqref{f3q}, we conclude that the considered cosmological scenario successfully describes the evolutionary phase of the Universe from a decelerated matter era to the accelerated expansion epoch in a stable way.

The work presented in this chapter shows a late-time accelerated expansion epoch of the Universe originating from the decelerated matter epoch utilizing the phase-space trajectories. As is well known, the Universe began with the big bang singularity, then it passed through radiation and matter epoch, and now it is in accelerating expansion phase. Thus, outcomes of the present manuscript describe a piece of the evolutionary phase of the Universe, particularly the late-time expansion phase. As a future perspective, it is interesting to find a suitable cosmological model with an unstable critical point representing the big bang singularity, two saddle points representing radiation and matter epoch, and a stable critical point showing a de Sitter accelerated expansion.


\chapter{Concluding remarks and future perspectives} 

\label{Chapter7} 

\lhead{Chapter 7. \emph{Concluding remarks and future perspectives}} 

 \clearpage
 
In this thesis, we have explored various cosmological models of acceleration that effectively capture the dynamics of the late time evolution of the Universe, with the broader aim of deepening our understanding of the accelerated expansion phenomenon. A summary of the discussions, analyses, and key outcomes presented in each chapter is provided below.

\section{Concluding remarks}

In Chapter \ref{Chapter1}, we began our introductory phase with the description of the accelerating Universe. Then, we briefly discussed the fundamentals of general relativity and its alternative formalism such as TEGR and STEGR. In addition, we have discussed several cosmological solutions to GR. Moreover, we highlight the pros and cons of the standard model of cosmology. Lastly, we have introduced the fundamentals of some important non-Riemannain spacetime geometry, such as modified teleparallel and symmetric teleparallel gravity and its extensions. 
 
In Chapter \ref{Chapter2}, we analyzed the cosmological implications of $f(T,\mathcal{T})$ theory by considering the squared-torsion model $f(T,\mathcal{T})= \alpha \mathcal{T}+ \beta T^2$, where $\alpha$ and $\beta$ are free parameters. We derived the solutions to the modified Friedmann equations, expressing the Hubble parameter as a function of the redshift $z$. Furthermore, we utilized recent observational datasets such as Hubble, BAO, SNeIa and the joint analysis to constrain the model's free parameters. Based on the constrained model parameters, we discovered a diverse range of intriguing cosmological behaviors. Notably, our analysis revealed the evolution of the deceleration parameter, which explicitly transitions from a decelerating phase to an accelerating one, effectively accounting for the late-time expansion of the Universe. Additionally, the effective EoS ($\omega_{eff}$) and the total EoS ($\omega$) behave in a similar fashion, demonstrating that the cosmic fluid has the characteristics of quintessence DE. Moreover, we found that the present values of $q_{0}$, $\omega_{0}$ and $z_{t}$ are in good agreement with SNeIa and Hz + BAO + SNeIa datasets.

In Chapter \ref{Chapter3}, we explore the inflationary scenario within the framework of torsion-trace coupling gravity, which is based on a Lagrangian density derived from a function $f(T,\mathcal{T})$. This function depends on the torsion $T$ and the trace of the energy-momentum tensor $\mathcal{T}$. Our analysis revealed the evolution of the deceleration parameter, which transitions from deceleration to acceleration, providing a plausible explanation for the late-time Universe. By applying the slow-roll conditions to a specific $f(T,\mathcal{T})$ model, we computed several key inflationary observables,  such as the ratio of tensor to scalar perturbations $(r)$, the spectral index of scalar perturbations $(n_s)$, the running of the spectral index $(\alpha_s)$, the tensor spectral $n_T$, and the number of e-folds parameter. The numerical results obtained are well aligned with current observational data. So far, many studies have been done in the background of the torsion-based modified theory focusing on the current acceleration scenario of the Universe, and its' late-time acceleration \cite{Harko/2014a, AAQID/2023, dsg/2016}, whereas in this chapter, we aimed to study the early-inflationary scenario and successfully presented it. We have not only successfully presented the slow-roll inflationary scenario, but also presented a way to constrain the parameters of the $f(T,\mathcal{T})$ cosmological model. 

In Chapter \ref{Chapter4}, we attempted to explore the dark sector of the Universe i.e. DM and DE. We considered an extended form of the EoS for DM, widely known as the EBEC EoS for DM,  a comprehensive model combining normal DM and the quantum ground state. Now to describe another prominent dark component, i.e. undetected DE, we consider the modified theories of gravitation within a flat spacetime geometry, dependent solely on non-metricity, particularly, we consider the power law $f(Q)$ Lagrangian $f(Q)= \gamma \left(\frac{Q}{Q_0}\right)^n$, where $\gamma$ and $n$ are free parameters. We presented the corresponding Friedmann-like equations and the continuity equation for both dark components along with an interacting term. The interaction term is directly proportional to the product of the Hubble parameter and the energy density of DM. In other words, it signifies the energy exchange between the dark sector of the Universe. We obtained the analytical solution of the corresponding equations, i.e. the Hubble function in terms of the redshift. In addition, to find the best-fit values of the parameters of the assumed theoretical model, we utilized Bayesian analysis to estimate posterior probability using the likelihood function and the MCMC sampling technique. The constraints obtained on the free parameter space with $68 \%$ confidence limit are $H_0=72^{+0.11}_{-0.12}$, $n=-2.9^{+0.066}_{-0.067}$, $\beta=0.83^{+0.12}_{-0.11}$, $\eta=-2.5^{+0.06}_{-0.061}$, and $c=0.17^{+0067}_{-0.0068}$. In addition, to examine the robustness of our MCMC analysis, we performed a statistical evaluation using the AIC and BIC. We obtained $\Delta AIC=0.85$ and $\Delta BIC=15.5$, so it is evident from the $\Delta AIC$ value that there is strong evidence in favor of the assumed theoretical $f(Q)$ model. We presented the evolutionary profile of the deceleration parameter and the energy density and found that the assumed model shows a transition from the decelerated epoch to the accelerated expansion phase of the de Sitter type with the redshift transition $z_t=0.288^{+0.031}_{-0.029}$. Moreover, the present value of the deceleration parameter is obtained as $q(z=0)=q_0=-0.56^{+0.04}_{-0.03}$ ($68 \%$ confidence limit), which is quite consistent with cosmological observations. Further, we found the expected positive behavior of the effective energy density. More on the energy density, we obtained $\Omega_0 = \frac{(1-2n)\gamma}{Q_0} $. For the STEGR case i.e. $n=1$ and $\frac{\gamma}{Q_0}=-1$, we obtained the expected value $\Omega_0=1$ that represents the matter dominated phase. Furthermore, for the parameter value $n=-2.9$ we acquired $\Omega_0 = \frac{6.8 \gamma}{Q_0} $, which aligns with the observed value $\Omega_0 \in [0.25, 0.35] $ for the range of viable parameters $\frac{\gamma}{Q_0} \in [19, 27] $. Lastly, we investigated the thermodynamical stability of the assumed theoretical model by examining the sound speed parameter and observed that the contribution of the normal DM and the DM halo existing in a quantum ground state can be analyzed separately by estimating the value of the parameter $\alpha$ using the specific value of the intensity parameter $b^2$ and the obtained constrained value of $\eta$. We found that the considered theoretical $f(Q)$ model can efficiently address the late-time expansion phase of the Universe with the observed transition epoch in a stable way. 

In Chapter \ref{Chapter5}, we introduced a further extension of symmetric teleparallel gravity by broadening the gravity Lagrangian with an arbitrary function of $f(Q,\mathcal{T_{\mu\nu}} \mathcal{T^{\mu\nu}})$. We derived the FLRW equations for a flat, homogeneous, and isotropic spacetime. To deepen our understanding of the cosmological framework within this theory, we found the analytical solution for the barotropic fluid case $p=\omega \rho$ for the model $f(Q, \mathcal{T_{\mu\nu}} \mathcal{T^{\mu\nu}})  = Q + \eta(\mathcal{T_{\mu\nu}} \mathcal{T^{\mu\nu}})$. Furthermore, we constrained the parameters of the derived solution
$H(z)$ By incorporating CC data, BAO measurements, and the latest SN samples. This was achieved using the MCMC sampling technique combined with Bayesian statistical analysis. The approach allowed us to rigorously determine the parameter values and their uncertainties, ensuring consistency with observational constraints. The obtained constraints on the parameters of considered cosmological settings are listed in the Table \eqref{table1v}. In addition, we found the deceleration parameter depicts the observed accelerating phenomenon with the transition epoch. The present value of the deceleration parameter along with the transition redshift is listed in the Table \eqref{table1v}. Moreover, the effective matter-energy density show expected positive behavior in the entire domain of redshift. Further, we employed the Om diagnostic test to assess the behavior of supporting DE. We found that the behavior of Om diagnostic parameter favors the quintessence type DE model. Thus, our investigation successfully describe late time expansion phase of the Universe. However, we would like to note that as the square gravity is inherently dependent on the choice of $\mathcal{L}$, it can not accommodate scalar fields like inflation fields as $p_{\phi}=\frac{1}{2}\dot{\phi}^2-V(\phi)$ that is very different from $\mathcal{L}_{fluid}=p$. This, incorporating the scalar field into $\mathcal{T}^2$ and $\mathcal{T}$ gravity, come along with more complex issues as discussed the same in case of $f(R,\mathcal{T})$ gravity \cite{Harko/2011}. \\

In Chapter \ref{Chapter6}, we delved into the dark sector of the Universe, specifically focusing on DM and DE. We examined an extended version of the EoS for DM, commonly referred to as the EBEC EoS for DM, given as $p=\alpha \rho + \beta \rho^2$, along with the modified $f(Q)$ Lagrangian. Specifically, we focus on the power law $f(Q)$ Lagrangian given by $f(Q)= \gamma Q^2$, where $\gamma$ represents the free parameter of the model. We derived the corresponding motion equations and the continuity relation for both dark fluid component, incorporating an interaction term. The interacting term considered $\mathcal{Q} = 3bH \rho$ indicates an energy exchange from DE to DM, where $b > 0$ is the strength of the energy transfer. We introduced a set of dimensionless phase space variables. These phase space variables allow us to reframe the dynamics of the given cosmological system into an autonomous system. Using these phase-space variables, we obtained an autonomous two-dimensional nonlinear system. Furthermore, by putting $x'=0$ and $y'=0$, we obtained a single critical point of the system $(x_c,y_c)=(0,0)$. In order to probe the behavior of this critical point, we linearize the autonomous system by considering small perturbations near the critical points as $(x,y) \longrightarrow (x_c+\delta x, y_c+\delta y)$, and then investigate that these perturbations $(\delta x, \delta y)$ grow or decay, via a linear system of autonomous differential equations obtained by using the Taylor series expansion with the variable $(\delta x, \delta y)$ around the critical point $(x_c,y_c)$ and neglecting the tiny quadratic expressions. We calculated the eigenvalues of the Jacobian matrix as $\lambda_1 = -\frac{3}{4} (\alpha +1) $ and $\lambda_2 = \frac{3}{2} (-\alpha +6 b-1) $. We found that the critical point $(x_c,y_c)=(0,0)$ exhibits a stable behavior for the parameter constraints $\alpha > -1$ and $b < \frac{\alpha+1}{6}$. We choose the parameter values as $\alpha=-0.95$, $b=0.001 < \frac{\alpha+1}{6} = 0.0083$, and $\beta=-1$, and then present the corresponding phase space portrait. We obtained the phase space trajectories converges to the critical point $(x_c,y_c)=(0,0)$, having $q=-0.9625$ and $\omega =-0.975$. Thus, the critical points $(x_c,y_c)=(0,0)$ indicate a stable behavior of the given physical system. Further, we predicted that the DE density dominates, whereas the matter energy density vanishes with expansion of the Universe in the far future. Lastly, we conclude that the considered cosmological scenario successfully describes the evolutionary phase of the Universe from a decelerated matter era to the accelerated expansion epoch in a stable way.

\section{Future perspectives}

The present investigation is completely focused on the physical capabilities of the non-metricity and torsion-based modified gravity to reconstruct the late-time scenario. Numerous investigations along with the present thesis have shown that this modified theory can be efficient in describing the accelerating phase by bypassing the need of $\Lambda$, as well as in resolving the tension of $H_0$. From a future perspective, it would be interesting to study the early Universe behavior of this modified gravity, particularly the Big Bang nucleosynthesis constraints as well as the formation of large-scale structures.






\addtocontents{toc}{\vspace{2em}} 

\backmatter


\label{References}
\lhead{\emph{References}}

\cleardoublepage
\pagestyle{fancy}

\label{Publications}
\lhead{\emph{List of publications}}

\chapter{List of publications}
\section*{Thesis publications}
\begin{enumerate}

\item Simran Arora, \textbf{Aaqid Bhat},  P.K. Sahoo, \textit{Squared Torsion $f(T,\mathcal{T})$ Gravity and its Cosmological Implications}, \textcolor{blue}{Fortschritte der Physik} \textbf{71}, 2200162 (2023).

\item \textbf{Aaqid Bhat},  Sanjay Mandal,  P.K. Sahoo, \textit{Slow roll-inflation in $f(T,\mathcal{T})$ modified gravity}, \textcolor{blue}{Chinese Physics C} \textbf{47}, 125104 (2023).

\item \textbf{Aaqid Bhat}, Raja Solanki, P.K. Sahoo, \textit{Extended Bose-Einstein condensate dark matter in $f(Q)$ gravity}, \textcolor{blue}{General Relativity and Gravitation} \textbf{56}, 63 (2024).

\item \textbf{Aaqid Bhat},  P.K. Sahoo, \textit{Cosmology in energy-momentum squared symmetric teleparallel gravity}, \textcolor{blue}{Physics Letters B} \textbf{858}, 139608 (2024).

\item Purnima Zala,  Chetan Likhar, \textbf{Aaqid Bhat}, and  P.K. Sahoo, \textit{Stability analysis of $f(Q)$ cosmological model along with EBEC dark matter equation of state}, \textcolor{blue}{Physics of the Dark Universe} \textbf{47}, 101772 (2025).

\end{enumerate}
\section*{Other publications}
\begin{enumerate}

\item Raja Solanki, \textbf{Aaqid Bhat}, P.K. Sahoo, \textit{Bulk viscous cosmological model in $f(T,\mathcal{T})$ modified gravity}, \textcolor{blue}{Astroparticle Physics} \textbf{163}, 103013 (2024).

\item Sai Swagat Mishra, \textbf{Aaqid Bhat}, P.K. Sahoo, \textit{Probing baryogenesis in f(Q) gravity}{Europhysics Letters} \textbf{146}, 29001 (2024).

\end{enumerate}
\cleardoublepage
\pagestyle{fancy}

\lhead{\emph{Paper presented at conferences}}

\chapter{Paper presented at conferences}
\label{Paper presented at conferences}

\begin{enumerate}
\item Presented research paper entitled “\textit{Extended Bose-Einstein condensate dark matter in $f(Q)$ gravity}" at the conference “\textbf{31st International Conference of
International Academy of Physical Sciences (CONIAPS XXXI) On Emerging Trends in Physical Sciences}” organized by Organized by School of Studies in Chemistry and School of Studies in Physics and Astrophysics Pt. Ravishankar Shukla University, Raipur (C.G.), during the period $20^{th}-21^{th}$ December, $2024$.

\item Presented research paper entitled “\textit{Probing baryogenesis in f(Q) gravity}" at the conference “\textbf{International Conference on Recent Developments in Research}” organized by Department of Mathematics and B.Ed. Tamralipta Mahavidyalaya, West Bengal during the period $18^{th}-19^{th}$ March, $2025$.

\end{enumerate}

\cleardoublepage
\pagestyle{fancy}
\lhead{\emph{Biography}}

\chapter{Biography}

\section*{Brief biography of the candidate:}
\textbf{Mr. Aaqid Mohi ud din Bhat} obtained his bachelor's degree in computer science in 2016 and a Master's degree in mathematics from the University of Delhi, Delhi, in 2019. During his Masters, he qualified for GATE in 2019 with 98.42 percentile. He was working as an JRF at the National Institute of Technology Srinagar, Jammu and Kashmir, from January 2020 to April 2022. He qualified for the Council of Scientific and Industrial Research (CSIR), the National Eligibility Test (NET), for the Lectureship (LS), in 2021. He has published $10$ research articles in renowned international journals during his Ph.D. research career. He has presented his research at several National and International conferences.

\section*{Brief biography of the supervisor:}

\textbf{Prof. P.K. Sahoo} received his Ph.D. degree from Sambalpur University, Odisha, India, in 2004. He is currently serving as a Professor in the Department of Mathematics, Birla Institute of Technology and Science-Pilani, Hyderabad Campus. He also holds the position of Head of the Department during the period from October 2020 to September 2024. He has conducted several academic and scientific events in the department, including the 89th Annual Conference of the Indian Mathematical Society in 2023. He contributed to BITS through five sponsored research projects from the University Grants Commission (UGC 2012-2014), DAAD Research Internships in Science and Engineering (RISE) Worldwide (2018, 2019, 2023 and 2024), Council of Scientific and Industrial Research (CSIR 2019-2022), National Board for Higher Mathematics (NBHM 2022-2025), Science and Engineering Research Board (SERB), Department of Science and Technology (DST 2023-2026). He is an expert reviewer of Physical Science Projects, SERB, DST, Govt. of India, and University Grants Commission (UGC) research schemes. 

\textbf{Prof. Sahoo} has published more than 250 research articles in various renowned national and international journals. He has participated in many international and national conferences, most of which he has presented his work as an invited speaker. He has collaborated in various research projects at the national and international levels. He has been placed among the top $2\%$ scientists of the world according to the survey by researchers from Stanford University, USA, in Nuclear and Particle Physics for five consecutive years.   He serves as an expert reviewer and editorial member for a number of reputed scientific journals and is also a Ph.D. examiner at several universities. He has been awarded a visiting professor fellowship at the Transilvania University of Brasov, Romania. He is also the recipient of the Science Academics Summer Research Fellowship, a UGC Visiting Fellow, a Fellow of the Institute of Mathematics and its Applications (FIMA), London, and a Fellow of the Royal Astronomical Society (FRAS), London, elected as a foreign member of the Russian Gravitational Society. Prof. Sahoo is also a COST (CA21136) member: Addressing observational tensions in cosmology with systematics and fundamental physics. As a visiting scientist, he had the opportunity to visit the European Organization for Nuclear Research (CERN), Geneva, Switzerland, a well-known research center for scientific research.

\end{document}